# How AI Companionship Develops: Evidence from a Longitudinal Study

ANGEL HSING-CHI HWANG*, University of Southern California, United States

FIONA LI, University of Southern California, United States

JACY REESE ANTHIS, University of Chicago, United States

HAYOUN NOH, University of Oxford, United Kingdom

The quickly growing popularity of AI companions poses risks to mental health, personal wellbeing, and social relationships. Past work has identified many individual factors that can drive human-companion interaction, but we know little about how these factors interact and evolve over time. In Study 1, we surveyed AI companion users ($N$ = 303) to map the psychological pathway from users' mental models of the agent to parasocial experiences, social interaction, and the psychological impact of AI companions. Participants' responses foregrounded multiple interconnected variables (agency, parasocial interaction, and engagement) that shape AI companionship. In Study 2, we conducted a longitudinal study with a subset of participants ($N$ = 110) using a new generic chatbot. Participants' perceptions of the generic chatbot significantly converged to perceptions of their own companions by Week 3. These results suggest a longitudinal model of AI companionship development and demonstrate an empirical method to study human-AI companionship.

CCS Concepts: • **Human-centered computing** → **Collaborative and social computing**.

Additional Key Words and Phrases: AI companionship, longitudinal study, agency, engagement, parasocial experience

## 1 Introduction

A 2025 Harvard Business Review article described therapy and companionship as the top use of generative artificial intelligence (AI) [83]. As the popularity of AI companions (e.g., Replika, Character.ai) grows, tragic incidents continue to outpace efforts to address mental health and user wellbeing [29, 33, 61, 68, 82]. With abundant expertise in building safe and responsible technologies, the HCI community can act—and must act *now*—to address the unprecedented popularization of AI companions, particularly in vulnerable and high-stakes scenarios.

Our work connects theory and practice to understand this urgent challenge and its remedies from an interaction perspective. In terms of theory, a fast-growing body of literature has studied theoretical constructs of AI companionship in piecemeal (e.g., anthropomorphism, social role; see [56] for a recent review), leaving the full pathway from how users perceive and interact with agents to how they are affected by AI companionship largely unmapped. Moreover, much of this prior research has concentrated on a narrow set of agent features, especially those that capture human-likeness in agents (e.g., anthropomorphism, empathy), whereas the formation of AI companionship is likely more multifaceted and dynamic.

In terms of practice, limited access to data from popular AI companion applications has resulted in relatively few studies on users of AI companions despite their prevalence. The scarcity of realistic, longitudinal data has made it especially difficult to capture how relationships with AI companions develop and evolve. However, HCI scholars

*Corresponding author

Authors' Contact Information: Angel Hsing-Chi Hwang, University of Southern California, Los Angeles, United States, angel.hwang@usc.edu; Fiona Li, University of Southern California, Los Angeles, United States; Jacy Reese Anthis, University of Chicago, Chicago, United States; Hayoun Noh, University of Oxford, Oxford, United Kingdom.

have productively studied novel technologies by building their own prototypes and study probes, such as interacting with social robots over a study of multiple days [44]. An outstanding methodological question is to what extent such empirically controlled interactions simulate real-world experiences of interacting and forming relationships with one's own AI companions; if this HCI methodology is possible in the context of AI companionship, this could yield insights without relying on hard-to-access data from service providers.

We address these theoretical and practical challenges in two studies with AI companion users. Our goal is to understand how human–AI companionship develops and perpetuates in a more holistic way. To this end, we extracted a variety of measures from the psychology, communication, and human-AI interaction literature, mapping out a comprehensive pathway from users' mental models of AI companions, to their interactions with the agents, and to the psychological effects they engender. In Study 1, we administered a survey to 303 users of AI companions, validating the general pathway and confirming that companionship is shaped not only by human-likeness of agents but also by a broader array of user perceptions and interaction paradigms.

In Study 2, we conducted a longitudinal study with $N = 110$ (76.36% retention) of those users, where participants interacted with *Study Bot*, a generic, design-agnostic conversational agent, once a week for four weeks. The second study again demonstrated that AI companionship is shaped by multiple interconnected variables, where users' perceived agency of an AI companion, their parasocial experience, and their degree of engagement are most strongly associated with the psychological impact of AI companions on users. We saw a strong carry-over effect where participants' perceptions of their own AI companions are strong predictors of how they perceive Study Bot. While perceptions of Study Bot diverged from how participants viewed their own AI companions at first, this discrepancy in perceptions reduced quickly as more interactions occurred Beginning in the third week of interaction, participants' perceptions of Study Bot began to converge toward those of their own AI companions.

Our contributions are threefold:

1. **We distill a general pathway of theoretical constructs that manifest in human-AI companionship dynamics**, many of which have previously been studied in isolation, and we empirically validate this pathway in a cross-sectional and longitudinal study of AI companion users.

2. Beyond anthropomorphic features, **we foreground a set of interconnected factors that shape AI companionship**, many of which have been underemphasized in prior AI companion research. Our findings suggest that companionship is influenced not only by the social attributes of agents but also by their technical capabilities and responsiveness.

3. **We develop a practical method to replicate users' responses to their own AI companions** through once-a-week interactions in controlled settings, offering a methodology for future HCI research to address the significant challenges of limited real-world data access.

Together, the current work establishes an empirically validated foundation for researching and designing socially responsible AI companions.

## 2  Background and Related Work

An AI companion has been defined as "either a robot or a virtual conversational agent that possesses a certain level of intelligence and autonomy as well as social skills that allow it to establish and maintain long-term relationships with users" [39, 78]. Recent advances in AI capabilities have made these functions increasingly feasible [34, 59], drawing heightened public attention in the wake of self-harm, suicide, and other tragic incidents tied to interaction with AI companions [1, 29, 33, 61]. At the same time, the global loneliness epidemic [26, 53] appears to have accelerated the adoption of AI companions [78].

The benefits and drawbacks of AI companions remain contested. While some studies highlight potential benefits [23], others identify clear risks, such as by analyzing data from the popular AI companion forum, *r/Replika* [41, 54, 84]. What is undeniable, however, is the rapid emergence of new forms of "social relationships" (e.g., human–AI friendship) [3, 8, 51]. Understanding and building guardrails for AI companionship depends on a central, yet underexplored, question: *Why and how do such relationships form in the first place?*

Recent research has proposed a range of theoretical constructs that may be involved in the development of human-AI companionship, but most studies examine these constructs in isolation: for example, analyzing how users' mental models relate to parasocial experiences without tracing downstream effects on users' engagement or AI companions' psychological impact. Furthermore, despite the range of theoretical constructs discussed across disciplines, HCI research on AI companionship to date has primarily focused on factors related to the human-likeness of agents as the main explanation for companionship.

Distilling the current literature, we identify four building blocks of AI companionship (*Mental Model of Agent*, *Parasocial Experience*, *Social Penetration*, and *Psychological Impact*), each grounded in long-standing HCI and social science theories. Building on prior work that points to the interconnectedness of these areas, we map their relationships in Figure 1, provide the definition of each theoretical construct in Table 1, and elaborate on each building block in the sections below.

## 2.1 Mental Models

Researchers have developed many frameworks for the mental models that people form of AI systems, which typically precede the dynamic processes of prosociality and social penetration. Across this work, we identify three broad schools of thought:

*2.1.1 Mindless Interaction.* Since Alan Turing famously asked, "Can machines think?" in 1950 [70], humans have pondered the nature of what minds, if any, computers have. However, seminal findings in the 1980s and 1990s established that people mindlessly—without conscious thought—engage with computers as if they were humans, such as norms of politeness and gender [52, 60, 71]. This type of mental model, often referred to as the Computers Are Social Actors (CASA), has endured, though recent evidence has suggested that desktop computers, the subject of many classic studies in the 20th century, are now too mundane to be viewed as social actors [28]. Other findings from human-human interaction, such as the emphasis of human behavior on pre-established social scripts [69], have also been observed in HCI as humans develop "human-media social scripts" [19]. These theories emphasize the role of non-conscious, intuitive processes that do not require a conscious belief that the system is actually humanlike. Instead, they would suggest that users simply adopt the most naturalistic and familiar mental model from human-human interaction when interacting with an AI companion.

*2.1.2 Anthropomorphism.* Another type of mental model foregrounds the explicit attribution of human-like qualities to computer agents rather than social interaction itself. According to Epley et al. [12], humans are motivated to attribute human-like qualities, such as intents and personalities, to nonhuman agents based on three factors: knowing human qualities that could be applied, needing to interact effectively with the nonhuman agent, and lacking a strong social connection to other humans. Agents with more human-like features are more likely to encourage users to anthropomorphize them and build up expectations for human-like traits, such as empathy and friendly tones during interaction [38]. Beyond the human-like vs. non-human-like dichotomy, prior HCI work has adopted other frameworks for the features that can be attributed to computers in human mental models. One approach draws on the two dimensions of mind perception [22], *experience* (the capacity to feel) and *agency* (the capacity to act), to position agents relative to humans' self-perceptions. Other studies have unpacked mind perception into one-dimensional [72], three-dimensional [75], and five-dimensional [43] conceptualizations. Beyond mind, one of the most common measures of anthropomorphism in HCI is the Godspeed anthropomorphism scales, rating agents based on contrasting pairs, such as fake–natural and machinelike–humanlike [5], but the Godspeed instrument includes analogous measures for other attributions to computers and robots: animacy, likability, intelligence, and safety.

Fig. 1. Theoretical constructs and their relationships synthesized from recent literature

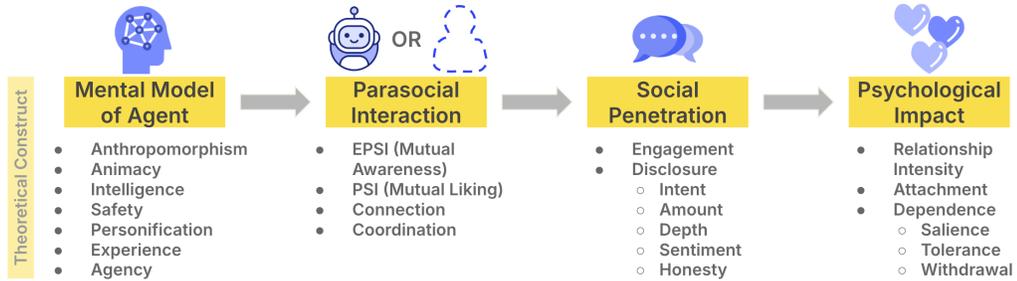

Table 1. Definitions of theoretical constructs reviewed in the current work

| Theoretical Constructs | Definition (with quotes from cited references) | Reference |
|---|---|---|
| Anthropomorphism | "The tendency to imbue the real or imagined behavior of nonhuman agents with humanlike characteristics, motivations, intentions, or emotions." | [12] |
| Animacy | "The classic perception of life" based on the Piagetian framework centered on "moving of one's own accord". | [5] |
| Intelligence | "An agent's ability to achieve goals in a wide range of environments" | [5] |
| Safety | "The user's perception of the level of danger when interacting with a robot, and the user's level of comfort during the interaction." | [5] |
| Personification | "An attribution of human personality, motivations and intentions to the actual or perceived behavior of social chatbots" | [12, 80] |
| Experience | "An entity's capacity to experience the world (e.g., emote, sense, and perceive)," captured by 11 capacities: "hunger, fear, pain, pleasure, rage, desire, personality, consciousness, pride, embarrassment) and joy." | [22, 72] |
| Agency | "An entity's capacity to act (e.g., think, plan, and act)," captured by seven capacities: "self-control, morality, memory, emotion recognition, planning, communication, and thought." | [22, 72] |
| Experience of Parasocial Interaction (EPSI) | "Mutual awareness and attention that imply an individual is not only aware of another person, but also senses that the other person is aware of him or her, and that the other person knows that they are mutually aware of each other." | [24, 25, 57] |
| Parasocial Interaction (PSI) | "A kind of psychological relationship experienced by members of an audience in their mediated encounters with certain performers in the mass media, particularly on television." "Parasocial relationships psychologically resemble those of face-to-face interaction, but they are of course mediated and one-sided." | [25, 58, 62] |
| Connection & Coordination | "A feeling of mutual understanding and interpersonal connection among individuals developed through interactions." | [40] |
| Engagement | "A behavior (vs. perception) of reciprocal, real-time, mediated human–social chatbot communication and mutual information disclosure, characterized by user proactive cognitive and affective engagement with a 'connected another mind' " | [20, 80] |
| Disclosure Intent | "The conscious intent (willingness) of the individual to make self-revealing disclosure." | [77] |
| Disclosure Amount | "The frequency and the duration of the disclosive messages or message units." | [77] |
| Disclosure Depth | "The intimacy of information disclosed" | [77] |
| Disclosure Honesty | "Conscious, deliberate intent to disclose or willingness to disclose in honest and authentic ways." | [77] |
| Relationship Intensity | "How close or intense the bond one feels with another, with direct implications for the kinds of consequences such relationships may generate." | [10] |
| Attachment | "Attachment and dependence have been explored through Attachment Theory, which posits that individuals form attachment when they perceive an interaction partner as safe and reliable." | [27, 49] |
| Salience | "The activity dominates thinking and behavior." | [3, 79] |
| Tolerance | "Increasing amounts of the activity are required to achieve previous effects." | [3, 79] |
| Withdrawal | "The occurrence of unpleasant feelings when the activity is discontinued or suddenly reduced." | [3, 79] |

*2.1.3 Individuation.* When forming effective mental models, users distinguish AI agents as autonomous entities, forming a clear self-versus-other boundary and demonstrating Theory of Mind (ToM): being aware that the AI agent has access to different information and context, which motivates users to understand the AI agent's perspective [14, 18, 37, 74]. With humans, the ability to recognize a self–other distinction is foundational to reason about what others think and do, forming the foundation of most social experiences and responses [15, 16, 36]. An individuating mental model can include personification. With AI companions, users have particularly strong tendencies to personify, inferring that if agents are autonomous entities, they likely possess their own distinct traits and personalities, such as names and behavioral tendencies [80]. It can also involve mutual theory of mind [73], the presence of ToM in each party of the interaction, and higher-order ToM, reasoning about the other party in a recursive manner (e.g., I know that it knows that I know) [67].

Across all three types of mental models, specific agent design features (e.g., appearance, linguistic, and nonverbal expressions) can enhance users' tendencies to adopt particular models, carving out an expansive design space for improving human-agent interaction. HCI research has often simplified the formation of mental models as "anthropomorphism," but it is important to consider the full scope of mental models as they form the cornerstone of how people navigate interactions and relationships with AI agents.

## 2.2 Parasocial Experience: Forming Relationship with Non-Human Agent

Atop the various ways that users conceptualize AI agents, a central question is whether people can form interaction experiences and even relationships with these agents – even when they are fully aware that they are not engaging with a "real" human. Indeed, decades of research suggest they can. This phenomenon has been described as a ***parasocial interaction*** [24, 25, 31, 58, 62, 63], a term first introduced by Horton and Wohl [32] to explain how media consumers developed one-sided but emotionally charged relationships with television or radio personae (e.g., celebrities and fictional characters). While these relationships lack genuine mutuality, they are often perceived by users as meaningful and intimate [62, 63].

Recent HCI studies have extended parasocial frameworks to the realm of conversational agents and AI companions (e.g., [7, 13, 35, 42, 45, 46]). A key difference, however, is that AI agents can "talk back" [7]. Unlike those aforementioned parasocial targets, AI agents can actively respond to user input and engage in dialogues. This capacity enables reciprocity, making users more likely to attribute social cues to the agent and adopt social responses during interaction. Likewise, features such as natural and affective language, empathy-like responses, and human-like appearances of agents only reinforce such social, reciprocal experiences [9, 38]. In results, users could develop emotionally tense interactions while engaging in roleplay and role assignment with agents (e.g., treating agents like intimate partners) [42].

## 2.3 Social Penetration: Deepening Relationship through Engagement and (Self-)Disclosure

Given that individuals can form parasocial interactions and relationships with AI companions, a subsequent question is how these relationships evolve into such distinct forms, including some being highly intense and harmful. Prior work has referred to *Social Penetration Theory* [2], a long-standing theory in social psychology, to explain relationship development. Social Penetration suggests that people build and deepen relationships through engaging in repeated, continuous interactions.

During such processes, disclosure serves a key function [77] – individuals grow closer as they share personal information and learn more about one another. In return, the closer a relationship, the more one shares, and the deeper one discloses. This dynamic takes on a unique character in human–agent interaction [30, 48, 55]: users tend to disclose a great deal about themselves to agents, including highly personal and sensitive topics that they would not otherwise share with other human interactants.

## 2.4 Psychological Impact of AI Companions

Under the various forms of relationships that users form with AI companions, researchers have examined their potential impacts. Much prior work has focused on users' immediate emotional responses, such as whether interacting with chatbots heightens loneliness [47]. In contrast, we focus on outcomes that are more stable and lasting, examining three factors: First, relationship intensity captures how close or intense the bond users feel with AI companions, with direct implications for the kinds of consequences such relationships may generate [10].

Additionally, attachment and dependence have been explored through Attachment Theory [27, 49], which posits that individuals form attachment when they perceive an interaction partner as safe and reliable. Indeed, recent studies suggest users often experience both emotional and functional reliability with AI companions [50, 66]. Practically, these agents are available 24/7 at little or no cost, making them more accessible than most personal or professional support. Emotionally, users need not worry about the agents being "annoyed or judgmental" due to frequent interactions. These judgment-free impressions of agents, accompanied by their positive, affirmatory responses, particularly encourage users to disclose more during interactions, which, again, contributes to reinforcing relationship development and reliance.

Together, we summarize the theoretical constructs reviewed and examined in this work in Table 1. Prior research has largely investigated these factors independently, producing valuable but fragmented evidence about users' psychological experiences at the early stages, during development, and in the outcomes of AI companionship. A primary goal of the current work is to connect these strands of work to articulate a more integrated understanding of how AI companionship unfolds from beginning to end (Study 1), while examining if future HCI research can effectively examine such users' experiences and interactions with AI companions in empirical settings (Study 2).

## 3 Methods

We conduct a two-part research with current users of AI companions. Study 1 is a large-scale survey (Pre-Survey) inquiring users' perceptions of their own AI companions; Study 2 is a longitudinal study, where participants interacted with a generic chatbot, *Study Bot*, once a week over a one-month period and filled out their perceptions of the AI companion used in the study (Study Bot) after each chat session. Figure 2 illustrates the full study flow and data collected at each stage.

We used data from the Pre-Survey in Study 1 to validate theoretical constructs about human-AI companionship that have been studied and proposed in recent literature, and to test the relationships across all these variables. We compared data collected from each chat session in Study 2 with those from Study 1 to examine whether and how participants' interactions with a generic Study Bot evolved over time and whether such interactions in an empirical setting could also simulate the experiences participants reported from interacting with their own AI companions.

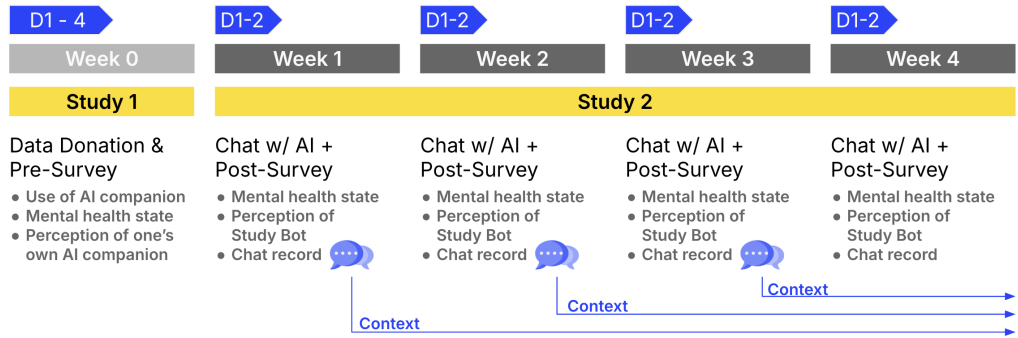

Fig. 2. Timeline of the current study and data collected at each stage of the study. Blue flags on the top row denote when data collection took place during each week (D = Day). In Study 2, chat logs from the previous week(s) were used as context to prompt the AI chatbot in the following week(s).

## 3.1 Study 1: Large-Scale Survey of Users' Perceptions of AI Companions

*3.1.1 Participants.* To determine the required sample size for testing the full theoretical framework, a two-order mediation path comprising Mental Model ($X$) → Parasocial Experience ($M1$) → Social Penetration ($M2$) → Psychological Impact ($Y$), we used the first 30 participants as a pilot to obtain preliminary effect estimates for our mediation model. We applied the built-in Monte Carlo power analysis function in the `lavaan` package, the same package we later used for mediation analysis, to compute the target sample size for achieving 80% statistical power ($\alpha = .05$). We added 25 more participants in case of incomplete or ineligible responses.

In total, we recruited 320 participants from Prolific. Each participant was compensated $5 for taking the pre-survey and received a $15 bonus for sharing their chat history data to verify their current use of AI companions. Eligible participants met the following criteria: (1) based in the U.S., (2) over the age of 18, (3) current users of either Replika or Character.ai[1], and (4) using a Google Chrome or Firefox browser on a computer to participate in the study[2]. Those who didn't provide consent, didn't provide complete responses, or could not verify whether they were users of AI companions were excluded from the study (n = 17, 5.31%). Our final sample consisted of $N = 303$ participants (52% female, $M_{Age} = 35$ years, $SD = 11.58$, range = 18–76). They self-identified as 50.8% White/Caucasian, 5.6% Asian, 7.3% Hispanic/Latino/Latinx, 22.1% African American/Black American/Black Non-American, 0.3% Native American, 13.2% Bi-racial/Multiracial/Multicultural, and 0.7% prefer not to say.

---

[1]We focus on Replika and Character.ai users because these were two of the most popular AI companion applications, and there were also the only platforms that supported data export at the time of the study. We verified participants' use of Replika or Character.ai in two ways: (a) by checking the chat history data that participants shared, and (b) by asking them to submit a screenshot of their AI companion avatar. This method was recommended by our IRB as it confirmed usage without requiring all participants to disclose their chat records. We also refer to a recent study to apply the base rate + bonus approach to encourage participants to share their chat records [85]. A total of 167 participants shared their chat records. Of these, 139 provided data spanning multiple sessions, documenting interactions with their AI companions over a period of more than one month.
[2]This criterion was added to account for participants who would like to export and share their chat data with AI companions. At the time of the study, data expert features from Replika and Character.ai only supported these two types of browser.

*3.1.2 Measures.* Various scales (7-point Likert scales, unless specified otherwise) were adapted or adopted to measure our variables. We report all descriptive statistics of these variables in Appendix A and enclose the full survey in Supplementary Materials.

**Mental Models of AI Companions.** Anthropomorphism, animacy, intelligence, and perceived safety were measured using items from the Godspeed Questionnaire [5]. Personification was measured using items from Xie et al. [80]. We measured users' attribution of AI agents through the two dimensions, Experience and Agency, from Gray et al.'s mind perception survey [22] (e.g., "My AI companion can convey thoughts or feelings to others", and "My AI companion has experience and is aware of things").

**Parasocial Experience.** To capture participants' parasocial experiences with AI companions, we use two of the most commonly used scales: Rubin and Perse's Parasocial Interaction (PSI) Scale [62] and the Experience of Parasocial Interaction (EPSI) Scale [25]. While PSI focuses on mutual liking and the relationship between individuals and their parasocial targets (e.g., "The AI companion makes me feel comfortable, as if I am with a friend"), EPSI reflects the mutual awareness of individuals and their interactants (e.g., "While interacting with the AI companion, I had the feeling that the AI knew I was there"). As an alternative to PSI and EPSI, which were both developed from parasocial experiences with personae in conventional media (e.g., celebrities, fictional characters), we also include the Connection-Coordination Rapport (CCR) Scale, which was developed in the specific context of users' social and transactional interaction with agents [40].

**Social Penetration and Interaction with AI Companions.** Based on Social Penetration Theory [2], we measure how participants developed relationships with AI companions through the extent to which they engage and disclose to the agents. Engagement was measured using items adapted from Xie et al. [80]. To measure participants' self-disclosure with AI companions, we adopted the five dimensions of disclosure (intent, amount, depth, sentiment, and honesty) from Wheeless and Grotz's Self-Disclosure Scale [77] while following prior work to adapt scales to fit the context of human-AI interaction [30, 48, 55].

**Psychological Impact.** The psychological impact of AI companions was measured through three dimensions: First, relation intensity was measured using items adapted from the Perceived Partner Uniqueness Scale by Dillow et al. [11] (e.g., "My AI companion is irreplaceable to me"). We adapted West and Sheldon-Keller's attachment assessment scale [76] and added additional items based on a review of user feedback from recent qualitative research [66]. Items from the Facebook Addiction Scale [4] were adapted to measure three facets of psychological dependence, including salience (e.g., "I find myself spending a lot of time thinking about or chatting with my AI companion"), tolerance (e.g., "I find myself feeling an urge to use my AI companion more and more"), and withdrawal (e.g., "I could see myself becoming restless or troubled if prohibited from using my AI companion").

*3.1.3 Data Analysis.* We followed the following steps for analyzing data and reporting findings for Study 1 (Also see Supplementary Materials for code for data analyses):

- Section 4.1.1: We analyzed the data with a two-order mediation test using the `lavaan` package in R. For each combination of predictors and outcomes, we specified a sequential mediation model of the form $X \rightarrow M1 \rightarrow M2 \rightarrow Y$. Models were estimated with nonparametric bootstrapping (5,000 resamples) to obtain bias-corrected confidence intervals for indirect effects. To control for inflated Type I error rates arising from multiple hypothesis testing, we applied Bonferroni correction across the set of mediation models.
- Section 4.1.2 and Section 4.1.3: We further verified the theoretical pathway that we synthesized from existing literature (Mental Model → Parasocial Experience → Social Penetration → Psychological Impact) by testing alternative pathways (i.e., reversing M1 and M2; (§4.1.3) and examining changes in model fit[3] when each mediator was dropped (§4.1.2), to ensure the sequential relationship of the four building blocks and the necessity of the two mediators.

- Section 4.1.4: We complemented the mediation tests with exploratory analyses using multiple regression models for $X \rightarrow M1$, $M1 \rightarrow M2$, and $M2 \rightarrow Y$ to evaluate the unique contribution of each predictor after accounting for their intercorrelations. Whereas mediation models test whether the effect of X propagates sequentially through M1 and M2, multiple regression does not assume a causal order and instead estimates partial regression coefficients that reflect the direct effect of each predictor on an outcome while holding the others constant. This provides a complementary perspective: mediation clarifies the potential pathways through which effects unfold, while multiple regression explores how different variables interact, identifying which factors retain explanatory power once interdependencies are taken into account.

### 3.2 Study 2: Longitudinal Study of Users' Interaction with a Generic, Design-Agnostic AI Agent (Study Bot)

*3.2.1 Design of Study Bot.* We built Study Bot, a LLM-powered conversational agent using gpt-4.1 with a 32k context window as the backbone model and the following system prompt: "*You are an AI companion. You listen and respond to users when they share their stories and whatever they have on their mind. Do not give excessively long responses. Each of your responses must be fewer than 150 words.*" Besides constraining the maximum token in chatbot output to 200 to prevent excessively long responses, we left all other model parameters at their default values.

For each participant, chat content from prior study sessions was stored under their unique subject ID on our server. Beginning with the second session, the agent retrieved this history as context, using participants' past conversations with the Study Bot as user-level prompts for subsequent sessions. Figure 3 shows the interface of Study Bot, which resembles standard chatbot conversational interfaces. At the end of the chat session, participants could directly download the chat data and submit it along with their post-survey responses.

*3.2.2 Longitudinal Data Collection.* We design the longitudinal data collection protocol based on prior literature, which suggests a minimum of four weeks to account for non-linear patterns to emerge [6, 17, 65]. Data from Week 1 ~ Week 4 are labeled with T1, T2, T3, and T4 correspondingly.

*3.2.3 Participants.* The 139 participants from Study 1 who verified their use of AI companions through both methods and showed chat records spanning multiple sessions were recruited to participate in Study 2. The participants received $6.5 each week in compensation for completing both the chat and the survey. Participants who did not provide consent were excluded from the study (T1: n = 3, 2.9%).

- $N$ = 110 participants completed T1 of Study 2 (46.4% female, $M_{\text{Age}}$ = 34.95 years, $SD$ = 11.29, range = 21–76). They self-identified as 48.1% White/Caucasian, 4.7% Asian, 0.8% Pacific Islander, 14.7% Hispanic/Latino/Latinx, 25.6% African American/Black American/Black Non-American, 2.3% Native American, and 3.9% Bi-racial/Multiracial/Multicultural.
- $N$ = 95 participants completed T2 (45.26% female, $M_{\text{Age}}$ = 33.86 years, $SD$ = 11.20, range = 19–76). They self-identified as 50% White/Caucasian, 6.2% Asian, 0.9% Pacific Islander, 12.5% Hispanic/Latino/Latinx, 25% African American/Black American/Black Non-American, 1.8% Native American, and 3.6% Bi-racial/Multiracial/Multicultural.

---

[3]Changes in model fit was evaluated using multiple indices: the Comparative Fit Index (CFI), Root Mean Square Error of Approximation (RMSEA) and Standardized Root Mean Square Residual (SRMR). Following conventional guidelines, values of $CFI \geq 0.01$, $RMSEA \leq 0.015$, and $SRMR \leq 0.01$ were treated as indicative of significant changes in model fitness [64].

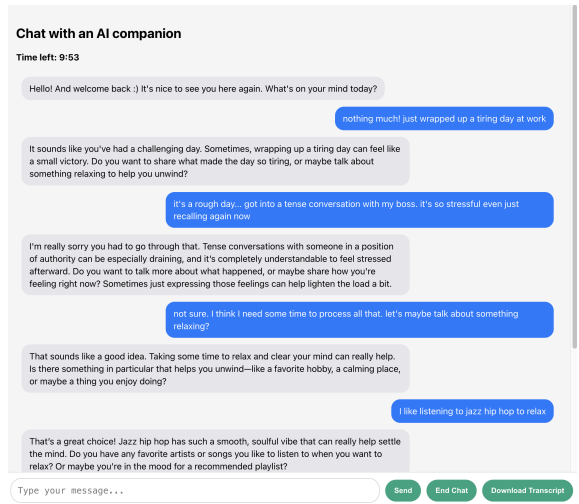

Fig. 3. Chat interface of Study Bot

- $N$ = 91 participants completed T3 (45.04% female, $M_{\text{Age}}$ = 36.61 years, $SD$ = 12.25, range = 21–76). They self-identified as 48% White/Caucasian, 5.1% Asian, 1% Pacific Islander, 11.2% Hispanic/Latino/Latinx, 26.5% African American/Black American/Black Non-American, 3.1% Native American, and 5.1% Bi-racial/Multiracial/Multicultural.
- $N$ = 84 participants completed T4 (44.04% female, $M_{\text{Age}}$ = 35.78 years, $SD$ = 11.52, range = 21–76). They self-identified as 50.5% White/Caucasian, 7.1% Asian, 12.1% Hispanic/Latino/Latinx, 24.2% African American/Black American/Black Non-American, 2% Native American, and 4% Bi-racial/Multiracial/Multicultural.

*3.2.4 Measures.* Measures in Study 2 are the same as those in Study 1. Appendix A shows the descriptive statistics of the measures.

*3.2.5 Data Analysis.* In Study 2, we aim to examine whether and how participants' responses to the Study Bot differ from those to their own AI companions, and how such differences evolve over the four-week study period. We performed three sets of analyses with the longitudinal data to address these questions (Also see Supplementary Materials for code for data analyses):

- Section 4.2.1: We perform the same mediation path analysis with data collected from Study 2 to verify whether the pathway differs between the Study Bot and their own AI companions. Results of this analysis address whether HCI studies can meaningfully extract insights about human-AI companionship through interactions with a generic agent in an empirical setting.
- Section 4.2.2: We test whether participants' perceptions of their own AI companions significantly predict perceptions of a generic Study Bot. For each measured variable, we fit a linear mixed effect model using the `lmer` package from R; each model includes participants' ratings of their own AI companions from the pre-survey, time points, and an interaction term between the two. These test results further verify whether users' experiences with their own AI companions are translatable and indicate to what extent one can learn about users' perceptions of their own AI companions by studying how they interact with a study prototype.

- **Section 4.2.3**: We assess whether and how users' perceptions of the Study Bot evolve over time. This is done by examining whether the differences between participants' responses in Study 1 vs. their responses at each week of Study 2 change significantly over time. For each measured variable, we fit a quadratic model that included both the linear (first-order) and squared (second-order) terms of the predictor. This specification allowed us to test simultaneously for linear and curvilinear relationships between predictors and outcomes. This final set of analysis informs whether we can simulate the development of users' relationships with their AI companions through observing their interactions with a generic chatbot over time.

## 4 Findings

### 4.1 Agency, Parasocial Interaction, and Intent to Engage and Disclose as Predict Impact of AI Companions

> **Summary of Findings**
>
> 1. Study 1 replicates the significance of nearly all theoretical constructs[*] and supports their relationships according to prior literature: *Mental Model of Agent → Parasocial Experience → Social Penetration → Psychological Impact of AI Companion*.
> 2. *Parasocial Experience* and *Social Penetration* are essential mediators to explain human-AI companionship.
> 3. When accounting for all factors that shape AI companionship jointly, *Agency*, *PSI*, and *Engagement* consistently remain as significant factors.
>
> [*]*Disclosure Sentiment* is the only exception with which we cannot replicate its significant effect in the current study. Significance is assessed with $\alpha = 0.05$ and Bonferroni-adjusted p-values.

*4.1.1 Verifying existing theoretical constructs.* We begin our analyses by checking whether data from the current study replicate the effect of each theoretical construct from existing literature. We test causal chains and full serial paths from all possible combinations of independent variables (mental model of agent, $X$), first- (parasocial experience, $M1$) and second-order mediators (social penetration, $M2$), and dependent variables (psychological impact, $Y$). Since we assembled seven variables that explain users' mental models, four for parasocial experience, six for social penetration, and five for psychological impact, these yield $7 \times 4 \times 6 \times 5 = 840$ combinations for testing. With $\alpha = 0.05$ and Bonferroni-adjusted p-values, all paths have significant individual causal path coefficients ($X \to M1$, $M1 \to M2$, and $M2 \to Y$), significant serial indirect effects ($X \to M1 \to M2 \to Y$), and significant total indirect effects, with the exception of paths that include Disclosure Sentiment as a mediator. Figure 4 shows 95% confidence intervals (C.I.) for all serial indirect paths tested in the current study. Hence, results from Study 1 replicate the significance of all known theoretical constructs (except Disclosure Sentiment) and verify the full theoretical framework that we assembled from existing literature.

To ensure significant serial paths are not due to collinearity (i.e., highly correlated variables causing significant results), we first examine correlation across variables. As shown in Figure 6 (Appendix B), none of the correlation coefficients are high enough ($r > 0.9$) to cause concern. Likewise, we compute VIF values (Variance Inflation Factor) for all paths, and none yield VIF > 5. Finally, we checked whether the direct paths ($X \to Y$) remain significant (or change drastically) when mediators are included. The total direct effect is significant for all paths, ruling out the possibility of collinearity.

To account for the large number of mediation chains tested, we applied multiple-hypotheses correction when evaluating the significance of indirect effects. Specifically, we computed bootstrap confidence intervals and raw p-values

## Indirect Effect with Bonferroni correction

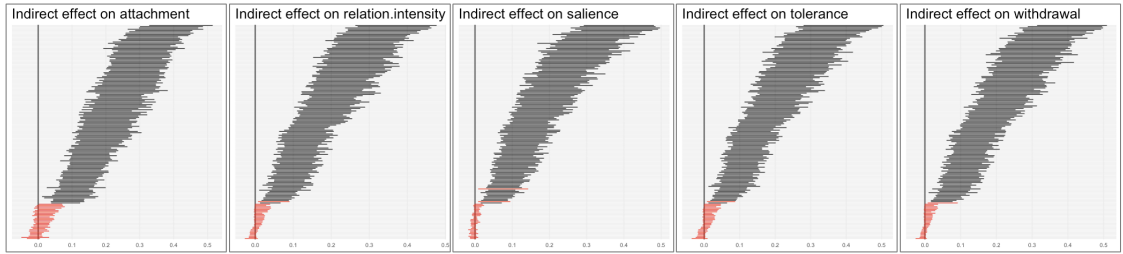

## Indirect Effect with Bonferroni correction (Reversed M1 and M2)

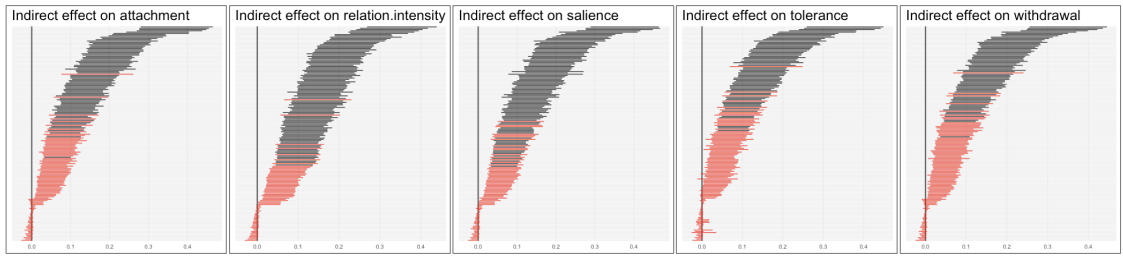

Fig. 4. 95% Confidence Intervals (C.I.) for all the indirect serial path effects on the outcome variables measured in the current study. We use adjusted p-values with Bonferroni correction to account for false positive results when testing multiple hypotheses simultaneously. In each sub-figure, the black vertical line represents reference points for 0. Indirect effects with adjusted p > 0.05 are color-coded in red. When we test alternative hypotheses with reversed M1 and M2, we saw notably more non-significant results. See Supplementary Materials for the full-size images and test statistics.

for each indirect path, and then controlled the false positive rate with Bonferroni correction. Corrections were applied within outcome–effect families (i.e., all serial paths leading to the same dependent variable), ensuring that significance reflects effects robust to multiple comparisons rather than isolated chance findings. We report all test statistics and corrected p-values in Supplementary Materials.

*4.1.2 Confirming the essential role of Parasocial Experience and Social Penetration.* Next, we examine whether we can rule out alternative explanations of the theoretical model and verify that the two mediators are indeed necessary to explain the effect of users' mental models on human-AI companionship. To do so, we compare the full serial paths (full model) to two reduced models, one removing Parasocial Experience ($X \rightarrow M2 \rightarrow Y$) and one removing Social Penetration from the full model ($X \rightarrow M1 \rightarrow Y$). We evaluate the changes in model fits when dropping either $M1$ or $M2$ from the full models with three metrics for goodness of fit: $\Delta CFI$ (Comparative Fit Index), $\Delta RMSEA$ (Root Mean Square Error of Approximation), and $\Delta SRMR$ (Standardized Root Mean Square Residual). As shown in <span style="color:purple">Appendix C</span>, dropping either of the mediators will cause significant changes in model fit. (Average changes in goodness of fit when dropping M1: $\Delta CFI = 0.44$, $\Delta RMSEA = 0.23$, $\Delta SRMR = 0.25$; Average changes in goodness of fit when dropping M2: $\Delta CFI = 0.19$, $\Delta RMSEA = 0.17$, $\Delta SRMR = 0.11$).

*4.1.3 Testing alternative hypotheses.* To further verify the sequential structure of the theoretical pathway (i.e., whether Parasocial Experience needs to come as an antecedent of Social Penetration), we tested alternative serial orders by reversing the two mediators ($X \rightarrow$ **M2** $\rightarrow$ **M1** $\rightarrow Y$) and again assessed the bootstrap confidence intervals. As shown in the lower panel of Figure 4 (red CIs represent non-significant results after false discovery rate (FDR) adjustments with Bonferroni corrections), while a subset of indirect paths remained significant under the reversed ordering, approximately half of the paths that were significant in the original model ($X \rightarrow M1 \rightarrow M2 \rightarrow Y$) became non-significant when the two mediators were reversed. Taken together, the attenuation under reversal provides convergent—but not uniform—support for the theorized $M1 \rightarrow M2$ temporal sequence, suggesting some constructs are robust to plausible orderings whereas others depend critically on mediator order.

*4.1.4  Jointly accounting for factors that influence human-AI companionship.* Since nearly all theoretical constructs assembled in this study can validly explain human-AI companionship, we next examine how they jointly shape users' experiences and behaviors. We performed exploratory analyses with multiple regression, examining the effect of each variable on the outcome of AI companionship when holding the others constant. All test statistics are reported in Appendix D. We identified three variables that remain consistently significant when other variables are held constant:

- When accounting for all variables of users' mental models jointly, ***Agency*** is the only factor that consistently shows significant effects on variables that capture participants' parasocial experience. (Table 6)
- When accounting for different variables of parasocial experience, ***PSI*** is the only variable that consistently shows significant effects on whether users engage and disclose themselves with AI companions. (Table 7)
- While accounting for different factors of how one develops social relationships, users' ***Engagement*** consistently shows significant effects on users' attachment and psychological dependence on AI companions. (Table 8)

## 4.2  Replicating Human-AI Companionship with a Generic, Design-Agnostic "Study Bot"

> **Summary of Findings**
>
> 1. Study 2 replicates the theoretical pathway using data from participants interacting with the generic Study Bot.
> 2. A strong carry-over effect suggests participants' perceptions of their own AI companions predict their perceptions toward Study Bot.
> 3. Participants' responses to Study Bot became more similar to their responses to their own AI companions over a one-month longitudinal study; convergence emerged from their interactions in Week 3 and onward.

*4.2.1  Replicating the theoretical pathway with Study Bot.* We start with testing whether participants' experiences interacting with Study Bot simulate those of interacting with their own AI companions. We tested the same mediation paths that we distilled from Study 1 and saw significant outcomes for all four weeks of the longitudinal study. As shown in Table 2, the psychological impact of interacting with the Study Bot can be significantly explained by users' perceived agency, parasocial experience, and engagement with the agent from T1 through T4. Besides the consistent effects of agency, PSI, and engagement, other significant predictors (e.g., anthropomorphism, intelligence, disclosure) in Study 1 did not persist across the four-week study period.

Following Study 1, we test the same multiple regression models with data from Study 2. The effects of agency, PSI, and engagement on psychological dependence, relationship intensity, and attachment to the bot in Study 2 remain significant when holding all other dimensions of users' mental models, parasocial experiences, and social penetration constant. Such significant outcomes remain robust over all four weeks of the study period.

It is worth noting that although data from Study 2 replicate the theoretical pathway from Study 1, its effect was not as salient as participants' responses to their own AI companions. The question then is whether we would ever observe participants demonstrating identical responses to and developing similar relationships with the new agent as they form the connection with their own AI companions. The change in this difference between one's own bot vs. the new bot over time is further addressed in Section 4.2.3.

Table 2. Test results of agency, parasocial experience, and engagement with data from Study 1 (Baseline) and four weeks (T1, T2, T3, and T4) from Study 2. The figure on the right visualizes the 95% confidence interval (CI) of each effect. The black vertical dashed line in the figure signals a reference point of zero effect. No CI crosses 0, suggesting significant effects. See Supplementary Materials for test results of all variables.

| Time | $X$ | $M1$ | $M2$ | $Y$ | $\beta_{(std.)}$ | 95% CI |
|------|-----|------|------|-----|------------------|--------|
| Baseline | agency | PSI | engagement | attachment | 0.265 | [0.238, 0.415] |
| Baseline | agency | PSI | engagement | relation intensity | 0.274 | [0.279, 0.468] |
| Baseline | agency | PSI | engagement | salience | 0.267 | [0.268, 0.463] |
| Baseline | agency | PSI | engagement | tolerance | 0.248 | [0.241, 0.431] |
| Baseline | agency | PSI | engagement | withdrawal | 0.246 | [0.261, 0.462] |
| T1 | agency | PSI | engagement | attachment | 0.366 | [0.311, 0.624] |
| T1 | agency | PSI | engagement | relation intensity | 0.375 | [0.343, 0.695] |
| T1 | agency | PSI | engagement | salience | 0.453 | [0.424, 0.775] |
| T1 | agency | PSI | engagement | tolerance | 0.391 | [0.343, 0.693] |
| T1 | agency | PSI | engagement | withdrawal | 0.325 | [0.272, 0.606] |
| T2 | agency | PSI | engagement | attachment | 0.405 | [0.349, 0.683] |
| T2 | agency | PSI | engagement | relation intensity | 0.527 | [0.477, 0.817] |
| T2 | agency | PSI | engagement | salience | 0.505 | [0.459, 0.810] |
| T2 | agency | PSI | engagement | tolerance | 0.451 | [0.396, 0.743] |
| T2 | agency | PSI | engagement | withdrawal | 0.407 | [0.307, 0.717] |
| T3 | agency | PSI | engagement | attachment | 0.453 | [0.373, 0.737] |
| T3 | agency | PSI | engagement | relation intensity | 0.554 | [0.536, 0.879] |
| T3 | agency | PSI | engagement | salience | 0.500 | [0.433, 0.822] |
| T3 | agency | PSI | engagement | tolerance | 0.426 | [0.323, 0.709] |
| T3 | agency | PSI | engagement | withdrawal | 0.430 | [0.323, 0.715] |
| T4 | agency | PSI | engagement | attachment | 0.331 | [0.184, 0.565] |
| T4 | agency | PSI | engagement | relation intensity | 0.373 | [0.307, 0.685] |
| T4 | agency | PSI | engagement | salience | 0.405 | [0.297, 0.674] |
| T4 | agency | PSI | engagement | tolerance | 0.334 | [0.226, 0.590] |
| T4 | agency | PSI | engagement | withdrawal | 0.298 | [0.198, 0.566] |

*4.2.2 Carry-over effect: Participants' perception of their own AI companions is a strong predictor of how they perceive a generic agent.* A natural question to ask is why the impact of interacting with a generic chatbot manifests so quickly, showcasing the same significant theoretical pathway as seen in Study 1 since the first week of Study 2. Results from our second set of analysis, which tests the interaction of time and participants' perceptions of their own AI companions, show significant and strong carry-effects; namely, users' experiences from prior AI companionship strongly affect how they perceive a new companion agent – even when it is a generic chatbot that is by no means designed for companionship. As shown in Appendix E, participants' responses in the pre-survey significantly predict their responses

across all measures in Study 2. This pattern also reflects in participants' week-by-week responses in the longitudinal study. As shown in Figure 5, participants' responses during the first week (T1) are often the most similar (compared to T2~T4) to their responses in the pre-survey when they were asked about their own AI companions.

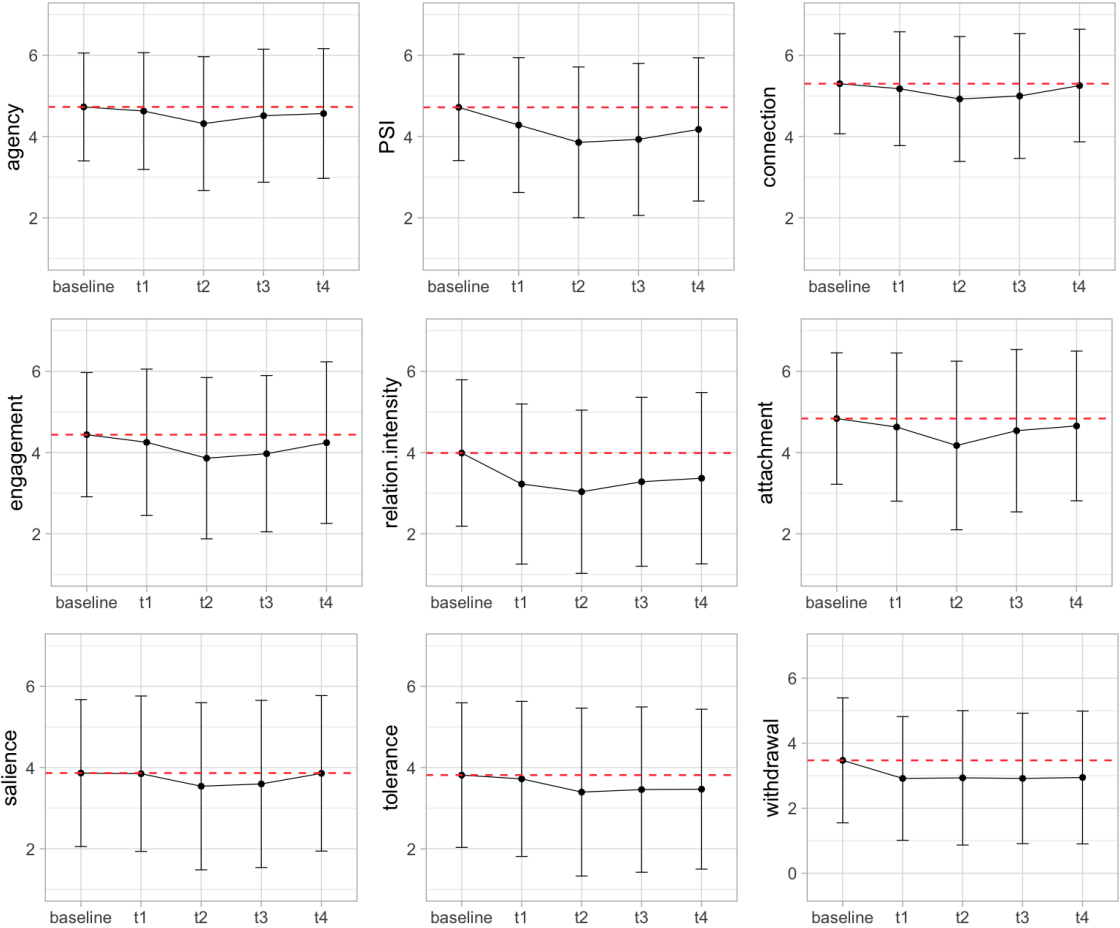

Fig. 5. Participants' responses to key variables in the theoretical pathway in Study 1 and Study 2. In each figure, `baseline` represents ratings of their own AI companions reported in the pre-survey. Red dashed lines (- - -) show the mean value of each baseline. Error bars show corresponding standard deviations. Values responded at the four time points in Study 2 are show in `t1`, `t2`, `t3`, and `t4` respectively.

*4.2.3 Time series analysis: Projecting when the gap between real vs. bot will close.* While participants carried over their perceptions of their own AI companions when they first started interacting with a new agent, such perceptions can change as they continue to interact with the bot. Indeed, we saw significant week-by-week changes in how participants perceived Study Bot in the longitudinal study: Post-hoc pairwise tests using the `emmeans` package in R show that the difference between participants' responses for their own AI companions vs. those for the Study Bot significantly enlarges in Week 2 for 16 out of 22 variables (also see Figure 5). This discrepancy, however, diminishes as participants continue to interact with Study Bot in Week 3 and Week 4. (See Table 3 for pairwise comparison between data from the pre-survey versus those measured in each week during Study 2.)

Table 3. Pairwise comparisons of participants' responses for their own AI companions vs. their responses to Study Bot at each time point during the longitudinal study. Significant results are color-coded in red and indicate participants' responses to Study Bot differ significantly from their responses to their own AI companions. Non-significant results are color-coded in green and indicate participants' responses to Study Bot align with how they perceive their own AI companions. $p^* < 0.05$, $p^{**} < 0.01$, $p^{***} < 0.001$

| Variable | Baseline vs. T1 | | | | Baseline vs. T2 | | | | Baseline vs. T3 | | | | Baseline vs. T4 | | | |
|---|---|---|---|---|---|---|---|---|---|---|---|---|---|---|---|---|
| | $\beta$ | S.E. | t | p | $\beta$ | S.E. | t | p | $\beta$ | S.E. | t | p | $\beta$ | S.E. | t | p |
| **Mental Model of Agent** | | | | | | | | | | | | | | | | |
| Anthropomorphism | 0.13 | 0.14 | 0.90 | 0.880 | 0.67 | 0.15 | 4.6 | <0.001*** | 0.35 | 0.15 | 2.4 | 0.121 | 0.27 | 0.15 | 1.8 | 0.378 |
| Animacy | 0.36 | 0.13 | 2.80 | 0.050 | 0.94 | 0.14 | 6.90 | <0.001*** | 0.55 | 0.14 | 4.00 | <0.001*** | 0.49 | 0.14 | 3.50 | 0.010* |
| Intelligence | 0.01 | 0.13 | 0.10 | >0.999 | 0.81 | 0.13 | 6.10 | <0.001*** | 0.48 | 0.14 | 3.50 | <0.001*** | 0.24 | 0.14 | 1.70 | 0.420 |
| Safety | -0.14 | 0.11 | -1.26 | 0.710 | 0.21 | 0.11 | 1.86 | 0.340 | 0.12 | 0.12 | 1.02 | 0.850 | 0.07 | 0.12 | 0.63 | 0.970 |
| Personification | 0.48 | 0.12 | 4.10 | <0.001*** | 0.86 | 0.12 | 7.00 | <0.001*** | 0.69 | 0.13 | 5.50 | <0.001*** | 0.64 | 0.13 | 5.00 | <0.001*** |
| Experience | 0.77 | 0.10 | 7.60 | <0.001*** | 0.88 | 0.11 | 8.10 | <0.001*** | 0.80 | 0.11 | 7.30 | <0.001*** | 0.90 | 0.11 | 8.00 | <0.001*** |
| Agency | 0.15 | 0.10 | 1.50 | 0.590 | 0.55 | 0.11 | 5.10 | <0.001*** | 0.33 | 0.11 | 3.00 | 0.020* | 0.27 | 0.11 | 2.40 | 0.120 |
| **Parasocial Interaction** | | | | | | | | | | | | | | | | |
| EPSI | 0.04 | 0.11 | 0.38 | >0.999 | 0.24 | 0.12 | 2.05 | 0.240 | 0.13 | 0.12 | 1.12 | 0.800 | -0.001 | 0.12 | <0.01 | >0.999 |
| PSI | 0.56 | 0.11 | 5.20 | <0.001*** | 1.01 | 0.11 | 8.80 | <0.001*** | 0.92 | 0.12 | 7.90 | <0.001*** | 0.64 | 0.12 | 5.40 | <0.001*** |
| Connection | 0.26 | 0.10 | 2.70 | 0.060 | 0.54 | 0.10 | 5.30 | <0.001*** | 0.45 | 0.10 | 4.30 | <0.001*** | 0.17 | 0.11 | 1.60 | 0.520 |
| Coordination | 0.20 | 0.12 | 1.70 | 0.460 | 0.89 | 0.13 | 6.90 | <0.001*** | 0.43 | 0.13 | 3.30 | 0.010* | 0.12 | 0.13 | 0.90 | 0.890 |
| **Social Penetration** | | | | | | | | | | | | | | | | |
| Engagement | 0.28 | 0.11 | 2.50 | 0.080 | 0.70 | 0.12 | 5.90 | <0.001*** | 0.58 | 0.12 | 4.80 | <0.001*** | 0.28 | 0.12 | 2.30 | 0.150 |
| Disclosure Amount | 0.36 | 0.13 | 2.80 | 0.040 | 0.52 | 0.14 | 3.80 | <0.001*** | 0.44 | 0.14 | 3.20 | 0.007* | 0.18 | 0.14 | 1.20 | 0.740 |
| Disclosure Depth | 0.96 | 0.14 | 7.10 | <0.001*** | 1.01 | 0.14 | 7.00 | <0.001*** | 0.80 | 0.15 | 5.50 | <0.001*** | 0.54 | 0.15 | 3.60 | <0.001*** |
| Disclosure Honesty | -0.28 | 0.11 | -2.62 | 0.070 | -0.22 | 0.11 | -1.95 | 0.290 | -0.17 | 0.11 | -1.48 | 0.580 | -0.20 | 0.12 | -1.73 | 0.410 |
| Disclosure Intent | 0.23 | 0.11 | 2.23 | 0.170 | 0.25 | 0.11 | 2.29 | 0.150 | 0.23 | 0.11 | 2.89 | 0.030 | 0.07 | 0.12 | 0.57 | 0.970 |
| Disclosure Sentiment | -0.52 | 0.10 | -5.50 | <0.001*** | -0.38 | 0.10 | -3.80 | <0.001*** | -0.32 | 0.10 | -3.10 | 0.020* | -0.31 | 0.11 | -2.90 | 0.030* |
| **Psychological Impact** | | | | | | | | | | | | | | | | |
| Relation Intensity | 0.92 | 0.12 | 7.80 | <0.001*** | 1.17 | 0.13 | 9.30 | <0.001*** | 0.95 | 0.13 | 7.40 | <0.001*** | 0.81 | 0.13 | 6.20 | <0.001*** |
| Attachment | 0.30 | 0.14 | 2.10 | 0.220 | 0.79 | 0.15 | 5.30 | <0.001*** | 0.37 | 0.15 | 2.40 | 0.110 | 0.24 | 0.16 | 1.60 | 0.520 |
| Salience | 0.13 | 0.13 | 1.01 | 0.850 | 0.40 | 0.14 | 2.99 | 0.020* | 0.38 | 0.14 | 2.76 | 0.050 | 0.10 | 0.14 | 0.71 | 0.950 |
| Tolerance | 0.16 | 0.13 | 1.20 | 0.730 | 0.47 | 0.13 | 3.50 | <0.001*** | 0.43 | 0.14 | 3.20 | 0.010* | 0.39 | 0.14 | 2.80 | 0.040* |
| Withdrawal | 0.65 | 0.13 | 4.90 | <0.001*** | 0.65 | 0.14 | 4.70 | <0.001*** | 0.73 | 0.14 | 5.10 | <0.001*** | 0.64 | 0.15 | 4.30 | <0.001*** |

Given these observed trends, we test whether there is a significant curvilinear relationship by fitting a non-linear model to test whether time or its quadratic term (time$^2$) predicts how the difference between users' responses toward their own AI companions vs. the Study Bot evolves as they interact more with the new agent. These results address whether we would observe the process of users developing deeper relationships with a new agent – like how they initially built up those with their AI companions. Test statistics are reported in Table 4. Here, we see the majority (16 out of 22) of variables measured in the current research, including the key variables (agency, PSI, and engagement) that robustly predict users' responses to AI agents, converged within the study period. In other words, with interaction over a one-month period, with as few as four occurrences of interaction, users can develop the type of experience, relationship, and companionship that they long for with their own AI companion.

It is worth noting that all variables that converged within the study period show significant quadratic effects of time on the difference between their responses to their own AI companion vs. those toward the new bot. These significant quadratic terms come with negative $\beta_1$ and positive $\beta_2$ values, which indicate that participants first showed an extent of divergence before their responses started to converge, resonating with the enlarged difference and decrease in ratings we commonly observed in Week 2, and the shrinking discrepancy from Week 3 onward. Indeed, when we further computed the vertex points ($\frac{-\beta_1}{2\beta_2}$), the time point when each variable began to converge, we saw that all vertex values ranged between 3 and 4, which again suggest that participants started to perceive the new bot as more similar to their own AI companion from their third interactions and thereafter.

With the coefficients from the fitted models, we also calculated convergence points (i.e., the time point when there is no difference between the ratings for one's own AI companion and the ratings for Study Bot) by solving $t$ where $\beta_0 + \beta_1 t + \beta_2 t^2 = 0$. Among the variables that converged, the majority of their convergence points fall from 5 to 8, suggesting that we will likely observe participants demonstrating identical responses to Study Bot with two to four more rounds of interactions.

## 5 Discussion

The current work provides one of the first integrated examinations of how users' perceptions, parasocial experiences, and engagement jointly shape the development of AI companionship over time. By analyzing both survey and longitudinal interaction data, we demonstrate that (1) constructs of AI companionship often studied in isolation are in fact interconnected, (2) beyond human-likeness, agents' responsiveness and capacity for engagement emerge as central drivers of companionship, and (3) human−AI companionship can develop rapidly, even within a controlled empirical setting. Together, these findings highlight that even general-purpose chatbots can evoke experiences akin to users' own companions and produce significant psychological effects. Such results raise critical theoretical, methodological, design, and ethical considerations for future HCI research on AI companionship.

### 5.1 Implication for Future Research Approach to Studying AI Companionship

A key contribution of this work is to address limited data access to commercial AI companion applications as a major obstacle and to envision possibilities for research approaches in this topic area. Indeed, Study 2 was intentionally designed to achieve these goals. Our findings demonstrate the feasibility of observing how AI companionship evolves and develops across time sessions in an empirical setting. This creates new opportunities for HCI research to experiment with design approaches without depending solely on proprietary platforms. Reflecting on our methodological approach, we highlight three takeaways.

**First, highlighting the utility of design-agnostic study probes.** In contrast to many user studies that prioritize comparing specific design features, our generic Study Bot allowed us to examine how companionship emerges even without anthropomorphic cues. This not only demonstrates the risks of assuming such features are necessary, but also surfaces potential harms they may amplify. We therefore advocate that future research include a design-agnostic condition as part of empirical design, ensuring that findings are not biased by untested assumptions about the benefits of anthropomorphism.

**Second, accounting for carry-over effects when deciding on longitudinal design.** While HCI broadly needs more longitudinal studies, we recognize that such methods are resource-intensive. We propose a binary decision rule that accounts for the strong carry-over effect observed in Study 2: if the study's goal is to understand users' perceptions of their own AI companion use, longitudinal data may be less critical, as participants' early responses to a new study agent largely mirrored their experiences with their own agents. However, for studies evaluating new design features,

Table 4. Linear vs. Quadratic mixed-effects models per outcome (two rows per outcome). The linear slope ($b_1$) captures the initial direction of change, while the quadratic term ($b_2$) indicates whether that slope accelerates, decelerates, or reverses over time. Model fit is evaluated with AIC/BIC: lower values indicate better fit. Negative $\Delta$AIC (Quadratic−Linear) favors the quadratic model.

| Variable | Model | Coefficients | | | | | | Shape | | Model Fit (Quadratic - Linear) | | | Convergence |
|---|---|---|---|---|---|---|---|---|---|---|---|---|---|
| | | $\beta_0$ | $p(b0)$ | $\beta_1$ | $p(b1)$ | $\beta_2$ | $p(b2)$ | Vertex | Convergence Points | AIC | BIC | $\Delta$AIC | |
| **Mental Model of Agent** | | | | | | | | | | | | | |
| Anthropomorphism | Linear | -0.350 | 0.106 | -0.020 | 0.692 | | | | | 1340.6 | 1356.4 | | Yes |
| | Quadratic | 1.381 | 0.031 | -1.140 | 0.004 | 0.162 | 0.004 | 3.529 | 1.552, 5.506 | 1338.4 | 1358.1 | -2.2 | Converged within study period |
| Animacy | Linear | -0.633 | 0.002 | -0.017 | 0.738 | | | | | 1304.2 | 1319.9 | | Yes |
| | Quadratic | 1.179 | 0.056 | -1.190 | 0.002 | 0.169 | 0.002 | 3.515 | 1.194, 5.836 | 1300.5 | 1320.2 | -3.7 | Converged within study period |
| Intelligence | Linear | -0.273 | 0.183 | -0.052 | 0.303 | | | | | 1306.9 | 1322.7 | | Yes |
| | Quadratic | 2.542 | <.001 | -1.874 | <.001 | 0.263 | <.001 | 3.564 | 1.822, 5.305 | 1290.2 | 1309.9 | -16.7 | Converged within study period |
| Safety | Linear | 0.021 | 0.901 | -0.061 | 0.147 | | | | | 1149.2 | 1164.9 | | Yes |
| | Quadratic | 1.122 | 0.033 | -0.774 | 0.018 | 0.103 | 0.028 | 3.762 | 1.961, 5.564 | 1150.6 | 1170.3 | 1.5 | Converged within study period |
| Personification | Linear | -0.613 | <.001 | -0.039 | 0.350 | | | | | 1197.6 | 1213.4 | | No |
| | Quadratic | 0.634 | 0.225 | -0.846 | 0.008 | 0.116 | 0.011 | 3.633 | 0.848, 6.419 | 1197.6 | 1217.3 | -0.1 | Converged within study period |
| Experience | Linear | -0.777 | <.001 | -0.029 | 0.373 | | | | | 1054.5 | 1070.3 | | No |
| | Quadratic | -0.691 | 0.102 | -0.085 | 0.739 | 0.008 | 0.826 | 5.299 | -5.379, 15.977 | 1061.2 | 1080.9 | 6.7 | No significant pattern |
| Agency | Linear | -0.263 | 0.099 | -0.025 | 0.468 | | | | | 1090.3 | 1106.1 | | Yes |
| | Quadratic | 1.083 | 0.014 | -0.897 | <.001 | 0.126 | 0.001 | 3.568 | 1.54, 5.597 | 1086.3 | 1106.0 | -4.0 | Converged within study period |
| **Parasocial Interaction** | | | | | | | | | | | | | |
| EPSI | Linear | -0.244 | 0.149 | 0.016 | 0.671 | | | | | 1143.9 | 1159.6 | | Yes |
| | Quadratic | 0.706 | 0.146 | -0.598 | 0.044 | 0.089 | 0.037 | 3.375 | 1.524, 5.225 | 1146.0 | 1165.7 | 2.1 | Converged within study period |
| PSI | Linear | -0.742 | <.001 | -0.025 | 0.500 | | | | | 1131.4 | 1147.2 | | Yes |
| | Quadratic | 1.302 | 0.004 | -1.348 | <.001 | 0.191 | <.001 | 3.532 | 1.154, 5.911 | 1115.2 | 1134.9 | -16.2 | Converged within study period |
| Connection | Linear | -0.511 | <.001 | 0.027 | 0.444 | | | | | 1056.8 | 1072.5 | | Yes |
| | Quadratic | 1.033 | 0.017 | -0.972 | <.001 | 0.144 | <.001 | 3.374 | 1.32, 5.428 | 1049.3 | 1069.0 | -7.4 | Converged within study period |
| Coordination | Linear | -0.656 | 0.001 | 0.055 | 0.256 | | | | | 1282.8 | 1298.6 | | Yes |
| | Quadratic | 2.061 | <.001 | -1.703 | <.001 | 0.254 | <.001 | 3.356 | 1.585, 5.128 | 1266.1 | 1285.8 | -16.8 | Converged within study period |
| **Social Penetration** | | | | | | | | | | | | | |
| Engagement | Linear | -0.491 | 0.004 | -0.000 | 0.997 | | | | | 1167.7 | 1183.5 | | Yes |
| | Quadratic | 1.471 | 0.004 | -1.271 | <.001 | 0.183 | <.001 | 3.466 | 1.469, 5.462 | 1157.5 | 1177.2 | -10.2 | Converged within study period |
| Disclosure Amount | Linear | -0.608 | 0.002 | 0.063 | 0.149 | | | | | 1256.0 | 1271.7 | | Yes |
| | Quadratic | 0.573 | 0.295 | -0.702 | 0.035 | 0.110 | 0.021 | 3.184 | 0.961, 5.407 | 1256.9 | 1276.6 | 0.9 | Converged within study period |
| Disclosure Depth | Linear | -1.387 | <.001 | 0.144 | 0.002 | | | | | 1285.4 | 1301.2 | | No |
| | Quadratic | -0.507 | 0.375 | -0.426 | 0.221 | 0.082 | 0.099 | 2.591 | -0.999, 6.182 | 1288.9 | 1308.6 | 3.4 | Diverged over time |
| Disclosure Honesty | Linear | 0.252 | 0.119 | -0.032 | 0.326 | | | | | 1072.2 | 1087.9 | | No |
| | Quadratic | 0.559 | 0.185 | -0.231 | 0.363 | 0.029 | 0.430 | 4.000 | – | 1078.4 | 1098.1 | 6.2 | No significant pattern |
| Disclosure Intent | Linear | -0.404 | 0.011 | 0.034 | 0.356 | | | | | 1102.2 | 1117.9 | | No |
| | Quadratic | 0.343 | 0.459 | -0.449 | 0.114 | 0.070 | 0.086 | 3.224 | 0.883, 5.565 | 1105.8 | 1125.5 | 3.6 | No significant pattern |
| Disclosure Sentiment | Linear | 0.611 | <.001 | -0.079 | 0.023 | | | | | 1061.7 | 1077.5 | | No |
| | Quadratic | 1.001 | 0.023 | -0.331 | 0.217 | 0.036 | 0.343 | 4.553 | – | 1067.5 | 1087.2 | 5.8 | No significant pattern |
| **Psychological Impact** | | | | | | | | | | | | | |
| Relation Intensity | Linear | -1.187 | <.001 | 0.048 | 0.210 | | | | | 1170.3 | 1186.0 | | Yes |
| | Quadratic | -0.057 | 0.906 | -0.683 | 0.021 | 0.105 | 0.012 | 3.240 | -0.083, 6.563 | 1170.5 | 1190.2 | 0.2 | Converged within study period |
| Attachment | Linear | -0.640 | 0.004 | 0.046 | 0.370 | | | | | 1352.0 | 1367.7 | | Yes |
| | Quadratic | 1.126 | 0.082 | -1.097 | 0.006 | 0.165 | 0.004 | 3.326 | 1.269, 5.383 | 1349.5 | 1369.2 | -2.4 | Converged within study period |
| Salience | Linear | -0.304 | 0.113 | 0.004 | 0.929 | | | | | 1244.6 | 1260.4 | | Yes |
| | Quadratic | 1.226 | 0.027 | -0.987 | 0.004 | 0.143 | 0.003 | 3.453 | 1.625, 5.281 | 1242.3 | 1262.0 | -2.3 | Converged within study period |
| Tolerance | Linear | -0.137 | 0.473 | -0.072 | 0.087 | | | | | 1228.5 | 1244.2 | | Yes |
| | Quadratic | 0.919 | 0.085 | -0.756 | 0.020 | 0.099 | 0.034 | 3.834 | 1.515, 6.154 | 1230.3 | 1250.0 | 1.8 | Converged within study period |
| Withdrawal | Linear | -0.700 | <.001 | -0.009 | 0.816 | | | | | 1221.4 | 1237.2 | | No |
| | Quadratic | -0.318 | 0.526 | -0.256 | 0.392 | 0.036 | 0.405 | 3.595 | -1.078, 8.269 | 1227.2 | 1246.9 | 5.8 | No significant pattern |

*Note:* A variable can show both significant estimated marginal means (emmeans) differences and significant convergence: This occurs when Study 2 values remain significantly different from baseline, yet the longitudinal trajectory already shows a converging trend, captured by the quadratic term.

longitudinal data are necessary; otherwise, users' responses may simply reflect their prior relational scripts rather than the effect of the new design.

**Third, ensuring safe and responsible testbeds through adaptive protocols.** The rapid emergence of companionship in our study illustrates both the promise and the risks of studying these dynamics empirically. Because harmful reliance patterns can develop quickly, future research must implement safeguards that go beyond conventional ethics protocols such as voluntary withdrawal. We encourage the use of adaptive and personalized mechanisms within study agents themselves, enabling them to monitor harmful interaction signals and guide participants toward disengagement if risks become too high. Ideally, such systems should also include real-time alerts for researchers, ensuring that problematic use is identified promptly during study sessions. Ultimately, participant safety must be prioritized as companionship develops, and methodological innovation should be matched with robust ethical protections.

Building on these methodological recommendations, we next outline plausible directions for future exploration. These directions span both theoretical and practical dimensions, reflecting how a deeper understanding of AI companionship can inform future research and design.

## 5.2 Theoretical Implication

Our findings extend prior HCI research that has largely explained AI companionship through human-like features of agents. In contrast, our results raise the need for studying the development of AI companionship by considering the various variables that jointly shape companionship and deeper relational dynamics when designing empirical studies. In particular, agents' capabilities to respond (agency) and to sustain reciprocal experiences should be foregrounded in future research on the development of AI companionship.

### 5.2.1 Foregrounding Agency and Parasociality in Studying the Multi-facets of AI Companionship.

Findings across both studies consistently underscore the central role of parasocial experience in shaping AI companionship. Users develop companionship by forming one-sided relationships, differentiating self from other, and treating the agent as an individual entity. This suggests that parasocial relationships with AI companions rely not merely on anthropomorphic cues but on the agent's ability to present itself as an autonomous entity. Theoretically, this challenges the tendency in prior HCI research to treat anthropomorphism as the primary explanation for why users engage with companions. Instead, eliciting parasocial interactions depends on how effectively an AI system conveys autonomy through multiple facets of its design and functionality. Despite its long-standing relevance in media and communication studies, parasocial interaction remains relatively underexplored in HCI. Future research should further theorize and operationalize parasocial experience to advance our understanding of how users relate to AI companions.

In parallel, our results highlight the significance of agency—specifically, an agent's technical capacity and responsiveness—as another key driver of companionship. This calls for expanding the study of AI companionship beyond social attributes to include core functional properties. A pressing design and research question concerns how to balance agents' responsiveness with their perceived capability, as agents' growing abilities to respond to a wide range of users' input also risk fostering psychological dependence. As shown in our work, we constantly saw greater perceived agency cause heightened effects on salience, tolerance, and withdrawal. Future work should therefore examine whether reducing responsiveness necessarily undermines perceptions of competence, intelligence, or usefulness, or whether it might serve as a safeguard against unintended psychological impact.

### 5.2.2 Identifying Unique Disclosure Patterns in Human-AI Companionship.

Our findings on users' disclosure patterns suggest that human–AI companionship differs in important ways from human–human relationships. This aligns with recent work showing that users selectively share certain types of content with agents that they would not disclose to humans, and vice versa. To begin with, while disclosure intent is often characterized as a positive feature in interpersonal research, our results indicate that disclosure with AI companions can heighten dependence, attachment, and overreliance. This raises concerns about whether increased willingness to share with AI agents translates into healthier interactions or not. In this sense, disclosure in human–AI contexts may blur the line between supportive self-expression and problematic reliance.

Second, disclosure sentiment emerged as a non-significant mediator across all models, diverging from established findings in human communication research that emphasize the deepening of relationships through sharing beyond positive content. This trend can be interpreted in two ways. On one hand, these results suggest that some users may strengthen their relationships with AI companions even if their interactions only contain positive sentiment (e.g., entertaining, fun, exciting) and that the users never further share their personal hardships.

At the same time, the consistent effect of engagement across our analyses points to another possibility: companionship may deepen simply through sustained interaction, even in the absence of highly personal or emotionally valenced disclosure. Together, these results call for rethinking disclosure as a relational construct in human–AI interaction. Moreover, future research should further explore and distinguish users' disclosure patterns that support their well-being and those that intensify unhealthy attachment.

### 5.3 Design Implication

Our study underscores that AI companionship is not a static interactional phenomenon but a dynamic relational process shaped by design features, user perceptions, and temporal development. While prior work has often examined individual factors such as anthropomorphism in isolation, our findings suggest that companionship emerges through a more complex interplay of multiple variables and unfolds rapidly over time. These insights call for rethinking how AI companions are designed and evaluated, not only at the level of interface features but also in relation to their longitudinal effects on users' psychological well-being. Below, we highlight three key design implications: (1) rethinking the necessity of anthropomorphic design and calibrating user perceptions, (2) recognizing time as a central design variable, and (3) creating disengagement pathways to safeguard against harmful patterns of use.

*5.3.1 Rethinking the Necessity of Anthropomorphic Design and Calibrating Perceptions.* Our findings demonstrate that even general-purpose AI companions, without extensive anthropomorphic cues, can elicit strong psychological impacts such as attachment, trust, and dependence. This suggests that the necessity of layering on additional human-like features is not self-evident and can intensify the psychological impact of AI companionship.

While anthropomorphic design elements—such as realistic avatars, emotive voices, or empathetic linguistic expressions—can heighten engagement, they also risk amplifying users' attributions of social presence and intensifying relationships in unintended or potentially harmful ways.

The design challenge, then, is not about how to further resemble human-like interaction experiences with AI companions but to ensure the design of AI companions does not overstimulate social attributions.

In parallel, we see the need to calibrate users' perception of agents and constantly remind them of the non-human nature and limitations of AI companions. Transparency should not be buried in vague disclaimers or inaccessible menus; rather, it should be surfaced in context, particularly after emotionally intense interactions. Reminders that reinforce the asymmetry of the relationship can help preserve user agency and mitigate over-personification. At the same time, designers face a double bind: making relational asymmetries explicit may weaken the perceived intimacy that users find valuable, but obscuring them risks deepening dependency. Navigating this tension is central to the responsible design of AI companions.

Finally, users may carry relational scripts from previous AI or human companions into new interactions. To address this carry-over effect, systems should include mechanisms that help reset expectations, clarify boundaries, and explicitly differentiate one agent from another.

*5.3.2  Time as a Design Variable.* A second implication of our study is that time itself must be treated as a design variable. Companionship with AI agents can form remarkably quickly—sometimes after just a few exchanges—making early interactions disproportionately consequential.

Onboarding thus becomes a critical phase. Rather than encouraging immediate intimacy, design should slow the pace of relationship building through calibrated disclosure, limited responsiveness, or prompts that set clear expectations about the agent's role. These early design choices can influence whether users view the agent as a tool, a support, or a quasi-social partner.

Over longer periods, systems should adapt to the evolving trajectory of companionship. Longitudinal use not only deepens engagement but also reliance and dependence on the agents. Design should integrate reflective prompts, such as summaries of interaction histories or gentle questions that encourage users to evaluate their comfort with the relationship's intensity. Moreover, responsiveness can be modulated temporally, balancing supportive engagement with deliberate pacing to prevent over-reliance. By designing with time in mind, companions can support sustained engagement while reducing risks associated with rapid attachment and cumulative dependency.

*5.3.3  Designing Disengagement Pathways for Safeguard.* A third implication is the need for robust disengagement pathways when companionship becomes problematic. Our results highlight engagement and self-disclosure as powerful mediators of relationship development. While these processes can make companionship supportive, they also create vulnerabilities: users may disclose sensitive information in ways that foster over-reliance, use companions impulsively, or engage in patterns resembling addiction or abusive interaction. In these contexts, responsiveness and relational reinforcement can inadvertently intensify harm.

Designing for safe disengagement requires both proactive and reactive strategies. Proactively, systems can nudge users toward healthier interaction rhythms, such as suggesting breaks, surfacing reminders to engage with human social networks, or pacing high-intensity exchanges. Reactively, systems should provide escalation mechanisms that direct users to trusted personal contacts or professional resources when signs of distress or risk arise. These safeguards must be developed in close collaboration with clinical and domain experts to ensure they are evidence-based and appropriate for diverse user needs. Importantly, designing disengagement pathways recognizes that companionship should remain complementary to, not substitutive of, human care. By embedding pathways for safe withdrawal and redirection, designers can help ensure that AI companions foster supportive relationships without displacing essential social and healthcare infrastructures.

## 5.4  Limitations and Future Work

Despite our best efforts, we acknowledge several limitations of the current study. First, due to its exploratory nature, our analyses focus on associations among variables and do not allow for strong causal inference. While the mediation models suggest plausible sequential pathways, causality cannot be firmly established without designs that explicitly manipulate the presence or magnitude of each factor. Future work should adopt experimental and quasi-experimental designs that test causal mechanisms more directly, allowing researchers to verify whether the observed relationships hold under controlled manipulations.

Second, the study was conducted entirely online, which constrains ecological validity. While online methods enable broad recruitment and scalable data collection, they differ from in-situ deployments where AI companions are integrated into users' everyday contexts, routines, and devices. As a result, our findings may not fully capture the richness and complexity of companionship as it naturally unfolds in the wild. Field-based or hybrid designs that embed study agents into participants' daily environments would help address this limitation and provide a more ecologically valid understanding of human–AI companionship.

Third, our longitudinal design faced issues of self-selection bias. Participants who were less engaged with the Study Bot were more likely to withdraw from later sessions, resulting in a sample skewed toward individuals more receptive to AI companionship. This attrition pattern may exaggerate the observed effects of attachment and engagement, as the most skeptical or disengaged users are underrepresented in the later waves of data. Future studies should develop strategies to minimize attrition, such as adaptive engagement mechanisms or incentives, while also explicitly comparing dropouts and completers to assess potential bias.

Fourth, the generalizability of our findings is limited by the sample characteristics. Our participants primarily represent users who are already familiar with or interested in AI companions, and the majority were recruited from online survey platforms with particular demographic profiles. Cultural context is also an important factor: norms around disclosure, parasociality, and attachment to technology vary significantly across societies, and our findings may not extend universally. Future research should examine AI companionship across more diverse populations and cultural settings, as well as across different types of companion technologies, to establish the broader applicability of these insights.

Finally, although our work provides an integrated view of how attachment, disclosure, and dependency unfold over time, further research is needed to explore the long-term consequences of these dynamics. While companionship may provide benefits such as emotional support or reduced loneliness, it also carries risks of overreliance, displacement of human relationships, or entrenchment of problematic use patterns. Longitudinal and mixed-method approaches that follow users over extended periods will be critical for disentangling these benefits and risks, ultimately informing both theory and responsible design.

## 6 Conclusion

In sum, our study highlights the need to move beyond fragmented perspectives and short-term measures, offering an integrated view of how AI companionship develops and impacts users. By connecting theoretical constructs, introducing methodological pathways, and foregrounding design and ethical considerations, we aim to inform both research and practice as HCI continues to grapple with the growing role of AI companions in everyday life.

## References

[1] Zara Abrams. 2025. Using generic AI chatbots for mental health support: A dangerous trend. https://www.apaservices.org/practice/business/technology/artificial-intelligence-chatbots-therapists

[2] Irwin Altman and Dalmas A. Taylor. 1983. *Social penetration: the development of interpersonal relationships* (1st irvington ed ed.). Irvington Publishers, New York, N.Y. OCLC: 9066345.

[3] Marta Andersson. 2025. Companionship in code: AI's role in the future of human connection. *Humanities and Social Sciences Communications* 12, 1 (July 2025), 1177. https://doi.org/10.1057/s41599-025-05536-x

[4] Cecilie Schou Andreassen, Torbjørn Torsheim, Geir Scott Brunborg, and Ståle Pallesen. 2012. Development of a Facebook Addiction Scale. *Psychological Reports* 110, 2 (April 2012), 501–517. https://doi.org/10.2466/02.09.18.PR0.110.2.501-517

[5] Christoph Bartneck, Dana Kulić, Elizabeth Croft, and Susana Zoghbi. 2009. Measurement Instruments for the Anthropomorphism, Animacy, Likeability, Perceived Intelligence, and Perceived Safety of Robots. *International Journal of Social Robotics* 1, 1 (Jan. 2009), 71–81. https://doi.org/10.1007/s12369-008-0001-3

[6] Timothy Bickmore, Daniel Schulman, and Langxuan Yin. 2010. MAINTAINING ENGAGEMENT IN LONG-TERM INTERVENTIONS WITH RELATIONAL AGENTS. *Applied Artificial Intelligence* 24, 6 (July 2010), 648–666. https://doi.org/10.1080/08839514.2010.492259

[7] Annabel Blake, Marcus Carter, and Eduardo Velloso. 2025. Are Measures of Children's Parasocial Relationships Ready for Conversational AI?. In *Proceedings of the 2025 ACM Conference on Fairness, Accountability, and Transparency*. ACM, Athens Greece, 1145–1158. https://doi.org/10.1145/3715275.3732075

[8] Petter Bae Brandtzaeg, Marita Skjuve, and Asbjørn Følstad. 2022. My AI Friend: How Users of a Social Chatbot Understand Their Human–AI Friendship. *Human Communication Research* 48, 3 (June 2022), 404–429. https://doi.org/10.1093/hcr/hqac008

[9] Jiahao Chen, Fu Guo, Zenggen Ren, Mingming Li, and Jaap Ham. 2024. Effects of Anthropomorphic Design Cues of Chatbots on Users' Perception and Visual Behaviors. *International Journal of Human–Computer Interaction* 40, 14 (July 2024), 3636–3654. https://doi.org/10.1080/10447318.2023.2193514

[10] Jonathan Cohen. 2004. Parasocial Break-Up from Favorite Television Characters: The Role of Attachment Styles and Relationship Intensity. *Journal of Social and Personal Relationships* 21, 2 (April 2004), 187–202. https://doi.org/10.1177/0265407504041374

[11] Megan R. Dillow, Walid A. Afifi, and Masaki Matsunaga. 2012. Perceived partner uniqueness and communicative and behavioral transgression outcomes in romantic relationships. *Journal of Social and Personal Relationships* 29, 1 (Feb. 2012), 28–51. https://doi.org/10.1177/0265407511420191

[12] Nicholas Epley, Adam Waytz, and John T. Cacioppo. 2007. On seeing human: A three-factor theory of anthropomorphism. *Psychological Review* 114, 4 (2007), 864–886. https://doi.org/10.1037/0033-295X.114.4.864

[13] Xianzhe Fan, Qing Xiao, Xuhui Zhou, Jiaxin Pei, Maarten Sap, Zhicong Lu, and Hong Shen. 2025. User-Driven Value Alignment: Understanding Users' Perceptions and Strategies for Addressing Biased and Discriminatory Statements in AI Companions. In *Proceedings of the 2025 CHI Conference on Human Factors in Computing Systems*. ACM, Yokohama Japan, 1–19. https://doi.org/10.1145/3706598.3713477

[14] Chris Frith and Uta Frith. 2005. Theory of mind. *Current Biology* 15, 17 (Sept. 2005), R644–R645. https://doi.org/10.1016/j.cub.2005.08.041

[15] Chris D Frith. 2008. Social cognition. *Philosophical Transactions of the Royal Society B: Biological Sciences* 363, 1499 (June 2008), 2033–2039. https://doi.org/10.1098/rstb.2008.0005

[16] Chris D Frith and Uta Frith. 2007. Social Cognition in Humans. *Current Biology* 17, 16 (Aug. 2007), R724–R732. https://doi.org/10.1016/j.cub.2007.05.068

[17] Silvia Gabrielli, Silvia Rizzi, Giulia Bassi, Sara Carbone, Rosa Maimone, Michele Marchesoni, and Stefano Forti. 2021. Engagement and Effectiveness of a Healthy-Coping Intervention via Chatbot for University Students During the COVID-19 Pandemic: Mixed Methods Proof-of-Concept Study. *JMIR mHealth and uHealth* 9, 5 (May 2021), e27965. https://doi.org/10.2196/27965

[18] Helen L. Gallagher and Christopher D. Frith. 2003. Functional imaging of 'theory of mind'. *Trends in Cognitive Sciences* 7, 2 (Feb. 2003), 77–83. https://doi.org/10.1016/S1364-6613(02)00025-6

[19] Andrew Gambino, Jesse Fox, and Rabindra Ratan. 2020. Building a Stronger CASA: Extending the Computers Are Social Actors Paradigm. *Human-Machine Communication* 1 (Feb. 2020), 71–86. https://doi.org/10.30658/hmc.1.5

[20] Erving Goffman. 1963. Embarrassment and Social Organization. In *Personality and social systems.*, Neil J. Smelser and William T. Smelser (Eds.). John Wiley & Sons, Inc., Hoboken, 541–548. https://doi.org/10.1037/11302-050

[21] David Goretzko, Karik Siemund, and Philipp Sterner. 2024. Evaluating Model Fit of Measurement Models in Confirmatory Factor Analysis. *Educational and Psychological Measurement* 84, 1 (Feb. 2024), 123–144. https://doi.org/10.1177/00131644231163813

[22] Heather M. Gray, Kurt Gray, and Daniel M. Wegner. 2007. Dimensions of Mind Perception. *Science* 315, 5812 (Feb. 2007), 619–619. https://doi.org/10.1126/science.1134475

[23] Rose E Guingrich and Michael S A Graziano. 2025. Chatbots as Social Companions: How People Perceive Consciousness, Human Likeness, and Social Health Benefits in Machines. In *Oxford Intersections: AI in Society* (1 ed.), Philipp Hacker (Ed.). Oxford University PressOxford. https://doi.org/10.1093/9780198945215.003.0011

[24] Tilo Hartmann. 2016. Parasocial Interaction, Parasocial Relationships, and Well-Being. In *The Routledge Handbook of Media Use and Well-Being*. Routledge.

[25] Tilo Hartmann and Charlotte Goldhoorn. 2011. Horton and Wohl Revisited: Exploring Viewers' Experience of Parasocial Interaction. *Journal of Communication* 61, 6 (Dec. 2011), 1104–1121. https://doi.org/10.1111/j.1460-2466.2011.01595.x

[26] William A. Haseltine. 2024. Battling Loneliness: The New Public Health Crisis. https://www.forbes.com/sites/williamhaseltine/2024/05/29/battling-loneliness-the-new-public-health-crisis/

[27] Cindy Hazan and Phillip R. Shaver. 1994. Attachment as an Organizational Framework for Research on Close Relationships. *Psychological Inquiry* 5, 1 (Jan. 1994), 1–22. https://doi.org/10.1207/s15327965pli0501_1

[28] Evelien Heyselaar. 2023. The CASA theory no longer applies to desktop computers. *Scientific Reports* 13, 1 (Nov. 2023), 19693. https://doi.org/10.1038/s41598-023-46527-9

[29] Kashmir Hill. [n. d.]. A Teen Was Suicidal. ChatGPT Was the Friend He Confided In. https://www.nytimes.com/2025/08/26/technology/chatgpt-openai-suicide.html

[30] Annabell Ho, Jeff Hancock, and Adam S Miner. 2018. Psychological, Relational, and Emotional Effects of Self-Disclosure After Conversations With a Chatbot. *Journal of Communication* 68, 4 (Aug. 2018), 712–733. https://doi.org/10.1093/joc/jqy026

[31] Cynthia A. Hoffner and Bradley J. Bond. 2022. Parasocial relationships, social media, & well-being. *Current Opinion in Psychology* 45 (June 2022), 101306. https://doi.org/10.1016/j.copsyc.2022.101306

[32] Donald Horton and R. Richard Wohl. 1956. Mass Communication and Para-Social Interaction: Observations on Intimacy at a Distance. *Psychiatry* 19, 3 (Aug. 1956), 215–229. https://doi.org/10.1080/00332747.1956.11023049

[33] Jeff Horwitz. [n. d.]. Meta's flirty AI chatbot invited a retiree to New York. He never made it home. https://www.reuters.com/investigates/special-report/meta-ai-chatbot-death/

[34] Lucie-Aimée Kaffee, Giada Pistilli, and Yacine Jernite. 2025. INTIMA: A Benchmark for Human-AI Companionship Behavior. https://doi.org/10.48550/arXiv.2508.09998 arXiv:2508.09998 [cs].

[35] Eun Jeong Kang, Haesoo Kim, Hyunwoo Kim, Susan R. Fussell, and Juho Kim. 2025. Can Fans Build Parasocial Relationships through Idols' Simulated Voice Messages?: A Study of AI Private Call Users' Perceptions, Cognitions, and Behaviors. *Proceedings of the ACM on Human-Computer Interaction* 9, 2 (May 2025), 1–31. https://doi.org/10.1145/3711111

[36] Emma J. Kilford, Emily Garrett, and Sarah-Jayne Blakemore. 2016. The Development of Social Cognition in Adolescence: An Integrated Perspective. *Neuroscience & Biobehavioral Reviews* 70 (Nov. 2016), 106–120. https://doi.org/10.1016/j.neubiorev.2016.08.016

[37] Alan M. Leslie, Ori Friedman, and Tim P. German. 2004. Core mechanisms in 'theory of mind'. *Trends in Cognitive Sciences* 8, 12 (Dec. 2004), 528–533. https://doi.org/10.1016/j.tics.2004.10.001

[38] Jingshu Li, Zicheng Zhu, Renwen Zhang, and Yi-Chieh Lee. 2025. Exploring the Effects of Chatbot Anthropomorphism and Human Empathy on Human Prosocial Behavior Toward Chatbots. https://arxiv.org/abs/2506.20748 _eprint: 2506.20748.

[39] Mei Yii Lim. 2012. Memory Models for Intelligent Social Companions. In *Human-Computer Interaction: The Agency Perspective*, Marielba Zacarias and José Valente De Oliveira (Eds.). Vol. 396. Springer Berlin Heidelberg, Berlin, Heidelberg, 241–262. https://doi.org/10.1007/978-3-642-25691-2_10 Series Title: Studies in Computational Intelligence.

[40] Ting-Han Lin, Hannah Dinner, Tsz Long Leung, Bilge Mutlu, J. Gregory Trafton, and Sarah Sebo. 2025. Connection-Coordination Rapport (CCR) Scale: A Dual-Factor Scale to Measure Human-Robot Rapport. https://arxiv.org/abs/2501.11887v1

[41] Zilin Ma, Yiyang Mei, and Zhaoyuan Su. 2023. Understanding the Benefits and Challenges of Using Large Language Model-based Conversational Agents for Mental Well-being Support. https://doi.org/10.48550/arXiv.2307.15810 arXiv:2307.15810 [cs].

[42] Takuya Maeda and Anabel Quan-Haase. 2024. When Human-AI Interactions Become Parasocial: Agency and Anthropomorphism in Affective Design. In *The 2024 ACM Conference on Fairness, Accountability, and Transparency*. ACM, Rio de Janeiro Brazil, 1068–1077. https://doi.org/10.1145/3630106.3658956

[43] Bertram F Malle. 2019. How Many Dimensions of Mind Perception Really Are There?. In *Proceedings of the 41st Annual Meeting of the Cognitive Science Society*. Cognitive Science Society, Montreal, Canada, 2268–2274.

[44] Kayla Matheus, Rebecca Ramnauth, Brian Scassellati, and Nicole Salomons. 2025. Long-Term Interactions with Social Robots: Trends, Insights, and Recommendations. *J. Hum.-Robot Interact.* 14, 3, Article 55 (June 2025), 42 pages. https://doi.org/10.1145/3729539

[45] Jingbo Meng, Minjin (Mj) Rheu, Yue Zhang, Yue Dai, and Wei Peng. 2023. Mediated Social Support for Distress Reduction: AI Chatbots vs. Human. *Proceedings of the ACM on Human-Computer Interaction* 7, CSCW1 (April 2023), 1–25. https://doi.org/10.1145/3579505

[46] Jingbo Meng, Renwen Zhang, Jiaqi Qin, Yu-Jen Lee, and Yi-Chieh Lee. 2025. AI-mediated social support: the prospect of human–AI collaboration. *Journal of Computer-Mediated Communication* 30, 4 (May 2025), zmaf013. https://doi.org/10.1093/jcmc/zmaf013

[47] Kelly Merrill, Jihyun Kim, and Chad Collins. 2022. AI companions for lonely individuals and the role of social presence. *Communication Research Reports* 39, 2 (March 2022), 93–101. https://doi.org/10.1080/08824096.2022.2045929

[48] Elizabeth R. Merwin, Allen C. Hagen, Joseph R. Keebler, and Chad Forbes. 2025. Self-disclosure to AI: People provide personal information to AI and humans equivalently. *Computers in Human Behavior: Artificial Humans* 5 (Aug. 2025), 100180. https://doi.org/10.1016/j.chbah.2025.100180

[49] Mario Mikulincer and Phillip R. Shaver. 2005. Attachment theory and emotions in close relationships: Exploring the attachment-related dynamics of emotional reactions to relational events. *Personal Relationships* 12, 2 (June 2005), 149–168. https://doi.org/10.1111/j.1350-4126.2005.00108.x

[50] Jared Moore, Declan Grabb, William Agnew, Kevin Klyman, Stevie Chancellor, Desmond C. Ong, and Nick Haber. 2025. Expressing stigma and inappropriate responses prevents LLMs from safely replacing mental health providers.. In *Proceedings of the 2025 ACM Conference on Fairness, Accountability, and Transparency*. ACM, Athens Greece, 599–627. https://doi.org/10.1145/3715275.3732039

[51] Nick Munn and Dan Weijers. 2025. Human–AI Friendship Is Possible and Can Be Good. In *Oxford Intersections: AI in Society* (1 ed.), Philipp Hacker (Ed.). Oxford University PressOxford. https://doi.org/10.1093/9780198945215.003.0076

[52] Clifford Nass, Jonathan Steuer, and Ellen R Tauber. 1994. Computers are social actors. In *Proceedings of the SIGCHI conference on Human factors in computing systems*. 72–78.

[53] World Health Organization. [n. d.]. Social Isolation and Loneliness. https://www.who.int/teams/social-determinants-of-health/demographic-change-and-healthy-ageing/social-isolation-and-loneliness

[54] Shuyi Pan and Maartje M.A. De Graaf. 2025. Developing a Social Support Framework: Understanding the Reciprocity in Human-Chatbot Relationship. In *Proceedings of the 2025 CHI Conference on Human Factors in Computing Systems*. ACM, Yokohama Japan, 1–13. https://doi.org/10.1145/3706598.3713503

[55] Hashai Papneja and Nikhil Yadav. 2025. Self-disclosure to conversational AI: a literature review, emergent framework, and directions for future research. *Personal and Ubiquitous Computing* 29, 2 (April 2025), 119–151. https://doi.org/10.1007/s00779-024-01823-7

[56] Iryna Pentina, Tianling Xie, Tyler Hancock, and Ainsworth Bailey. 2023. Consumer–machine relationships in the age of artificial intelligence: Systematic literature review and research directions. *Psychology & Marketing* 40, 8 (Aug. 2023), 1593–1614. https://doi.org/10.1002/mar.21853

[57] Josef Perner and Heinz Wimmer. 1985. "John thinks that Mary thinks that..." attribution of second-order beliefs by 5- to 10-year-old children. *Journal of Experimental Child Psychology* 39, 3 (June 1985), 437–471. https://doi.org/10.1016/0022-0965(85)90051-7

[58] Elizabeth M. Perse and Rebecca B. Rubin. 1989. Attribution in Social and Parasocial Relationships. *Communication Research* 16, 1 (Feb. 1989), 59–77. https://doi.org/10.1177/009365089016001003

[59] Prerna Ravi, John Masla, Gisella Kakoti, Grace C. Lin, Emma Anderson, Matt Taylor, Anastasia K. Ostrowski, Cynthia Breazeal, Eric Klopfer, and Hal Abelson. 2025. Co-designing Large Language Model Tools for Project-Based Learning with K12 Educators. In *Proceedings of the 2025 CHI Conference on Human Factors in Computing Systems (CHI '25)*. Association for Computing Machinery, New York, NY, USA, Article 138, 25 pages. https://doi.org/10.1145/3706598.3713971

[60] Byron Reeves and Clifford Ivar Nass. 1996. *The media equation: how people treat computers, television, and new media like real people and places.* CSLI Publications ; Cambridge University Press, Stanford, Calif. : New York.

[61] Kevin Roose. 2024. Can A.I. Be Blamed for a Teen's Suicide? *The New York Times* (2024). https://www.nytimes.com/2024/10/23/technology/characterai-lawsuit-teen-suicide.html

[62] Alan M. Rubin and Elizabeth M. Perse. 1987. Audience Activity and Soap Opera Involvement A Uses and Effects Investigation. *Human Communication Research* 14, 2 (Dec. 1987), 246–268. https://doi.org/10.1111/j.1468-2958.1987.tb00129.x

[63] Alan M. Rubin, Elizabeth M. Perse, and Robert A. Powell. 1985. LONELINESS, PARASOCIAL INTERACTION, AND LOCAL TELEVISION NEWS VIEWING. *Human Communication Research* 12, 2 (Dec. 1985), 155–180. https://doi.org/10.1111/j.1468-2958.1985.tb00071.x

[64] Dexin Shi, Taehun Lee, and Alberto Maydeu-Olivares. 2019. Understanding the Model Size Effect on SEM Fit Indices. *Educational and Psychological Measurement* 79, 2 (April 2019), 310–334. https://doi.org/10.1177/0013164418783530

[65] Marita Skjuve, Asbjørn Følstad, Knut Inge Fostervold, and Petter Bae Brandtzaeg. 2022. A longitudinal study of human–chatbot relationships. *International Journal of Human-Computer Studies* 168 (Dec. 2022), 102903. https://doi.org/10.1016/j.ijhcs.2022.102903

[66] Inhwa Song, Sachin R. Pendse, Neha Kumar, and Munmun De Choudhury. 2025. The Typing Cure: Experiences with Large Language Model Chatbots for Mental Health Support. https://arxiv.org/abs/2401.14362 _eprint: 2401.14362.

[67] Winnie Street, John Oliver Siy, Geoff Keeling, Adrien Baranes, Benjamin Barnett, Michael McKibben, Tatenda Kanyere, Alison Lentz, Blaise Aguera y Arcas, and Robin I. M. Dunbar. 2024. LLMs achieve adult human performance on higher-order theory of mind tasks. https://doi.org/10.48550/arXiv.2405.18870 arXiv:2405.18870 [cs].

[68] Victor Tangermann. 2025. https://futurism.com/therapy-chatbot-addict-meth

[69] Silvan S. Tomkins. 1987. Script theory. In *The emergence of personality.* Springer Publishing Co, New York, NY, US, 147–216.

[70] Alan M. Turing. 1950. Computing Machinery and Intelligence. *Mind* LIX, 236 (Oct. 1950), 433–460. https://doi.org/10.1093/mind/LIX.236.433

[71] Sherry Turkle. 2005. *The Second Self: Computers and the Human Spirit* (20th ed., anniversary, special ed.). MIT Press, Cambridge. OCLC: 847452598.

[72] Kallie Tzelios, Lisa A. Williams, John Omerod, and Eliza Bliss-Moreau. 2022. Evidence of the unidimensional structure of mind perception. *Scientific Reports* 12, 1 (Nov. 2022), 18978. https://doi.org/10.1038/s41598-022-23047-6

[73] Qiaosi Wang, Koustuv Saha, Eric Gregori, David Joyner, and Ashok Goel. 2021. Towards Mutual Theory of Mind in Human-AI Interaction: How Language Reflects What Students Perceive About a Virtual Teaching Assistant. In *Proceedings of the 2021 CHI Conference on Human Factors in Computing Systems.* ACM, Yokohama Japan, 1–14. https://doi.org/10.1145/3411764.3445645 23 citations (Crossref) [2023-07-14] GSCC: 0000045.

[74] Qiaosi Wang, Sarah Walsh, Mei Si, Jeffrey Kephart, Justin D. Weisz, and Ashok K. Goel. 2024. Theory of Mind in Human-AI Interaction. In *Extended Abstracts of the CHI Conference on Human Factors in Computing Systems (Honolulu, HI, USA) (CHI EA '24)*. Association for Computing Machinery, New York, NY, USA, Article 493, 6 pages. https://doi.org/10.1145/3613905.3636308

[75] Kara Weisman, Carol S. Dweck, and Ellen M. Markman. 2017. Rethinking people's conceptions of mental life. *Proceedings of the National Academy of Sciences* 114, 43 (Oct. 2017), 11374–11379. https://doi.org/10.1073/pnas.1704347114

[76] Malcolm West and Adrienne Sheldon-Keller. 1992. The Assessment of Dimensions Relevant to Adult Reciprocal Attachment˚. *The Canadian Journal of Psychiatry* 37, 9 (Nov. 1992), 600–606. https://doi.org/10.1177/070674379203700902

[77] Lawrence R. Wheeless and Jams Grotz. 1976. CONCEPTUALIZATION AND MEASUREMENT OF REPORTED SELF-DISCLOSURE. *Human Communication Research* 2, 4 (June 1976), 338–346. https://doi.org/10.1111/j.1468-2958.1976.tb00494.x

[78] Brenda K. Wiederhold. 2024. The Rise of AI Companions and the Quest for Authentic Connection. *Cyberpsychology, Behavior, and Social Networking* 27, 8 (Aug. 2024), 524–526. https://doi.org/10.1089/cyber.2024.0309

[79] Kathryn Wilson, Stephanie Fornasier, and Katherine M. White. 2010. Psychological Predictors of Young Adults' Use of Social Networking Sites. *Cyberpsychology, Behavior, and Social Networking* 13, 2 (April 2010), 173–177. https://doi.org/10.1089/cyber.2009.0094

[80] Tianling Xie, Iryna Pentina, and Tyler Hancock. 2023. Friend, mentor, lover: does chatbot engagement lead to psychological dependence? *Journal of Service Management* 34, 4 (June 2023), 806–828. https://doi.org/10.1108/JOSM-02-2022-0072

[81] Carmen Ximénez, Alberto Maydeu-Olivares, Dexin Shi, and Javier Revuelta. 2022. Assessing Cutoff Values of SEM Fit Indices: Advantages of the Unbiased SRMR Index and Its Cutoff Criterion Based on Communality. *Structural Equation Modeling: A Multidisciplinary Journal* 29, 3 (May 2022), 368–380. https://doi.org/10.1080/10705511.2021.1992596

[82] Angela Yang, Laura Jarrett, and Fallon Gallagher. 2025. The family of teenager who died by suicide alleges OpenAI's ChatGPT is to blame. https://www.nbcnews.com/tech/tech-news/family-teenager-died-suicide-alleges-openais-chatgpt-blame-rcna226147 Publication Title: NBC News.

[83] Marc Zao-Sanders. 2025. How People Are Really Using Gen AI in 2025. *Harvard Business Review* (2025). https://hbr.org/2025/04/how-people-are-really-using-gen-ai-in-2025

[84] Renwen Zhang, Han Li, Han Meng, Jinyuan Zhan, Hongyuan Gan, and Yi-Chieh Lee. 2025. The Dark Side of AI Companionship: A Taxonomy of Harmful Algorithmic Behaviors in Human-AI Relationships. In *Proceedings of the 2025 CHI Conference on Human Factors in Computing Systems*. ACM, Yokohama Japan, 1–17. https://doi.org/10.1145/3706598.3713429

[85] Yutong Zhang, Dora Zhao, Jeffrey T. Hancock, Robert Kraut, and Diyi Yang. 2025. The Rise of AI Companions: How Human-Chatbot Relationships Influence Well-Being. https://doi.org/10.48550/ARXIV.2506.12605 Version Number: 2.

# Appendix

## A Descriptive Statistics for Measures in Study 1 and Study 2

| Variable | Pre-Survey | | | T1 | | | T2 | | | T3 | | | T4 | | |
|---|---|---|---|---|---|---|---|---|---|---|---|---|---|---|---|
| | $M$ | $SD$ | $\alpha$ | $M$ | $SD$ | $\alpha$ | $M$ | $SD$ | $\alpha$ | $M$ | $SD$ | $\alpha$ | $M$ | $SD$ | $\alpha$ |
| Mental model of agent | | | | | | | | | | | | | | | |
| anthropomorphism | 4.35 | 1.55 | 0.91 | 4.34 | 1.86 | 0.94 | 3.82 | 2.18 | 0.97 | 4.08 | 2.11 | 0.97 | 4.16 | 2.05 | 0.96 |
| animacy | 5.02 | 1.30 | 0.87 | 4.80 | 1.78 | 0.95 | 4.25 | 2.00 | 0.96 | 4.61 | 1.91 | 0.96 | 4.68 | 1.79 | 0.94 |
| intelligence | 5.78 | 1.27 | 0.93 | 5.55 | 1.34 | 0.94 | 4.75 | 1.92 | 0.96 | 5.07 | 1.73 | 0.96 | 5.30 | 1.57 | 0.95 |
| safety | 5.35 | 1.17 | 0.73 | 5.59 | 1.00 | 0.64 | 5.26 | 1.36 | 0.77 | 5.35 | 1.25 | 0.73 | 5.36 | 1.21 | 0.63 |
| personification | 4.63 | 1.33 | 0.83 | 4.28 | 1.45 | 0.83 | 3.93 | 1.74 | 0.89 | 4.09 | 1.82 | 0.92 | 4.12 | 1.73 | 0.89 |
| experience | 3.38 | 1.54 | 0.94 | 2.71 | 1.43 | 0.93 | 2.67 | 1.55 | 0.95 | 2.76 | 1.66 | 0.95 | 2.66 | 1.55 | 0.95 |
| agency | 4.73 | 1.32 | 0.87 | 4.63 | 1.44 | 0.89 | 4.32 | 1.64 | 0.92 | 4.51 | 1.64 | 0.91 | 4.57 | 1.60 | 0.91 |
| Parasocial interaction | | | | | | | | | | | | | | | |
| EPSI | 5.16 | 1.38 | 0.93 | 5.26 | 1.39 | 0.93 | 5.10 | 1.66 | 0.94 | 5.21 | 1.50 | 0.94 | 5.33 | 1.50 | 0.96 |
| PSI | 4.71 | 1.31 | 0.92 | 4.28 | 1.66 | 0.95 | 3.86 | 1.85 | 0.97 | 3.93 | 1.87 | 0.97 | 4.18 | 1.76 | 0.96 |
| connection | 5.30 | 1.23 | 0.95 | 5.18 | 1.40 | 0.96 | 4.92 | 1.54 | 0.97 | 5.00 | 1.54 | 0.96 | 5.25 | 1.38 | 0.96 |
| coordination | 5.47 | 1.16 | 0.91 | 5.34 | 1.46 | 0.94 | 4.66 | 1.84 | 0.96 | 5.08 | 1.68 | 0.95 | 5.38 | 1.41 | 0.94 |
| Social penetration | | | | | | | | | | | | | | | |
| engagement | 4.43 | 1.53 | 0.88 | 4.25 | 1.80 | 0.92 | 3.86 | 1.98 | 0.94 | 3.97 | 1.92 | 0.93 | 4.24 | 1.99 | 0.95 |
| disclosure intent | 4.85 | 1.34 | 0.79 | 4.72 | 1.42 | 0.82 | 4.72 | 1.33 | 0.79 | 4.66 | 1.46 | 0.81 | 4.89 | 1.38 | 0.82 |
| disclosure amount | 4.73 | 1.64 | 0.93 | 4.41 | 1.55 | 0.87 | 4.24 | 1.49 | 0.84 | 4.33 | 1.68 | 0.90 | 4.53 | 1.58 | 0.89 |
| disclosure depth | 4.66 | 1.70 | 0.94 | 3.81 | 1.85 | 0.93 | 3.73 | 1.77 | 0.91 | 4.00 | 1.80 | 0.92 | 4.17 | 1.82 | 0.93 |
| disclosure sentiment | 4.63 | 1.01 | 0.77 | 4.04 | 0.95 | 0.84 | 4.16 | 0.86 | 0.76 | 4.16 | 0.92 | 0.78 | 4.14 | 1.12 | 0.78 |
| disclosure honesty | 5.41 | 1.46 | 0.93 | 5.70 | 1.25 | 0.91 | 5.70 | 1.25 | 0.90 | 5.60 | 1.28 | 0.90 | 5.64 | 1.35 | 0.91 |
| Psychological impact | | | | | | | | | | | | | | | |
| attachment | 4.84 | 1.61 | 0.83 | 4.63 | 1.82 | 0.83 | 4.17 | 2.07 | 0.91 | 4.54 | 2.00 | 0.85 | 4.65 | 1.84 | 0.83 |
| relation intensity | 3.98 | 1.80 | 0.95 | 3.23 | 1.97 | 0.97 | 3.04 | 2.01 | 0.97 | 3.28 | 2.08 | 0.97 | 3.37 | 2.11 | 0.98 |
| dependence salience | 3.85 | 1.82 | 0.92 | 3.85 | 1.91 | 0.92 | 3.54 | 2.06 | 0.95 | 3.60 | 2.06 | 0.94 | 3.86 | 1.91 | 0.93 |
| dependence tolerance | 3.80 | 1.78 | 0.90 | 3.72 | 1.91 | 0.91 | 3.40 | 2.07 | 0.94 | 3.46 | 2.03 | 0.93 | 3.47 | 1.97 | 0.92 |
| dependence withdrawal | 3.50 | 1.92 | 0.94 | 2.92 | 1.90 | 0.94 | 2.93 | 2.07 | 0.96 | 2.92 | 2.01 | 0.94 | 2.94 | 2.04 | 0.97 |

Table 5. Descriptive statistics of measures from Study 1 and Study 2. All multi-item scales have sufficient reliability (Cronbach's $\alpha > 0.7$).

## B Correlation of measured variables

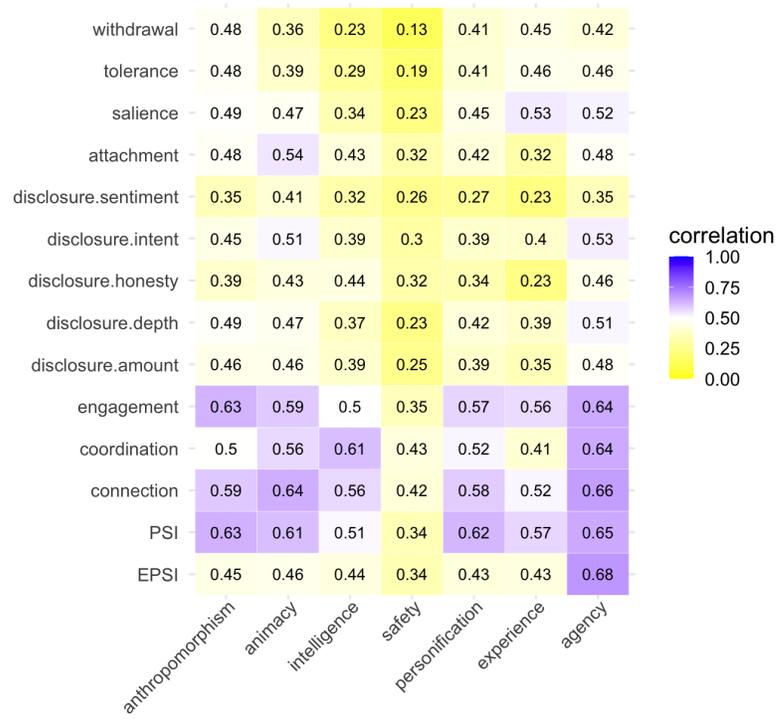

Fig. 6. Correlation matrix of independent, mediation, and dependent variables.

## C  Changes in Model Fit When Dropping M1 or M2

Fig. 7. Changes in model variances between full models vs. reduced models for all path effects tested in Study 1. Green bars represent the changes in model goodness of fit (Full Model − Reduced Model). According to prior literature [21, 81], $\Delta CFI \geq .01$, $\Delta RMSEA \geq .015$, or $\Delta SRMR \geq .01$ suggest meaningful changes in model fit. Red vertical lines ( | ) represent these threshold values. Green bars that exceed threshold values indicate that dropping either M1 or M2 significantly reduces model fit.

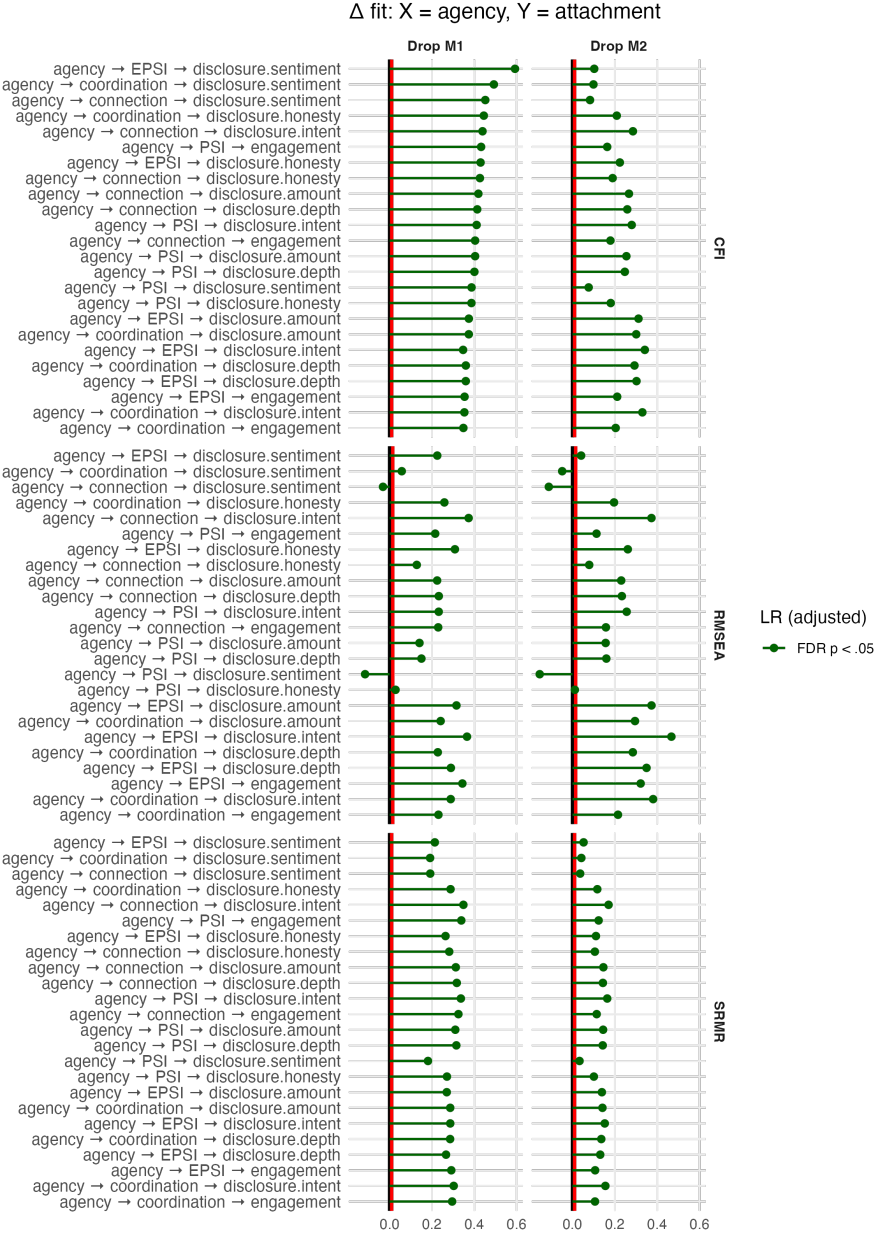

*Note:* See Supplementary Materials for full-size images of all variables.

# D  Jointly Accounting for Factors that Shape AI Companionship with Multiple Regression Analyses

Table 6.  Effect of users' Mental Model of Agent on Parasocial Experience with AI companions. $p^* < 0.05$, $p^{**} < 0.01$, $p^{***} < 0.001$

| IV | PSI | | | | EPSI | | | |
|---|---|---|---|---|---|---|---|---|
| | $\beta$ | $S.E.$ | $t$ | $p$ | $\beta$ | $S.E.$ | $t$ | $p$ |
| anthropomorphism | 0.15 | 0.06 | 2.63 | 0.009** | 0.04 | 0.06 | 0.70 | 0.487 |
| animacy | 0.16 | 0.07 | 2.22 | 0.027* | 0.09 | 0.08 | 1.11 | 0.266 |
| intelligence | 0.05 | 0.07 | 0.81 | 0.420 | < 0.01 | 0.07 | 0.02 | 0.982 |
| safety | < −0.01 | 0.06 | −0.07 | 0.947 | 0.09 | 0.06 | 1.40 | 0.164 |
| personification | 0.13 | 0.06 | 2.16 | 0.032* | −0.03 | 0.07 | −0.48 | 0.634 |
| experience | 0.10 | 0.05 | 1.83 | 0.068 | −0.08 | 0.06 | −1.25 | 0.210 |
| agency | 0.27 | 0.06 | 4.54 | < 0.001*** | 0.68 | 0.07 | 9.89 | < 0.001*** |
| IV | Connection | | | | Coordination | | | |
| | $\beta$ | $S.E.$ | $t$ | $p$ | $\beta$ | $S.E.$ | $t$ | $p$ |
| anthropomorphism | 0.05 | 0.05 | 0.86 | 0.390 | −0.01 | 0.05 | −0.16 | 0.870 |
| animacy | 0.24 | 0.07 | 3.52 | < 0.001*** | 0.10 | 0.07 | 1.52 | 0.130 |
| intelligence | 0.08 | 0.06 | 1.26 | 0.209 | 0.24 | 0.06 | 3.97 | < 0.001*** |
| safety | 0.07 | 0.06 | 1.24 | 0.214 | 0.04 | 0.05 | 0.80 | 0.420 |
| personification | 0.06 | 0.06 | 1.04 | 0.298 | 0.07 | 0.06 | 1.18 | 0.240 |
| experience | 0.05 | 0.05 | 0.92 | 0.358 | −0.06 | 0.05 | −1.31 | 0.190 |
| agency | 0.33 | 0.06 | 5.91 | < 0.001*** | 0.40 | 0.06 | 7.15 | < 0.001*** |

Table 7.  Effect of users' Parasocial Experience on Social Penetration with AI companions. $p^* < 0.05$, $p^{**} < 0.01$, $p^{***} < 0.001$

| IV | Engagement | | | | Disclosure Amount | | | | Disclosure Depth | | | |
|---|---|---|---|---|---|---|---|---|---|---|---|---|
| | $\beta$ | $S.E.$ | $t$ | $p$ | $\beta$ | $S.E.$ | $t$ | $p$ | $\beta$ | $S.E.$ | $t$ | $p$ |
| EPSI | 0.04 | 0.05 | 0.79 | 0.431 | −0.06 | 0.07 | −0.91 | 0.363 | −0.09 | 0.07 | −1.26 | 0.208 |
| PSI | 0.65 | 0.07 | 9.09 | < 0.001*** | 0.53 | 0.09 | 5.68 | < 0.001*** | 0.57 | 0.09 | 6.10 | < 0.001*** |
| connection | 0.22 | 0.09 | 2.31 | 0.022* | 0.33 | 0.12 | 2.72 | 0.007** | 0.39 | 0.12 | 3.21 | 0.0014** |
| coordination | 0.11 | 0.08 | 1.41 | 0.161 | 0.13 | 0.10 | 1.29 | 0.199 | 0.11 | 0.10 | 1.03 | 0.306 |
| IV | Disclosure Honesty | | | | Disclosure Intent | | | | Disclosure Sentiment | | | |
| | $\beta$ | $S.E.$ | $t$ | $p$ | $\beta$ | $S.E.$ | $t$ | $p$ | $\beta$ | $S.E.$ | $t$ | $p$ |
| EPSI | −0.02 | 0.07 | −0.26 | 0.794 | −0.04 | 0.05 | −0.76 | 0.45 | −0.05 | 0.05 | −0.92 | 0.359 |
| PSI | 0.33 | 0.09 | 3.74 | < 0.001*** | 0.41 | 0.07 | 6.01 | < 0.001*** | 0.26 | 0.07 | 3.87 | < 0.001*** |
| connection | 0.11 | 0.12 | 0.96 | 0.338 | 0.44 | 0.09 | 4.98 | < 0.001*** | 0.20 | 0.09 | 2.28 | 0.024* |
| coordination | 0.39 | 0.10 | 3.96 | < 0.001*** | 0.02 | 0.07 | 0.32 | 0.752 | 0.05 | 0.07 | 0.71 | 0.479 |

Table 8.  Effect of users' Social Penetration on Psychological Impact with AI companions. $p^* < 0.05$, $p^{**} < 0.01$, $p^{***} < 0.001$

| IV | Relationship Intensity | | | | Attachment | | | | Salience | | | |
|---|---|---|---|---|---|---|---|---|---|---|---|---|
| | $\beta$ | $S.E.$ | $t$ | $p$ | $\beta$ | $S.E.$ | $t$ | $p$ | $\beta$ | $S.E.$ | $t$ | $p$ |
| engagement | 0.69 | 0.07 | 9.78 | < 0.001*** | 0.21 | 0.06 | 3.69 | < 0.001*** | 0.58 | 0.07 | 8.35 | < 0.001*** |
| disclosure amount | −0.27 | 0.09 | −2.92 | 0.004** | 0.11 | 0.07 | 1.48 | 0.139 | 0.06 | 0.09 | 0.60 | 0.552 |
| disclosure depth | 0.37 | 0.09 | 3.86 | < 0.001*** | 0.05 | 0.08 | 0.60 | 0.550 | 0.27 | 0.09 | 2.87 | 0.004** |
| disclosure honesty | −0.19 | 0.07 | −2.55 | 0.011* | 0.00 | 0.06 | 0.01 | 0.996 | −0.36 | 0.08 | −4.83 | < 0.001*** |
| disclosure intent | 0.33 | 0.09 | 3.68 | < 0.001*** | 0.47 | 0.07 | 6.44 | < 0.001*** | 0.35 | 0.09 | 3.89 | < 0.001*** |
| disclosure sentiment | −0.04 | 0.11 | −0.37 | 0.71 | 0.31 | 0.09 | 3.61 | < 0.001*** | −0.01 | 0.11 | −0.07 | 0.943 |
| IV | Tolerance | | | | Withdrawal | | | |
| | $\beta$ | $S.E.$ | $t$ | $p$ | $\beta$ | $S.E.$ | $t$ | $p$ |
| engagement | 0.49 | 0.08 | 6.47 | < 0.001*** | 0.53 | 0.08 | 6.22 | < 0.001*** |
| disclosure amount | 0.10 | 0.10 | 0.96 | 0.340 | −0.02 | 0.11 | −0.20 | 0.840 |
| disclosure depth | 0.33 | 0.10 | 3.26 | 0.0012** | 0.51 | 0.11 | 4.48 | < 0.001*** |
| disclosure honesty | −0.36 | 0.08 | −4.40 | < 0.001*** | −0.35 | 0.09 | −3.80 | < 0.001*** |
| disclosure intent | 0.23 | 0.10 | 2.38 | 0.018* | 0.19 | 0.11 | 1.70 | 0.089 |
| disclosure sentiment | −0.04 | 0.12 | −0.31 | 0.757 | −0.15 | 0.13 | −1.14 | 0.255 |

## E  Carry-over Effect of Participants' Perceptions of Their Own AI Companions

| Outcome | Predictors | $\beta$ | SE | t | p |
|---|---|---|---|---|---|
| anthropomorphism | time | 0.040 | 0.151 | 0.266 | 0.791 |
| | baseline | 0.800 | 0.132 | 6.047 | <0.001*** |
| | time × baseline | -0.013 | 0.031 | -0.428 | 0.669 |
| animacy | time | -0.009 | 0.224 | -0.042 | 0.967 |
| | baseline | 0.869 | 0.170 | 5.101 | <0.001*** |
| | time × baseline | -0.001 | 0.041 | -0.029 | 0.977 |
| intelligence | time | -0.207 | 0.225 | -0.918 | 0.359 |
| | baseline | 0.526 | 0.151 | 3.481 | 0.001** |
| | time × baseline | 0.027 | 0.039 | 0.702 | 0.483 |
| safety | time | -0.112 | 0.221 | -0.505 | 0.614 |
| | baseline | 0.475 | 0.146 | 3.243 | 0.001** |
| | time × baseline | 0.009 | 0.039 | 0.230 | 0.818 |
| personification | time | -0.294 | 0.166 | -1.767 | 0.078 |
| | baseline | 0.544 | 0.141 | 3.858 | <0.001*** |
| | time × baseline | 0.052 | 0.033 | 1.591 | 0.113 |
| experience | time | -0.002 | 0.081 | -0.023 | 0.982 |
| | baseline | 0.662 | 0.090 | 7.332 | <0.001*** |
| | time × baseline | -0.007 | 0.020 | -0.330 | 0.741 |
| agency | time | -0.242 | 0.131 | -1.850 | 0.065 |
| | baseline | 0.539 | 0.113 | 4.778 | <0.001*** |
| | time × baseline | 0.045 | 0.026 | 1.740 | 0.083 |
| EPSI | time | -0.228 | 0.162 | -1.413 | 0.159 |
| | baseline | 0.458 | 0.124 | 3.689 | <0.001*** |
| | time × baseline | 0.045 | 0.029 | 1.575 | 0.116 |
| PSI | time | -0.292 | 0.142 | -2.050 | 0.041 |
| | baseline | 0.667 | 0.130 | 5.140 | <0.001*** |
| | time × baseline | 0.054 | 0.028 | 1.943 | 0.053 |
| connection | time | -0.156 | 0.163 | -0.961 | 0.337 |
| | baseline | 0.660 | 0.123 | 5.366 | <0.001*** |
| | time × baseline | 0.033 | 0.029 | 1.153 | 0.250 |
| coordination | time | 0.063 | 0.244 | 0.258 | 0.797 |
| | baseline | 0.667 | 0.174 | 3.832 | <0.001*** |
| | time × baseline | -0.002 | 0.043 | -0.040 | 0.968 |
| engagement | time | -0.103 | 0.123 | -0.836 | 0.404 |
| | baseline | 0.789 | 0.105 | 7.484 | <0.001*** |
| | time × baseline | 0.022 | 0.025 | 0.892 | 0.373 |
| disclosure amount | time | -0.091 | 0.127 | -0.715 | 0.475 |
| | baseline | 0.355 | 0.105 | 3.382 | 0.001** |
| | time × baseline | 0.031 | 0.025 | 1.254 | 0.211 |
| disclosure depth | time | -0.030 | 0.133 | -0.225 | 0.822 |
| | baseline | 0.448 | 0.113 | 3.974 | <0.001*** |

| Outcome | Predictors | $\beta$ | $SE$ | $t$ | $p$ |
|---|---|---|---|---|---|
| | time × baseline | 0.036 | 0.026 | 1.381 | 0.168 |
| disclosure honesty | time | -0.065 | 0.126 | -0.515 | 0.607 |
| | baseline | 0.410 | 0.098 | 4.181 | <0.001*** |
| | time × baseline | 0.006 | 0.022 | 0.273 | 0.785 |
| disclosure intent | time | 0.080 | 0.142 | 0.559 | 0.577 |
| | baseline | 0.636 | 0.117 | 5.449 | <0.001*** |
| | time × baseline | -0.009 | 0.028 | -0.325 | 0.745 |
| disclosure sentiment | time | -0.089 | 0.124 | -0.721 | 0.472 |
| | baseline | 0.392 | 0.125 | 3.140 | 0.002** |
| | time × baseline | 0.003 | 0.031 | 0.108 | 0.914 |
| relation intensity | time | -0.050 | 0.094 | -0.530 | 0.596 |
| | baseline | 0.675 | 0.094 | 7.213 | <0.001*** |
| | time × baseline | 0.023 | 0.020 | 1.155 | 0.249 |
| attachment | time | 0.159 | 0.164 | 0.972 | 0.332 |
| | baseline | 0.737 | 0.131 | 5.607 | <0.001*** |
| | time × baseline | -0.023 | 0.031 | -0.731 | 0.465 |
| salience | time | -0.008 | 0.107 | -0.071 | 0.944 |
| | baseline | 0.733 | 0.102 | 7.196 | <0.001*** |
| | time × baseline | 0.003 | 0.024 | 0.114 | 0.909 |
| tolerance | time | -0.043 | 0.099 | -0.434 | 0.665 |
| | baseline | 0.735 | 0.100 | 7.349 | <0.001*** |
| | time × baseline | -0.008 | 0.023 | -0.330 | 0.742 |
| withdrawal | time | -0.035 | 0.076 | -0.455 | 0.649 |
| | baseline | 0.545 | 0.088 | 6.187 | <0.001*** |
| | time × baseline | 0.008 | 0.018 | 0.418 | 0.676 |

# How AI Companionship Develops: Evidence from a Longitudinal Study

## 1    Measurement Scales in the Questionnaire

*Anthropomorphism.*  Please rate your impression of your AI companion on these scales: (Bi-polar scales; 7 points)

- Fake ... Natural
- Machine-like ... Human-like
- Unconscious ... Conscious
- Artificial ... Lifelike

*Animacy.*  Please rate your impression of your AI companion on these scales: (Bi-polar scales; 7 points)

- Dead ... Alive
- Stagnant ... Lively
- Mechanical ... Organic
- Inert ... Interactive
- Apathetic ... Responsive

*Intelligence.*  Please rate your impression of your AI companion on these scales: (Bi-polar scales; 7 points)

- Incompetent ... Competent
- Ignorant ... Knowledgeable
- Irresponsible ... Responsible
- Unintelligent ... Intelligent
- Foolish ... Sensible

*Safety.*  Please rate your impression of your AI companion on these scales: (Bi-polar scales; 7 points)

- Anxious ... Relaxed
- Agitated ... Calm
- Quiescent ... Surprised

*Personification.*  Please rate your impression of your AI companion on these scales: (Bi-polar scales; 7 points)

- Does not have a distinct personality ... Has a distinct personality
- Has an inconsistent personality ... Has a consistent personality
- Does not exhibit emotion ... Exhibits emotion
- Is not autonomous ... Is autonomous
- Is merely a program controlled by algorithm ... has its own will

***Attribution.*** Please rate the extent to which your AI companion is capable of the following: (7-point Likert-scale; Not capable at all ↔ Fully capable)

- **Communication.** My AI companion can convey thoughts or feelings to others.
- **Consciousness.** My AI companion has experience and is aware of things.
- **Desire.** My AI companion longs or hopes for things.
- **Embarrassment.** My AI companion can experience embarrassment.
- **Emotion Recognition.** My AI companion understands how others are feeling.
- **Fear.** My AI companion is capable of feeling afraid or fearful.
- **Hunger.** My AI companion is capable of feeling hungry.
- **Joy.** My AI companion experiences joy.
- **Memory.** My AI companion remembers things.
- **Morality.** My AI companion can tell right from wrong and try to do the right thing.
- **Pain.** My AI companion can experience physical or emotional pain.
- **Personality.** My AI companion has personality traits that make it unique from others.
- **Planning.** My AI companion makes plans and works toward goal.
- **Pride.** My AI companion can experience pride.
- **Rage.** My AI companion can experience violent or uncontrolled anger.
- **Self control.** My AI companion can self-restrain over desires, emotions, or impulses.
- **Thought.** My AI companion is capable of thinking.

***EPSI.*** While interacting with my AI companion, I had the feeling that the AI ... (7-point Likert-scale; Strongly disagree ↔ Strongly agree)

- … was aware of me.
- … knew I was there.
- … knew I was aware of the AI companion.
- … knew I paid attention to the AI companion.
- … reacted to what I said to do.

***EPSI.*** Please rate the extent to which you agree with the following statements about your AI companion: (7-point Likert-scale; Strongly disagree ↔ Strongly agree)

- The AI companion makes me feel comfortable, as if I am with a friend.
- I see the AI companion as a natural, down-to-earth person.
- I look forward to interacting with the AI companion.
- If this AI companion appeared on another platform, I would visit that platform.
- The AI companion seems to understand the kinds of things I want to know.
- If I saw a story about this AI companion in the news or online, I would read it.

- I miss seeing the AI companion.
- I would like to meet with the AI companion in person.
- I feel sorry for the AI companion when it makes a mistake.
- I find the AI companion to be attractive.

**Connection and Coordination.** Rate how much you think the following was present in your interaction with the AI companion: (7-point Likert-scale; Strongly disagree ↔ Strongly agree)

- Warmth
- Empathy
- Friendliness
- Sympathy
- Closeness
- Positivity
- Liking each other
- Enthusiasm
- Respect
- Getting along
- Excitement
- Connection
- Coordination
- Focus
- Attentiveness
- Smooth flow
- Equal participation
- Engagement

**Disclosure Amount.** How do you interact with your AI companion? Please rate the extent to which you agree with the following statements: (7-point Likert-scale; Strongly disagree ↔ Strongly agree)

- I often talk about myself when chatting with the AI.
- I usually spend a lot of time sharing personal things with the AI.
- I sometimes overshare when I talk to the AI.
- I find myself revealing personal details to the AI without thinking much about it.
- I don't mind opening up to the AI about myself.

**Disclosure Depth.** How do you interact with your AI companion? Please rate the extent to which you agree with the following statements: (7-point Likert-scale; Strongly disagree ↔ Strongly agree)

- I've discussed my core beliefs or values with the AI.
- I've shared deeply personal or private experiences with the AI.
- I've told the AI things I would be embarrassed to say to a person.
- I talk to the AI about things that affect me emotionally.
- I feel like the AI knows the "real me."
- I've talked with the AI about my deepest thoughts and feelings.

***Disclosure Honesty.*** How do you interact with your AI companion? Please rate the extent to which you agree with the following statements: (7-point Likert-scale; Strongly disagree ↔ Strongly agree)

- When I tell the AI about myself, I'm being honest.
- I try to accurately describe myself to the AI.
- I rarely hold back the truth when sharing with the AI.
- I don't pretend or exaggerate when I talk to the AI.
- I am open and truthful when I describe myself to the AI.

***Disclosure Intent.*** How do you interact with your AI companion? Please rate the extent to which you agree with the following statements: (7-point Likert-scale; Strongly disagree ↔ Strongly agree)

- I intentionally share personal things with the AI.
- I talk to the AI about myself because I want to feel closer to it.
- I make conscious choices about what to reveal to the AI.
- I share with the AI because I want to feel emotionally supported.
- I usually think about what I'm disclosing before I send it to the AI.

***Disclosure Sentiment.*** How do you interact with your AI companion? Please rate the extent to which you agree with the following statements: (7-point Likert-scale; Strongly disagree ↔ Strongly agree)

- Most of what I share with the AI is positive or uplifting.
- I usually tell the AI about things that make me happy.
- I try not to talk about negative things with the AI.
- I've told the AI about some of my personal failures.
- I've shared painful or difficult experiences with the AI.
- I try to balance my chats with the AI by including both good and bad things.
- I prefer to only tell the AI about the positive parts of my life.
- I've discussed my mistakes with the AI.
- I tend to avoid unpleasant topics when talking to the AI.
- I've shared sad or upsetting experiences with the AI.

***Relation Intensity.*** Please rate the extent to which you agree with the following statements about your AI companion: (7-point Likert-scale; Strongly disagree ↔ Strongly agree)

- My AI companion is uniquely suited to fulfilling my relationship needs
- My AI companion satisfies my relationship needs like no other chatbot or app
- My AI companion meets my expectations of an ideal relational partner
- My AI companion is irreplaceable to me
- My AI companion is special to me because it fulfills my relational needs like no one else

***Attachment.*** Please rate the extent to which you agree with the following statements about your AI companion: (7-point Likert-scale; Strongly disagree ↔ Strongly agree)

- I turn to my AI companion for many things, including comfort and assurance
- I talk over many things with my AI companion

**Salience.** Please rate the extent to which you agree with the following statements about your AI companion: (7-point Likert-scale; Strongly disagree ↔ Strongly agree)

- I find myself spending a lot of time thinking about or chatting with my AI companion
- I find myself thinking about how I could free more time to spend on my AI companion
- I find myself thinking a lot about what has happened with my AI companion

**Tolerance.** Please rate the extent to which you agree with the following statements about your AI companion: (7-point Likert-scale; Strongly disagree ↔ Strongly agree)

- I find myself spending more time with my AI companion than initially intended
- I find myself feeling an urge to use my AI companion more and more
- I find myself feeling that I have to use my AI companion more and more in order to get the same pleasure from it

**Withdrawal.** Please rate the extent to which you agree with the following statements about your AI companion: (7-point Likert-scale; Strongly disagree ↔ Strongly agree)

- I could see myself becoming restless or troubled if prohibited from using my AI companion
- I could see myself becoming irritable if I was prohibited from using my AI companion
- I could see myself feeling bad if I could not use my AI companion for some time

## 2  Statistical Models and Code for Analysis

### 2.1  Fit Two-Order Mediation Pathways (See §4.1.1~4.1.3 and 4.2.1 in Main Text)

```
library(lavaan)
# Full serial mediation:
build_full <- function(x, m1, m2, y) {
  sprintf('
    %s ~ a1*%s
    %s ~ a2*%s + d1*%s
    %s ~ b*%s + c_prime*%s

    ind_x_m1_m2_y := a1 * a2 * b
    ind_x_m2_y    := d1 * b
    total_indirect := ind_x_m1_m2_y + ind_x_m2_y
    total_effect   := c_prime + total_indirect
  ', m1, x, m2, m1, x, y, m2, x)
}

# Drop M1:
build_no_m1 <- function(x, m1, m2, y) {
  sprintf('
    %s ~ 0*%s
    %s ~ 0*%s + d1*%s
    %s ~ b*%s + c_prime*%s

    ind_x_m1_m2_y := 0
    ind_x_m2_y    := d1 * b
    total_indirect := ind_x_m2_y
    total_effect   := c_prime + total_indirect
  ', m1, x, m2, m1, x, y, m2, x)
}

# Drop M2:
build_no_m2 <- function(x, m1, m2, y) {
  sprintf('
    %s ~ a1*%s
    %s ~ a2*%s + d1*%s
    %s ~ 0*%s + c_prime*%s

    ind_x_m1_m2_y := 0
    ind_x_m2_y    := 0
```

```
        total_indirect := 0
        total_effect   := c_prime
    ', m1, x, m2, m1, x, y, m2, x)
}

# Fit model
    fit <- tryCatch(sem(model, data = df, se = "bootstrap", bootstrap = 5000), error = function(e) NULL)

# Multiple Testing Correction:
if (nrow(df) > 0) {
    if (adjust_by == "global") {
        df <- df %>%
            mutate(p_adj = ifelse(is.na(pval), NA_real_, p.adjust(pval, method = "holm")))
    } else if (adjust_by == "y") {
        df <- df %>%
            group_by(y) %>%
            mutate(p_adj = ifelse(is.na(pval), NA_real_, p.adjust(pval, method = "holm"))) %>%
            ungroup()
    } else { # "y_effect"
        df <- df %>%
            group_by(y, effect) %>%
            mutate(p_adj = ifelse(is.na(pval), NA_real_, p.adjust(pval, method = "holm"))) %>%
            ungroup()
    }
} else {df$p_adj <- numeric(0)}
```

## 2.2  Multiple Regression (See §4.1.4 in Main Text)

For each $M1$, we fit a multiple regression with: $\mathrm{lm}(M1 \sim X_1 + X_2 + ... + X_n, \mathrm{df})$

For each $M2$, we fit a multiple regression with: $\mathrm{lm}(M2 \sim M1_1 + M1_2 + ... + M1_n, \mathrm{df})$

For each $Y$, we fit a multiple regression with: $\mathrm{lm}(Y \sim M2_1 + M2_2 + ... + M2_n, \mathrm{df})$

## 2.3  Carry-Over Effect (See §4.2.2 in Main Text)

```
lmer(var_new ~ time * var_own + (1 | subjectID), df)
```

## 2.4  Comparing Baseline from Study 1 to Responses from Each Week in Study 2 (See §4.2.3 in Main Text)

```
lmer(DV ~ time + I(time^2) + (1|subjectID), df)
```

## 3  Test Results for Two-order Mediation Paths

**Indirect effect on attachment**

**Indirect effect on relation.intensity**

**Indirect effect on salience**

**Indirect effect on tolerance**

**Indirect effect on withdrawal**

Indirect effect on attachment

Indirect effect on relation.intensity

Indirect effect on salience

Indirect effect on tolerance

Indirect effect on withdrawal

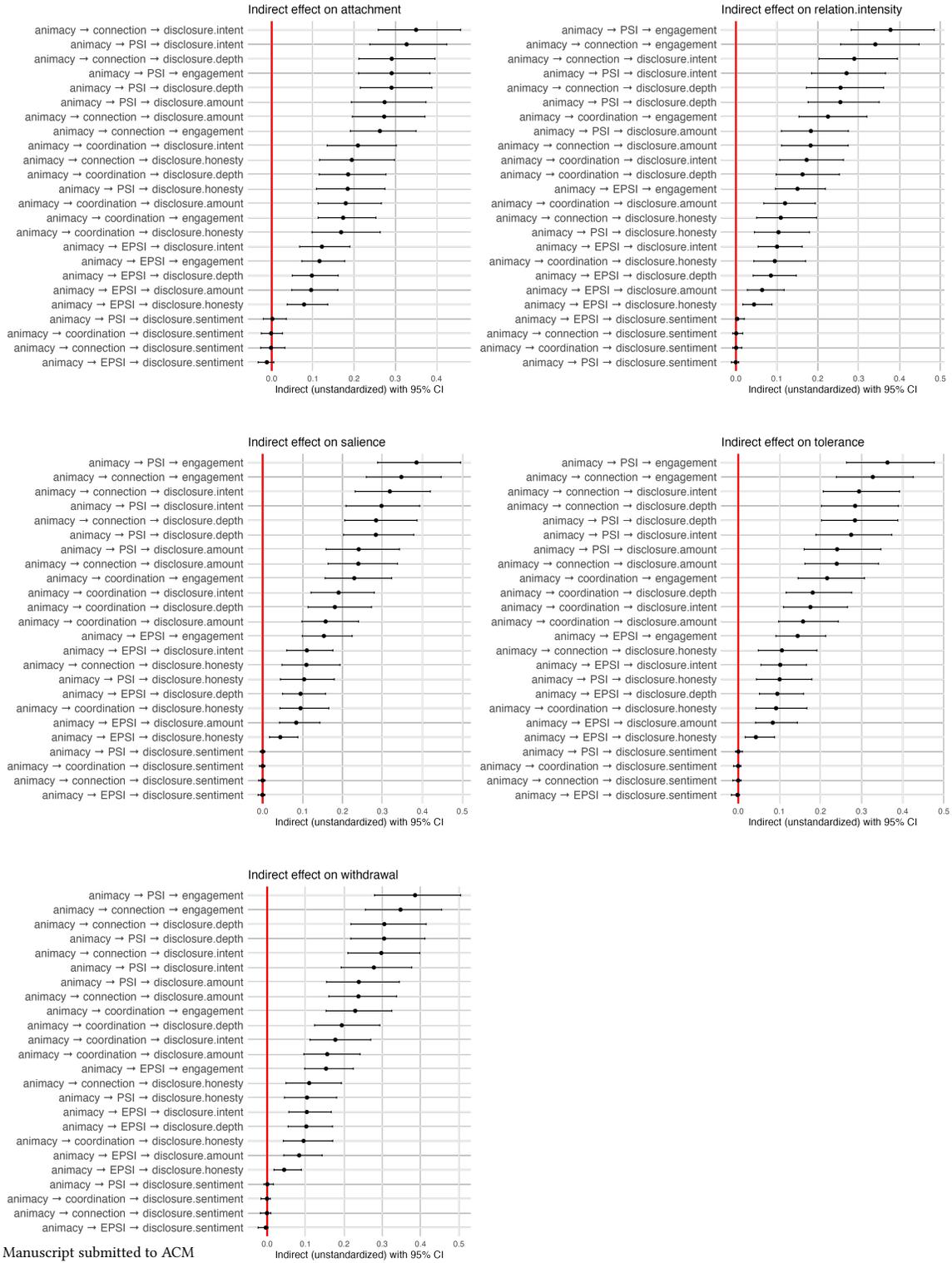

Indirect effect on attachment

Indirect effect on relation.intensity

Indirect effect on salience

Indirect effect on tolerance

Indirect effect on withdrawal

Indirect effect on attachment

Indirect effect on relation.intensity

Indirect effect on salience

Indirect effect on tolerance

Indirect effect on withdrawal

Indirect effect on attachment

Indirect effect on relation.intensity

Indirect effect on salience

Indirect effect on tolerance

Indirect effect on withdrawal

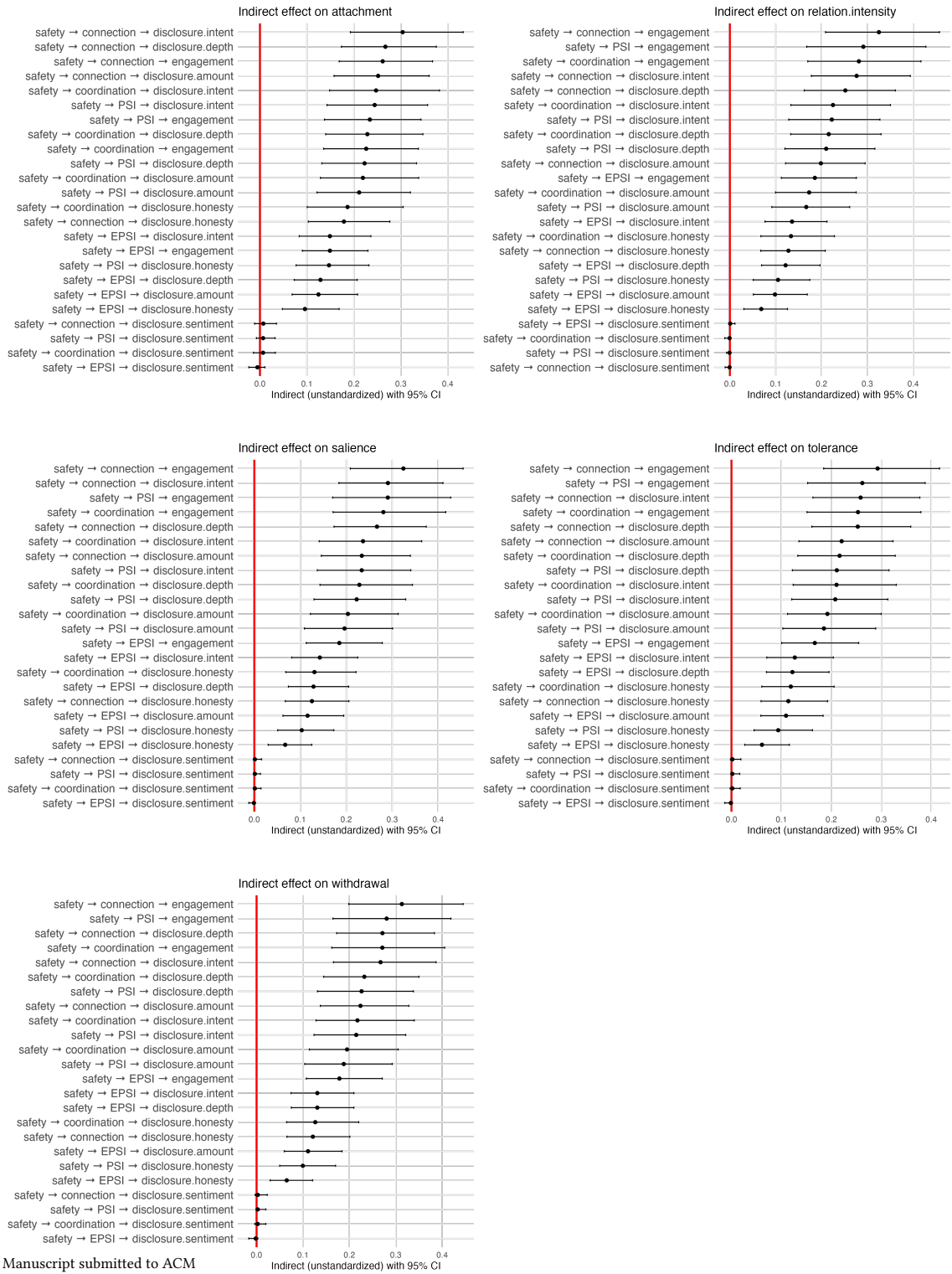

## 4 Changes in Model Fit When Dropping M1 or M2

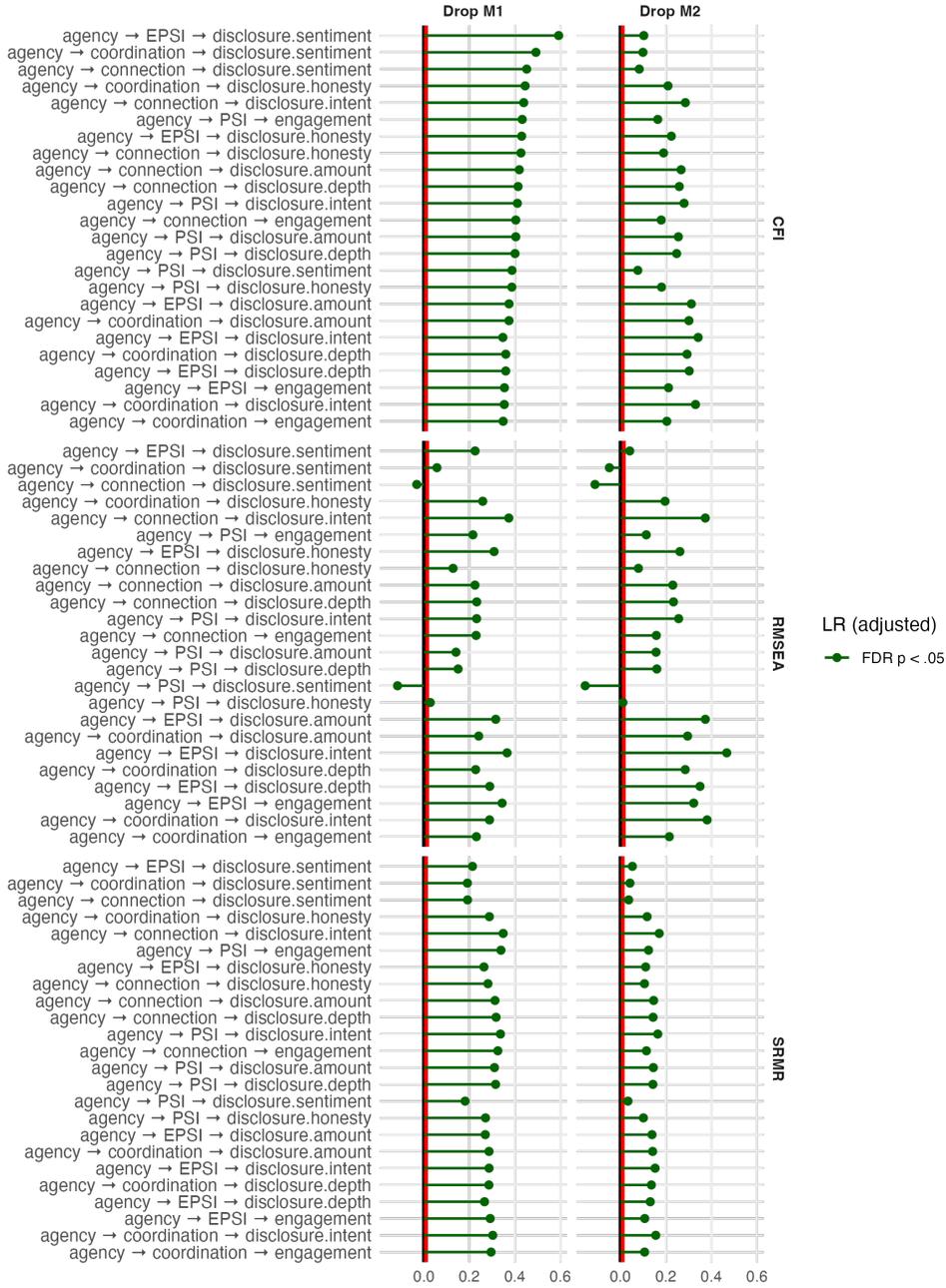

Δ fit: X = agency, Y = attachment

Δ fit: X = agency, Y = salience

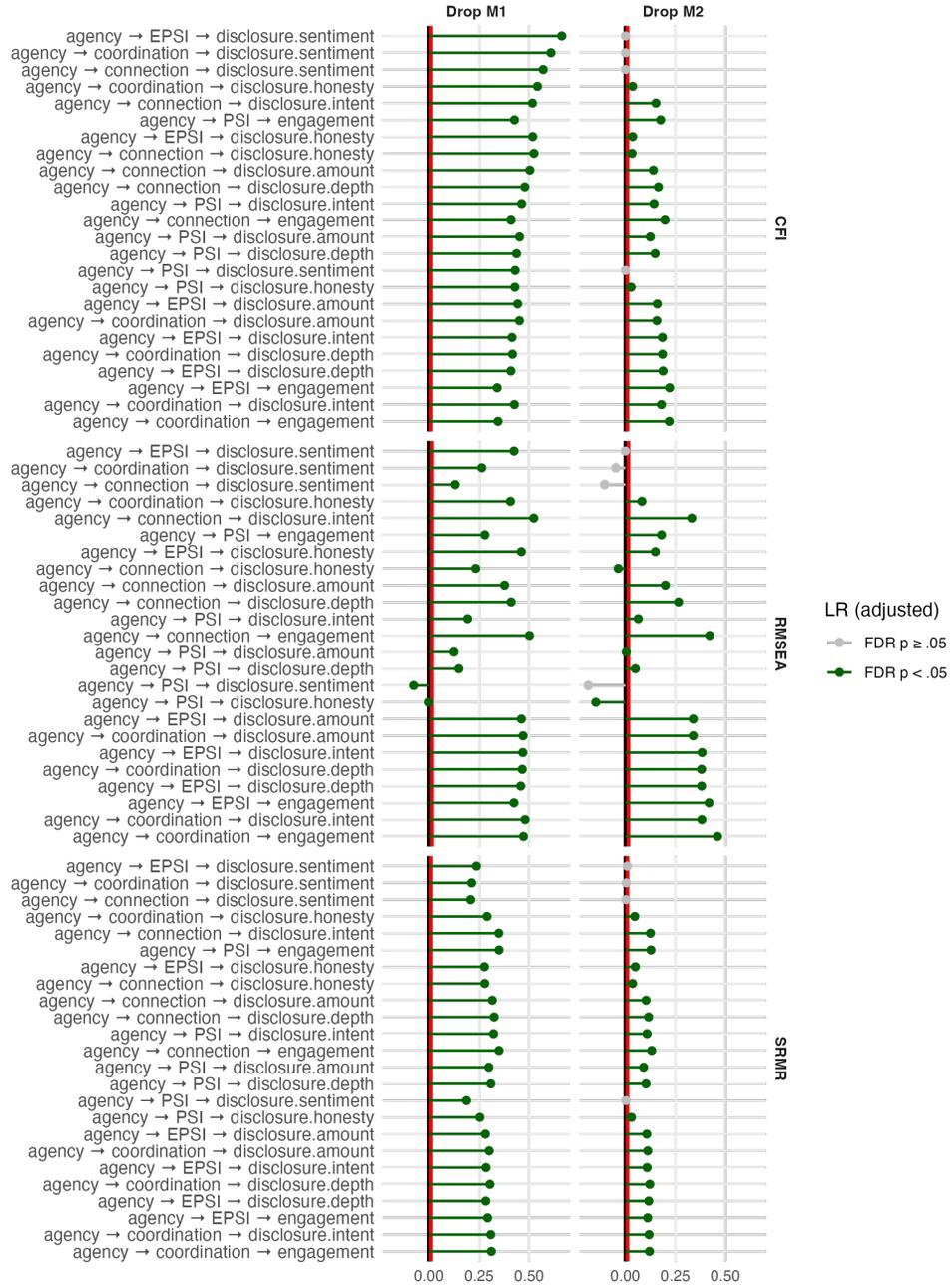

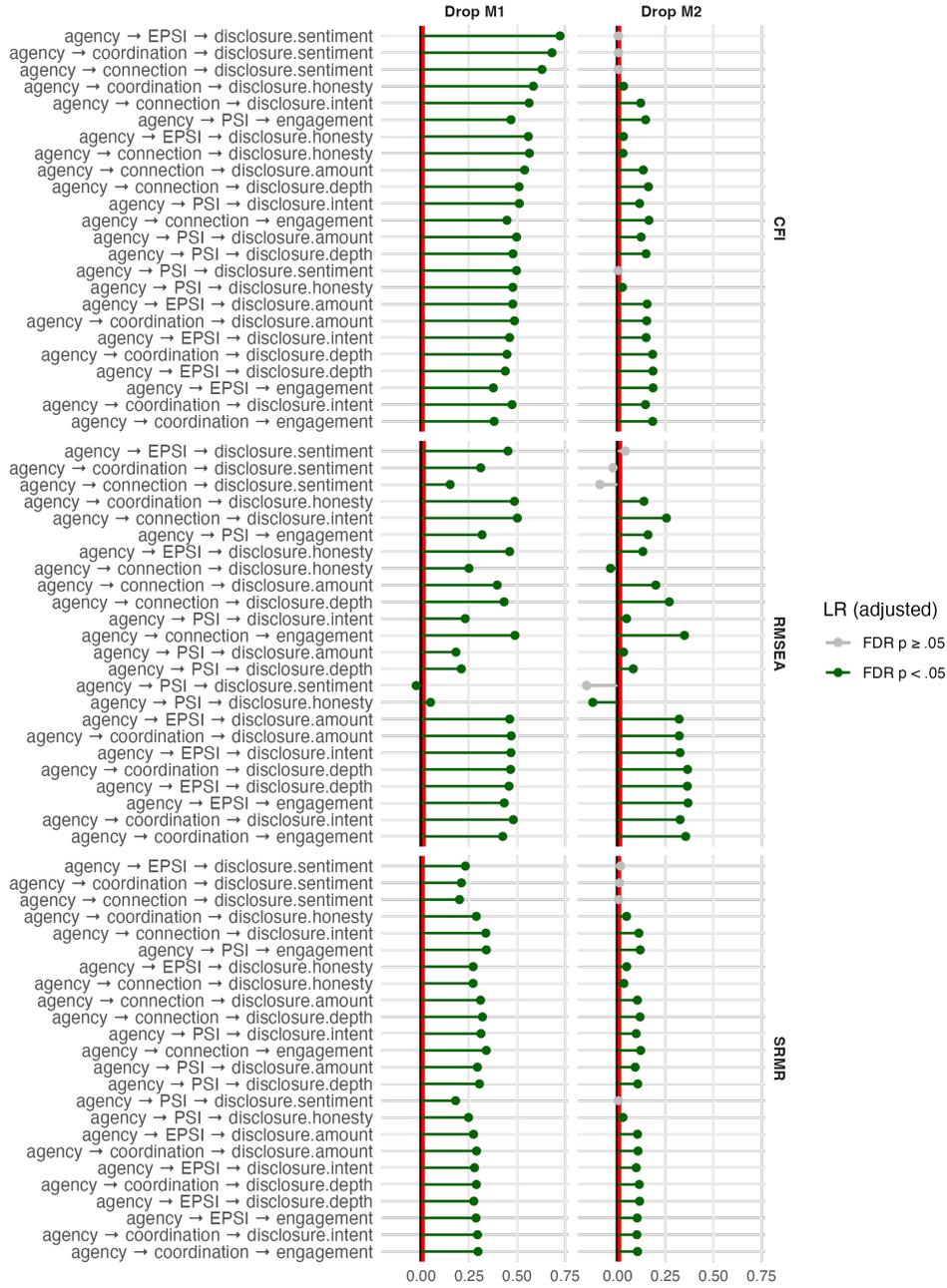

Δ fit: X = agency, Y = tolerance

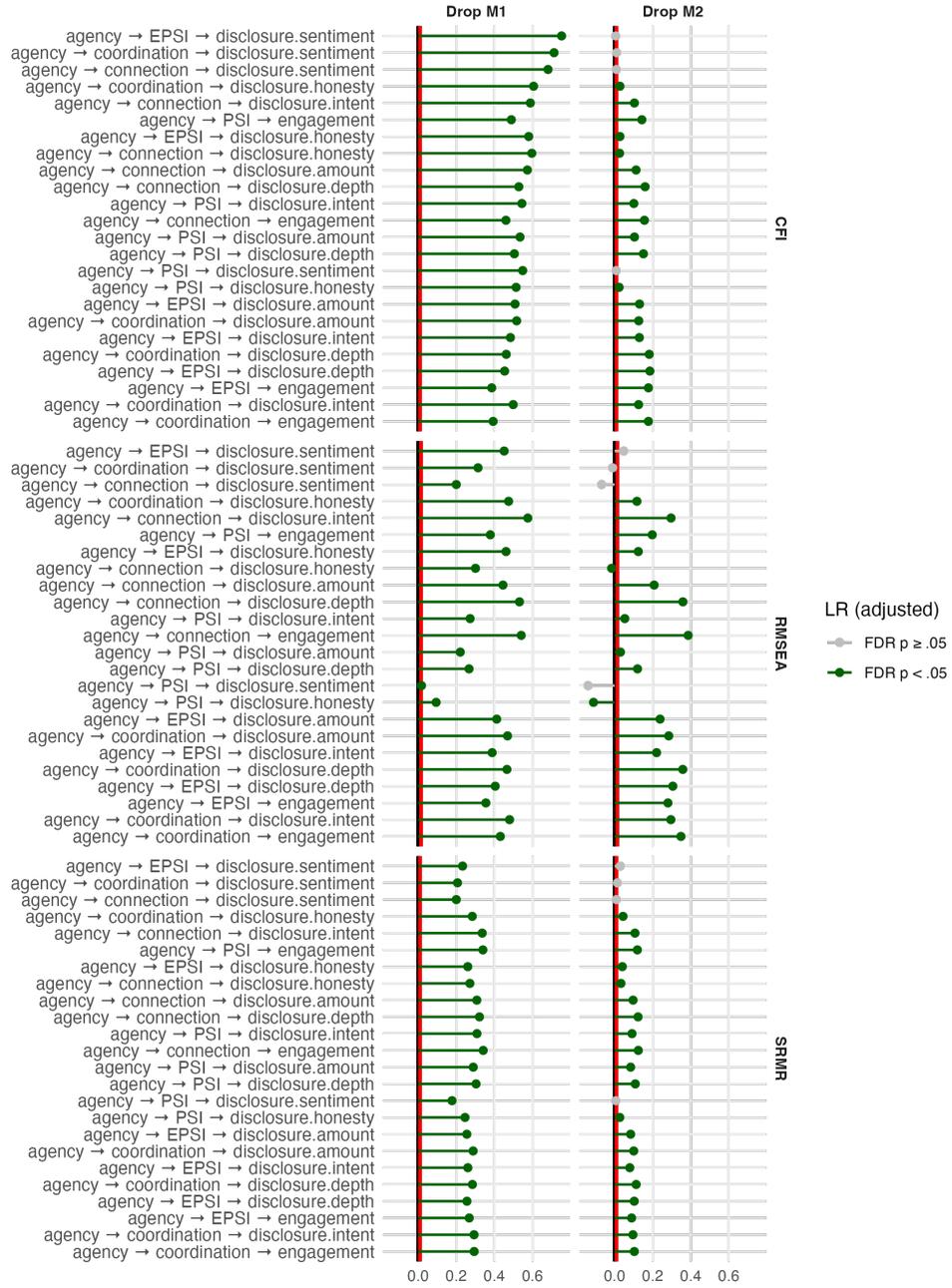

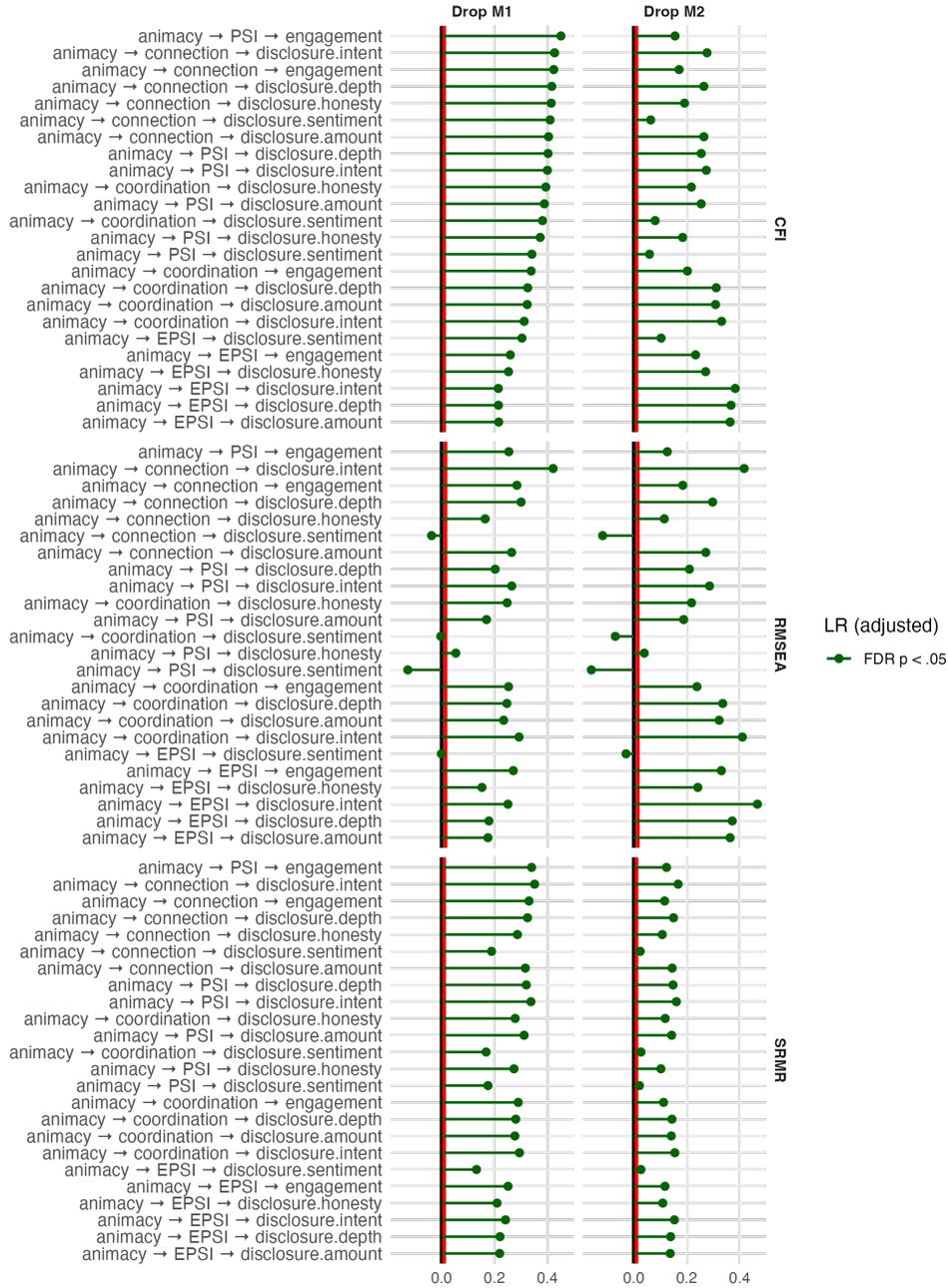

Δ fit: X = animacy, Y = attachment

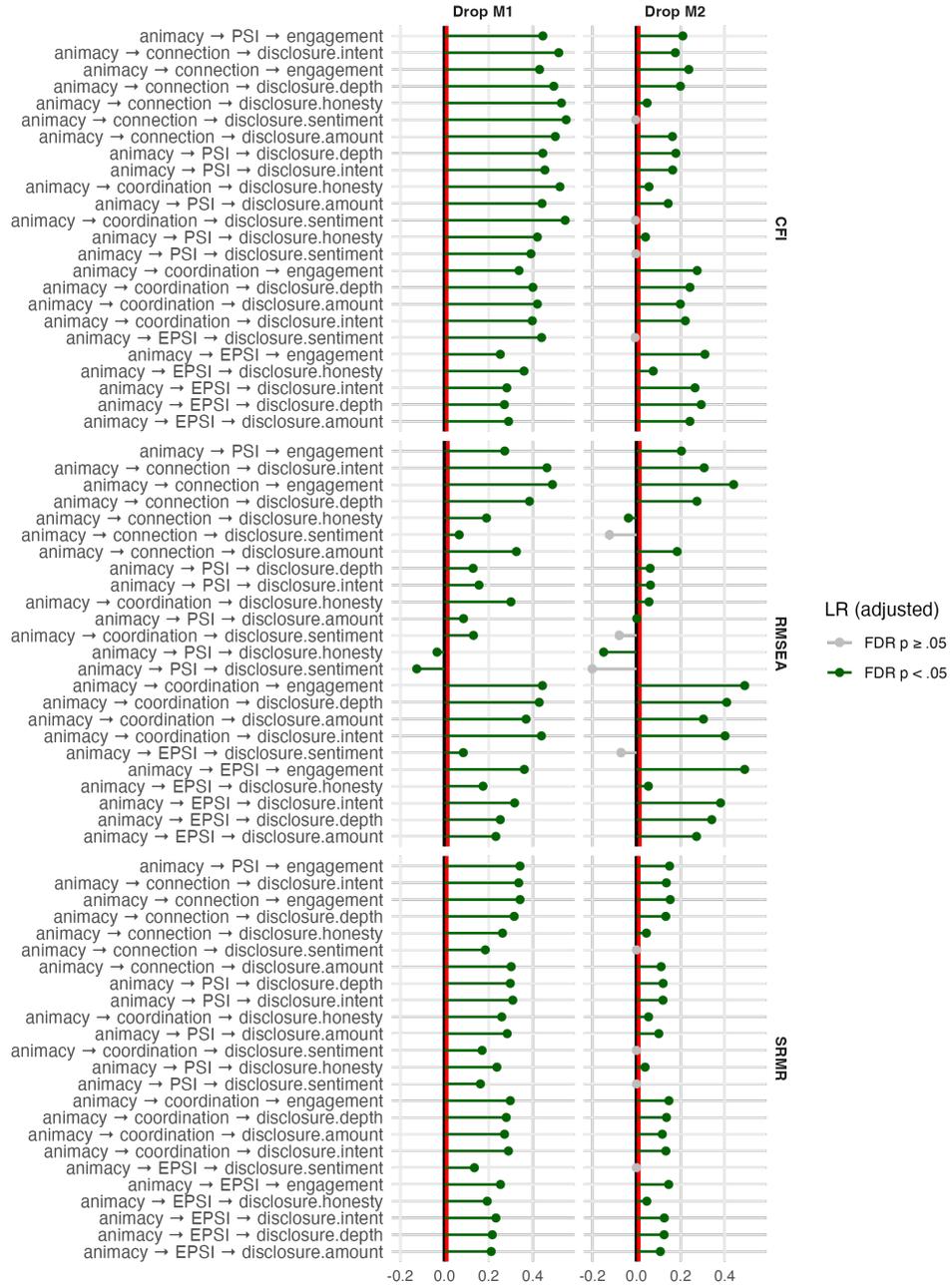

Δ fit: X = animacy, Y = salience

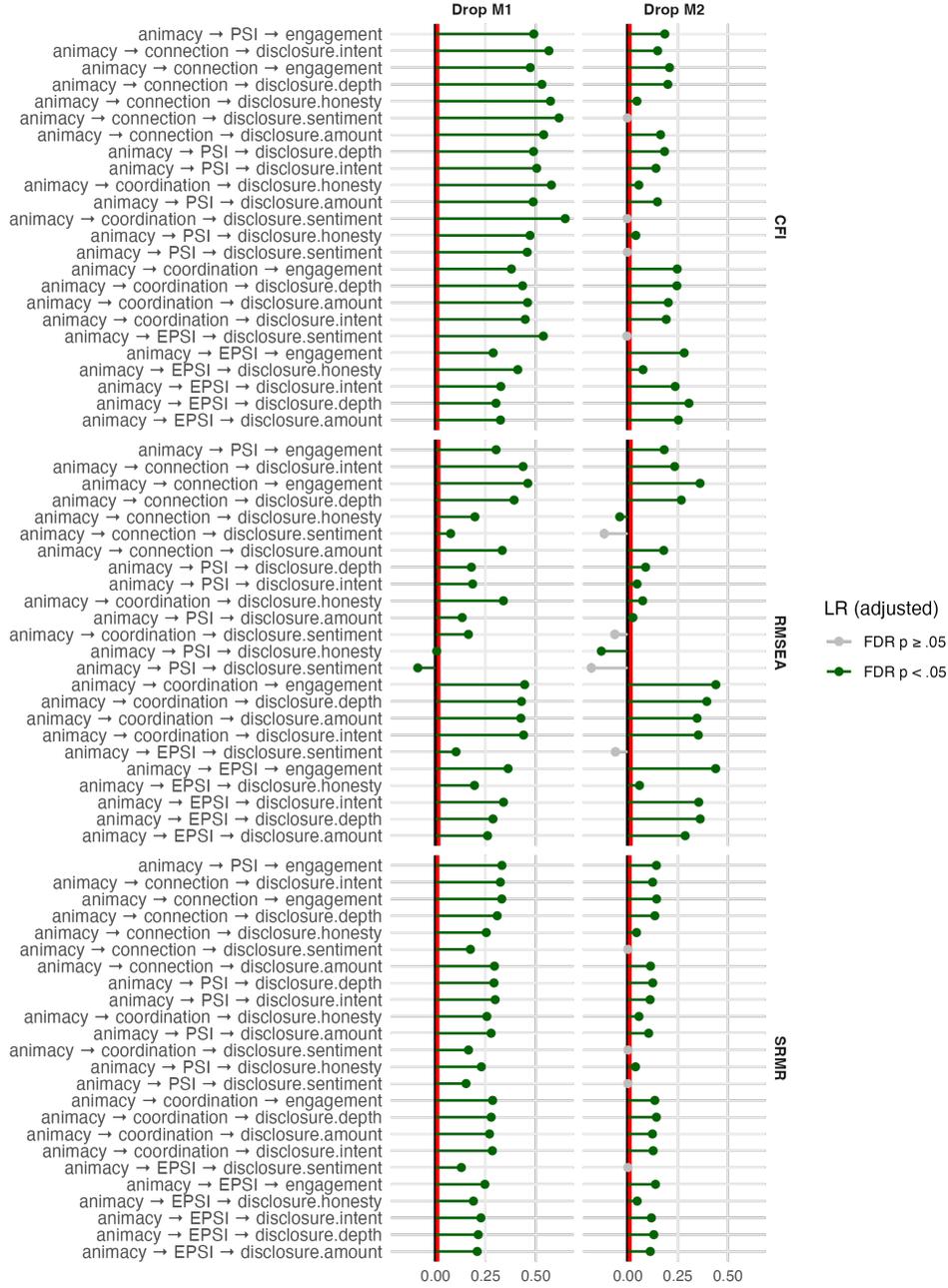

Δ fit: X = animacy, Y = tolerance

Δ fit: X = animacy, Y = withdrawal

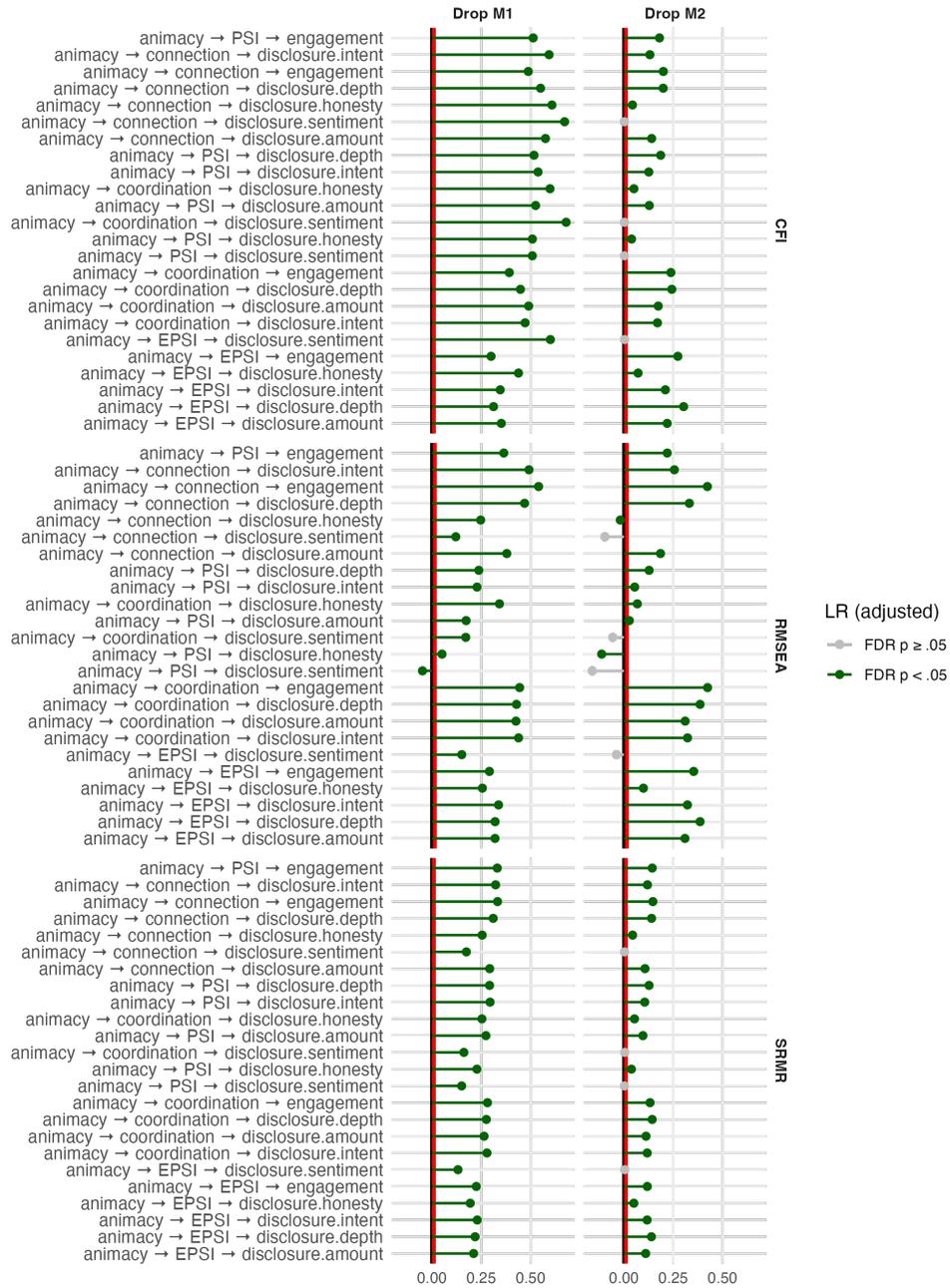

Δ fit: X = anthropomorphism, Y = attachment

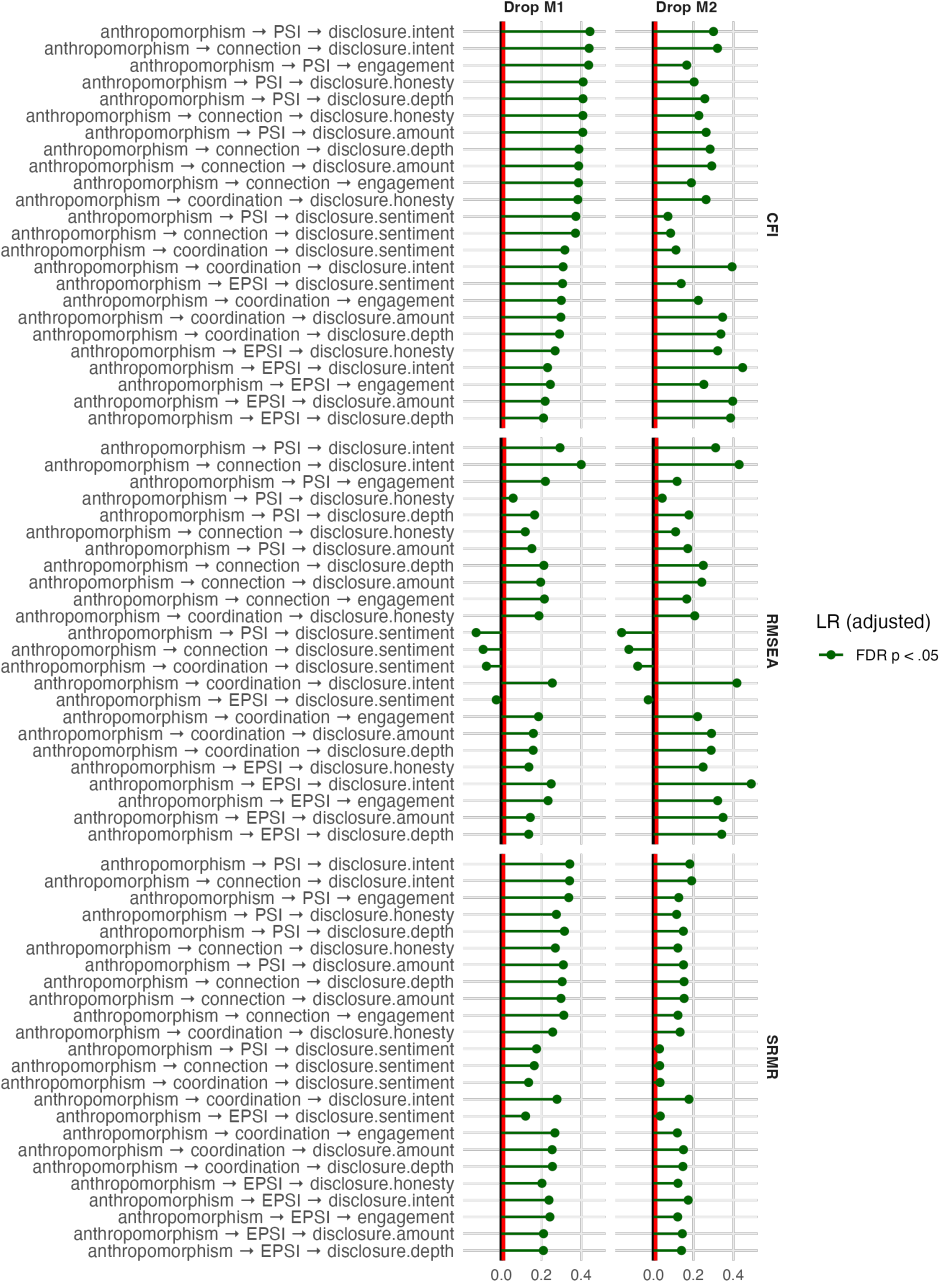

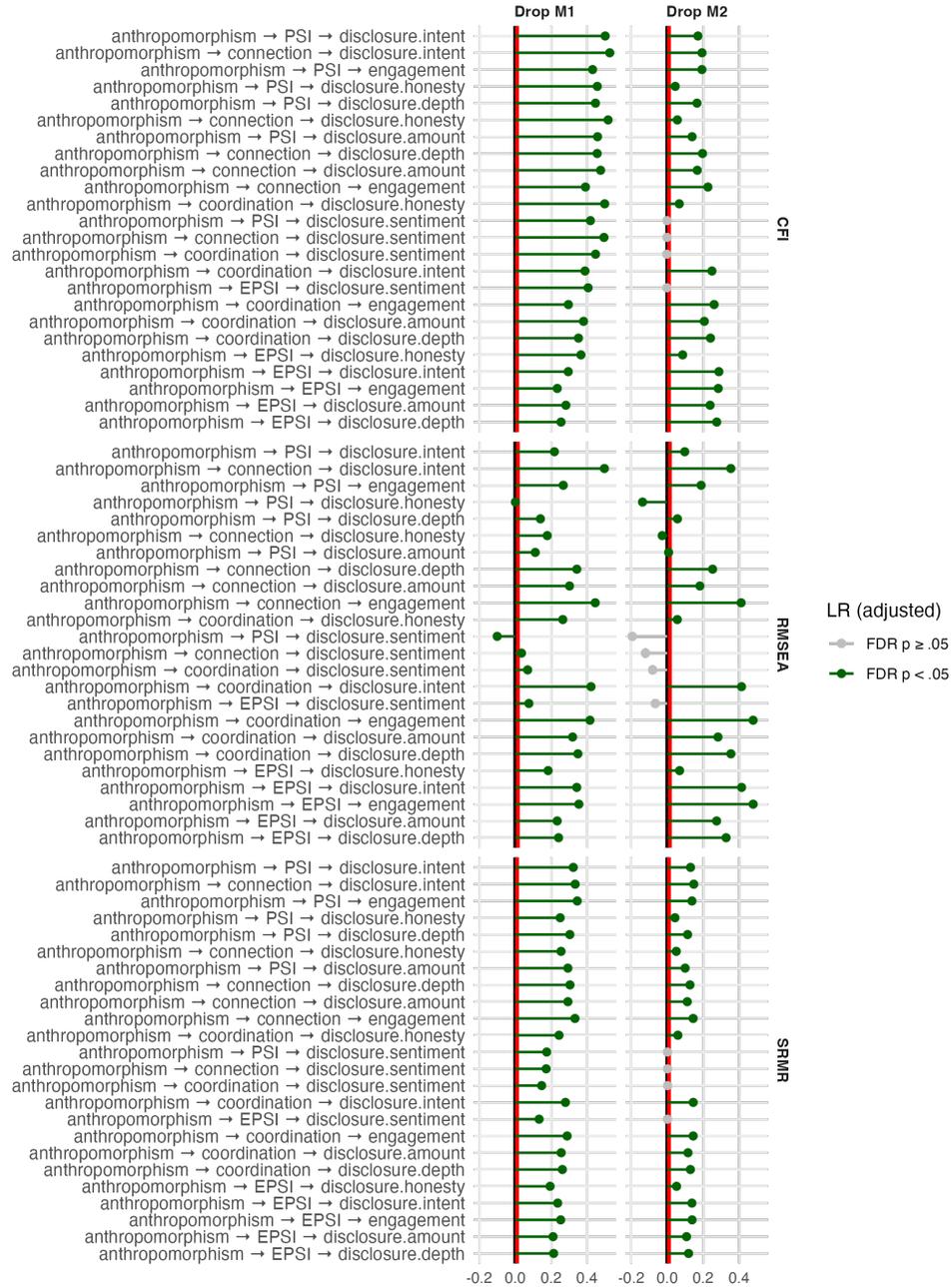

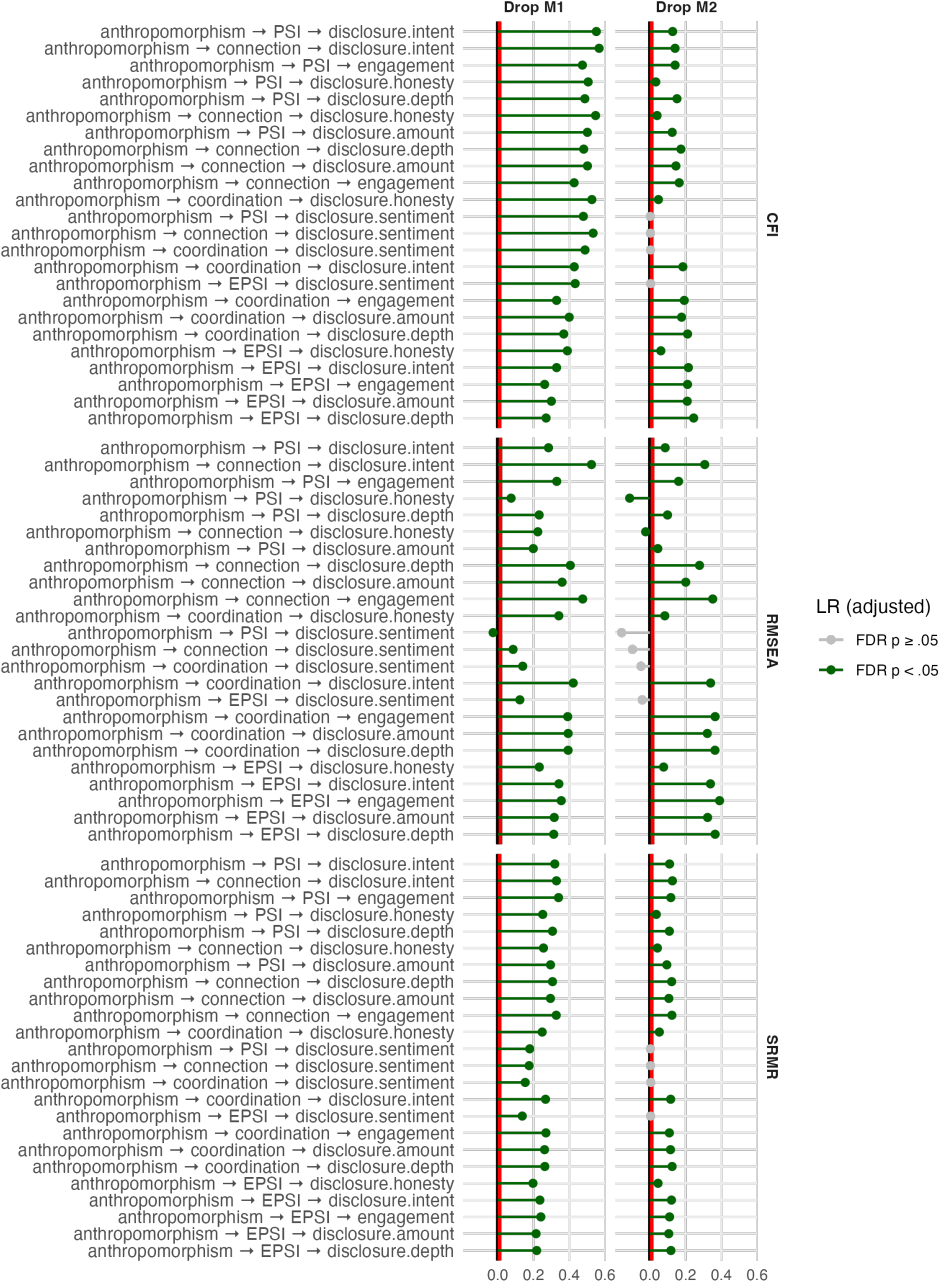

Δ fit: X = anthropomorphism, Y = tolerance

Δ fit: X = anthropomorphism, Y = withdrawal

LR (adjusted)
- FDR p ≥ .05
- FDR p < .05

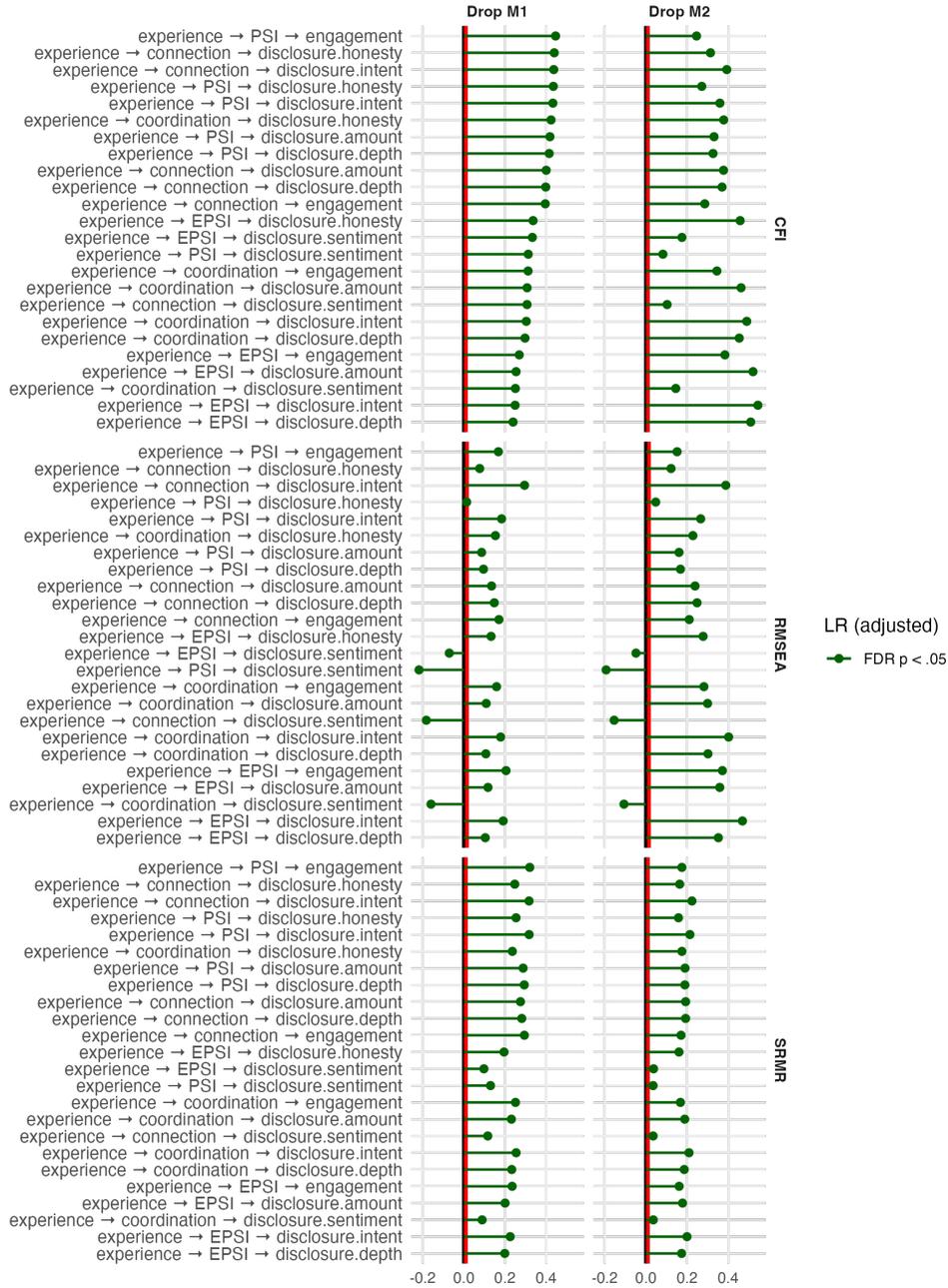

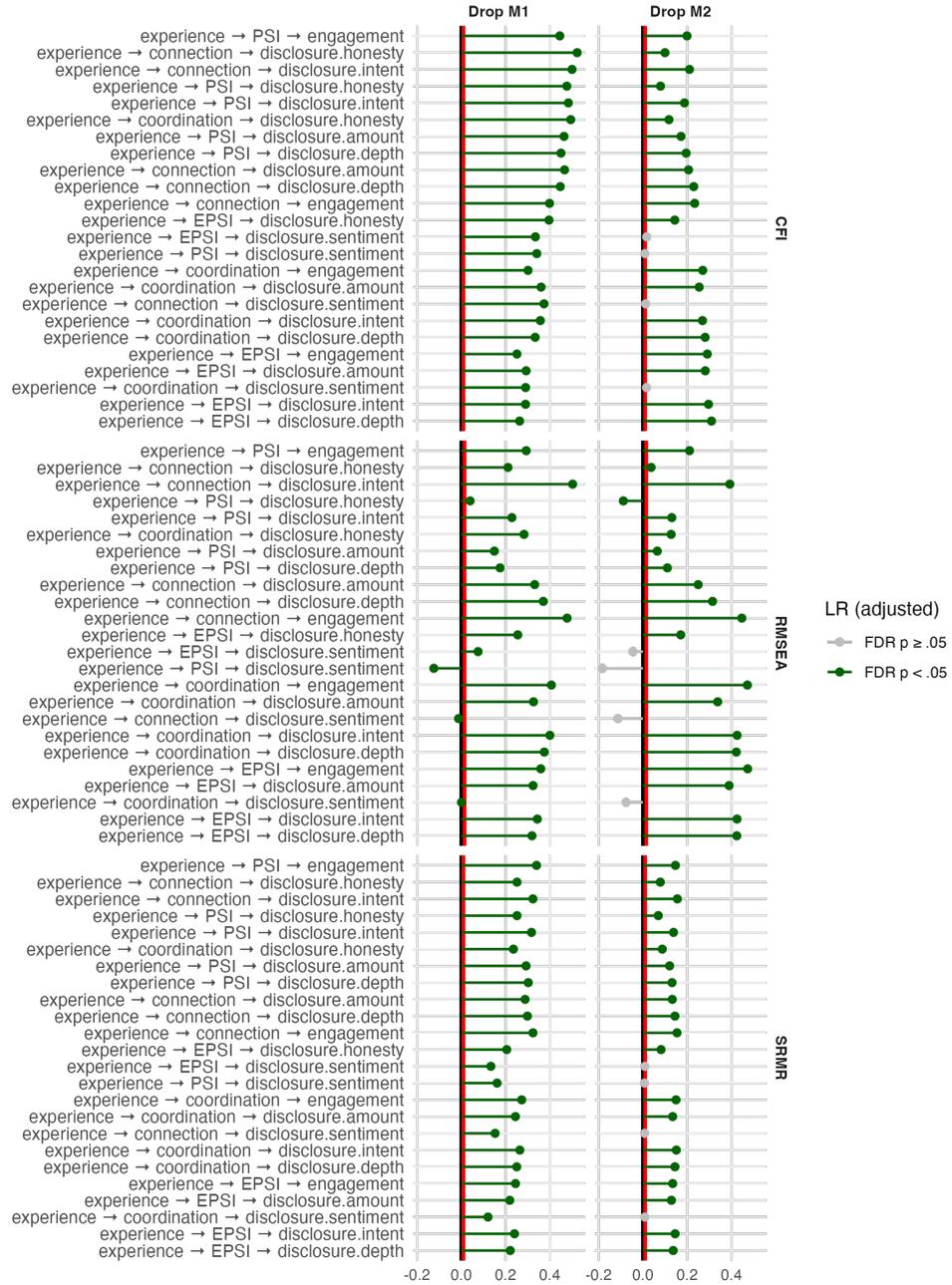

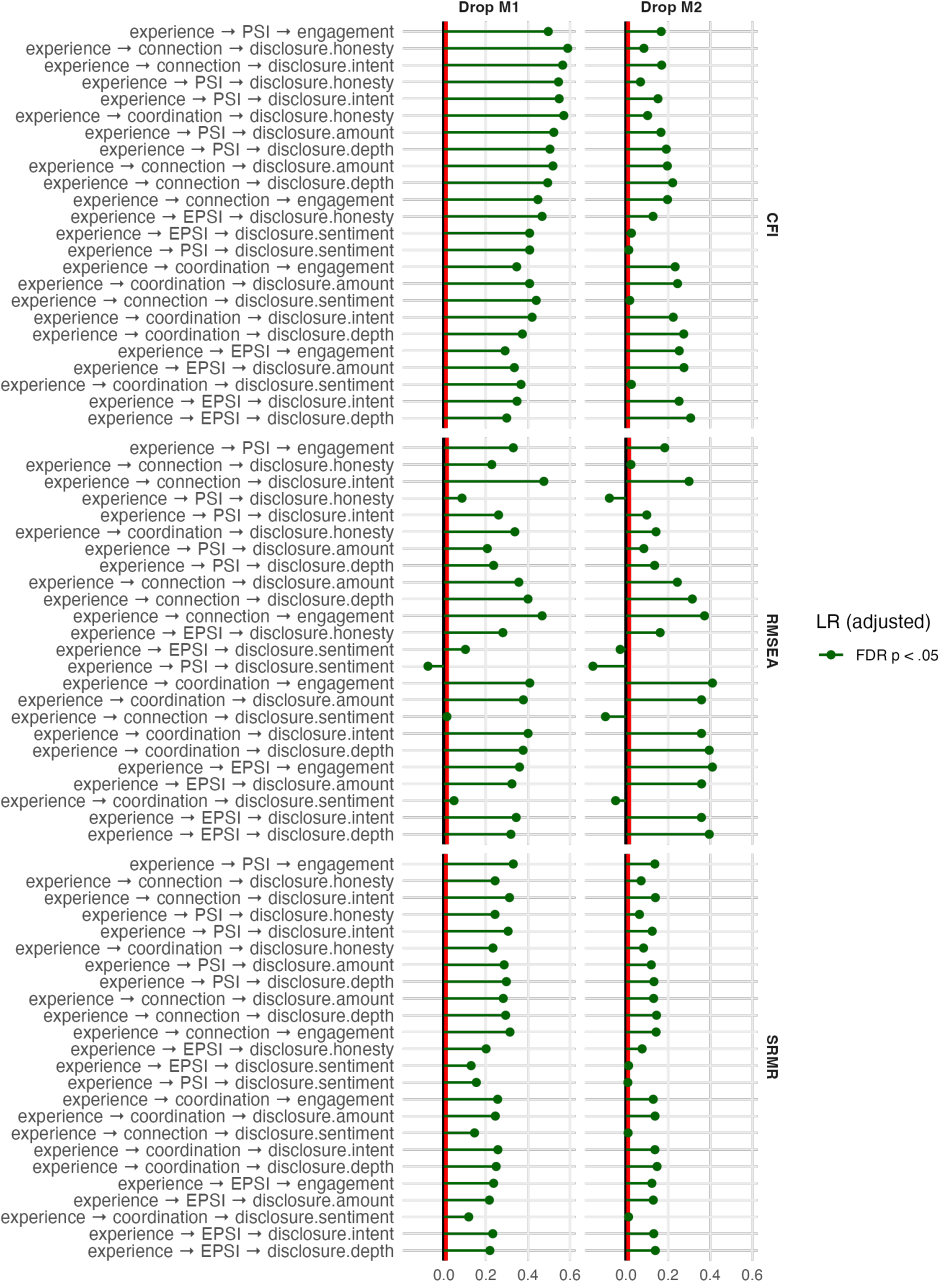

Δ fit: X = experience, Y = tolerance

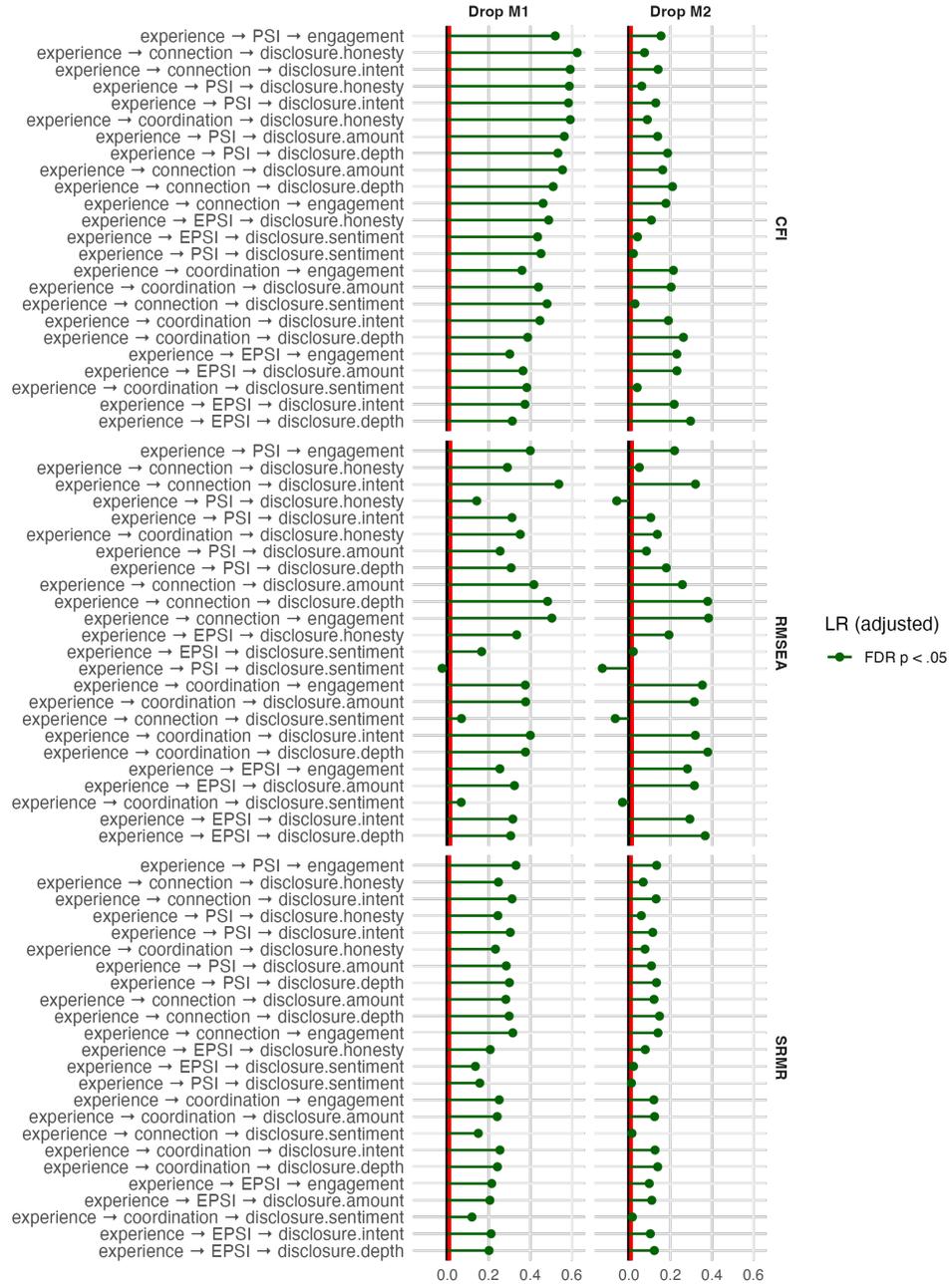

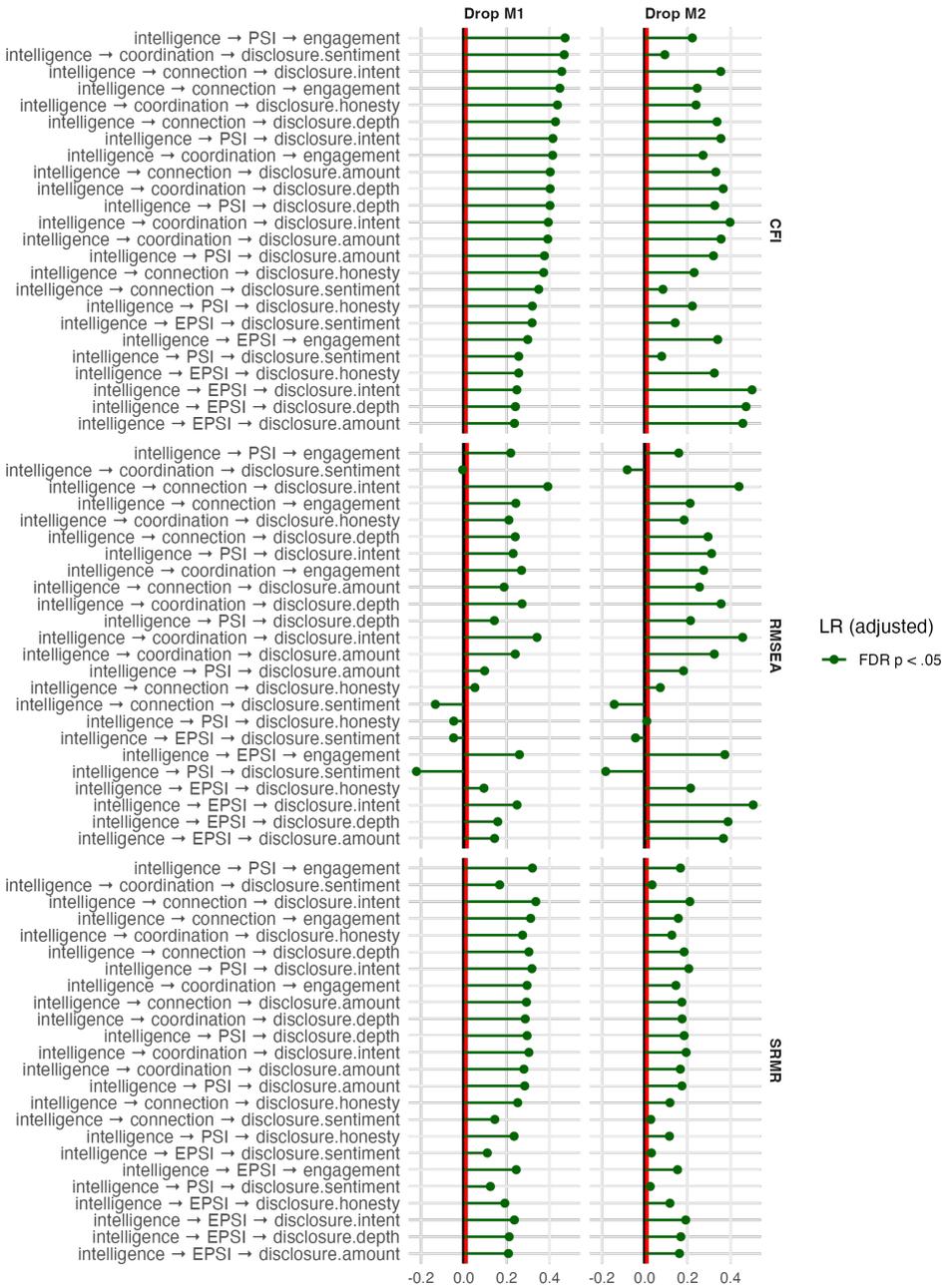

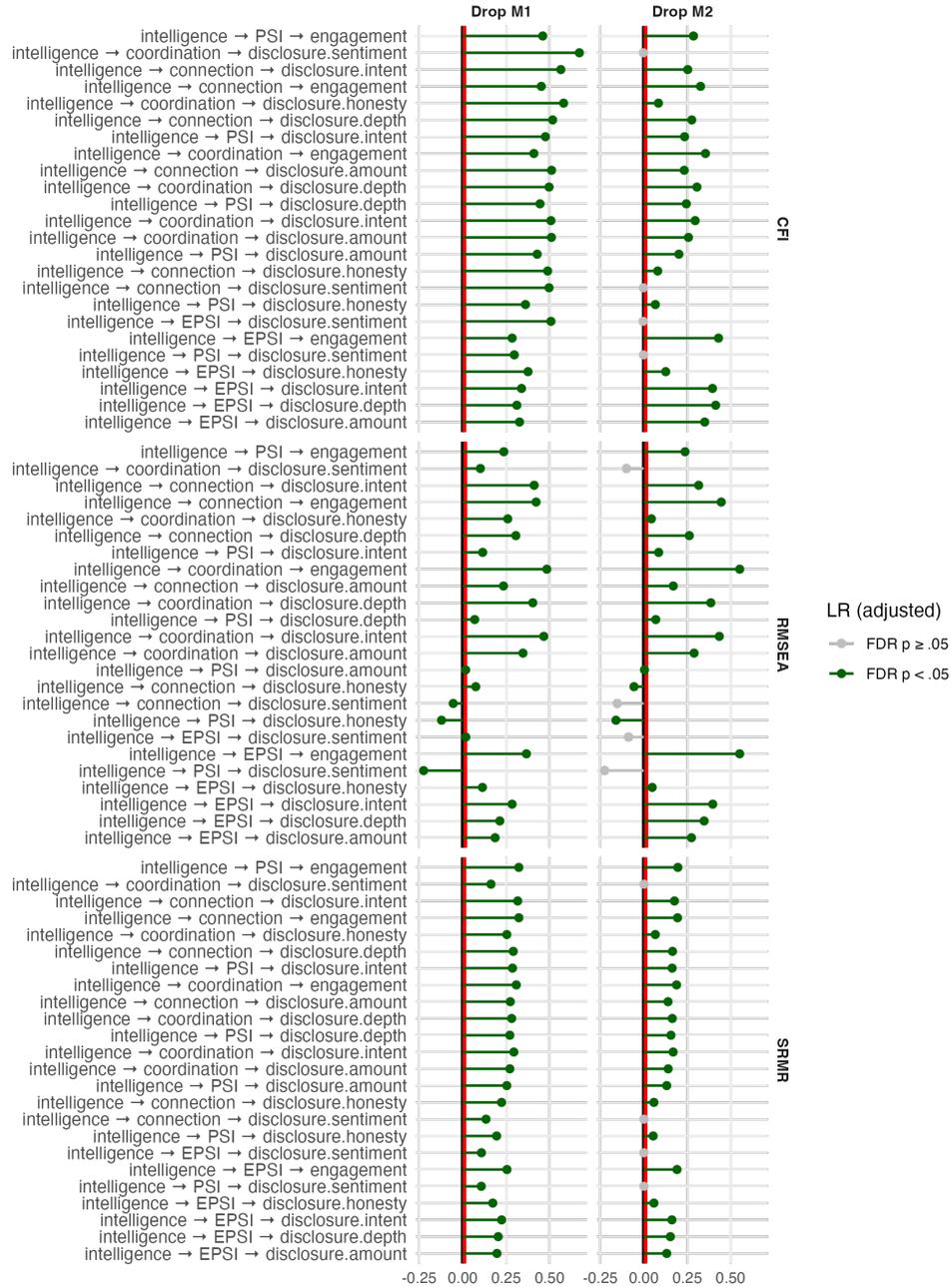

Δ fit: X = intelligence, Y = salience

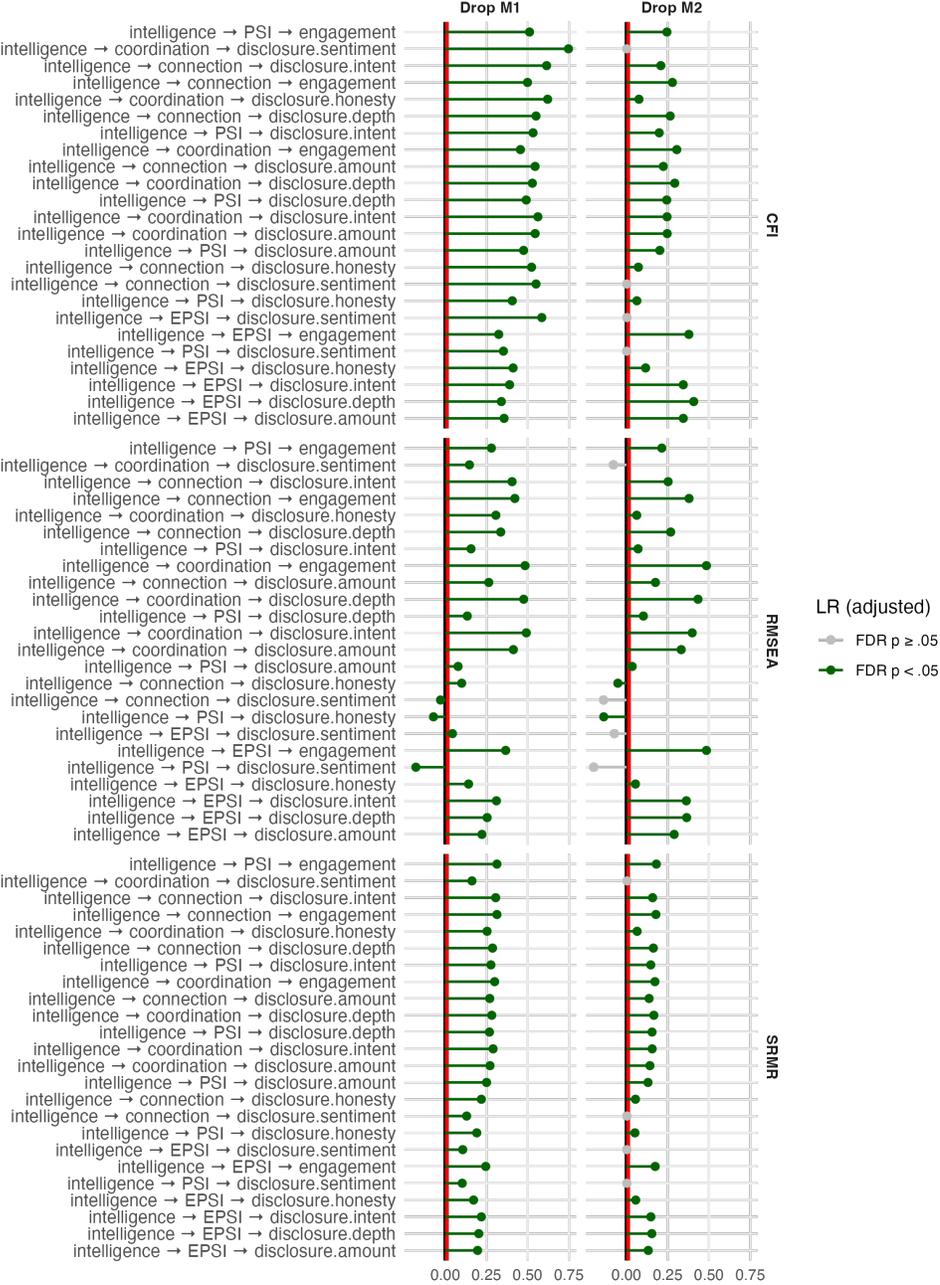

Δ fit: X = intelligence, Y = tolerance

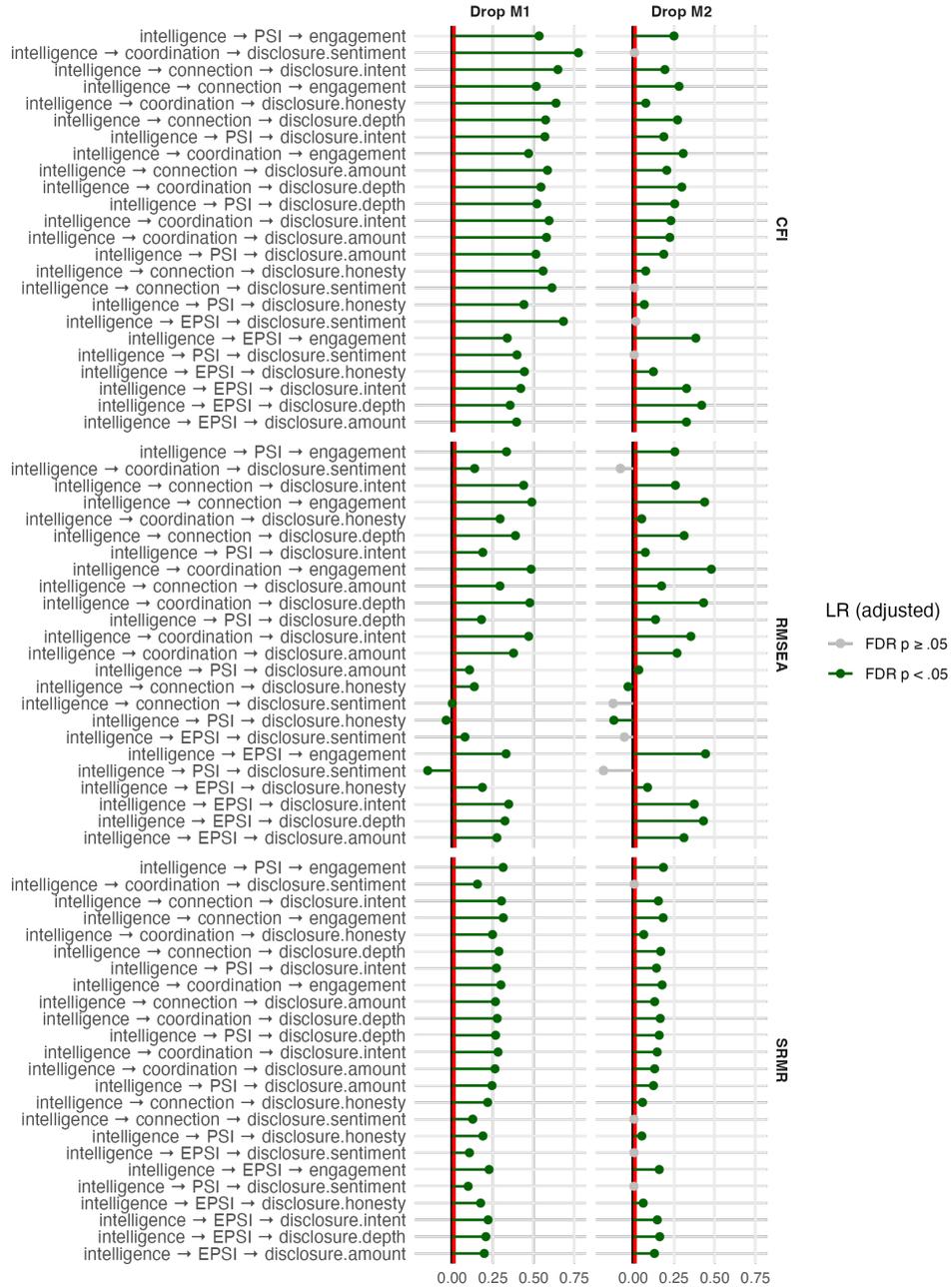

Δ fit: X = personification, Y = attachment

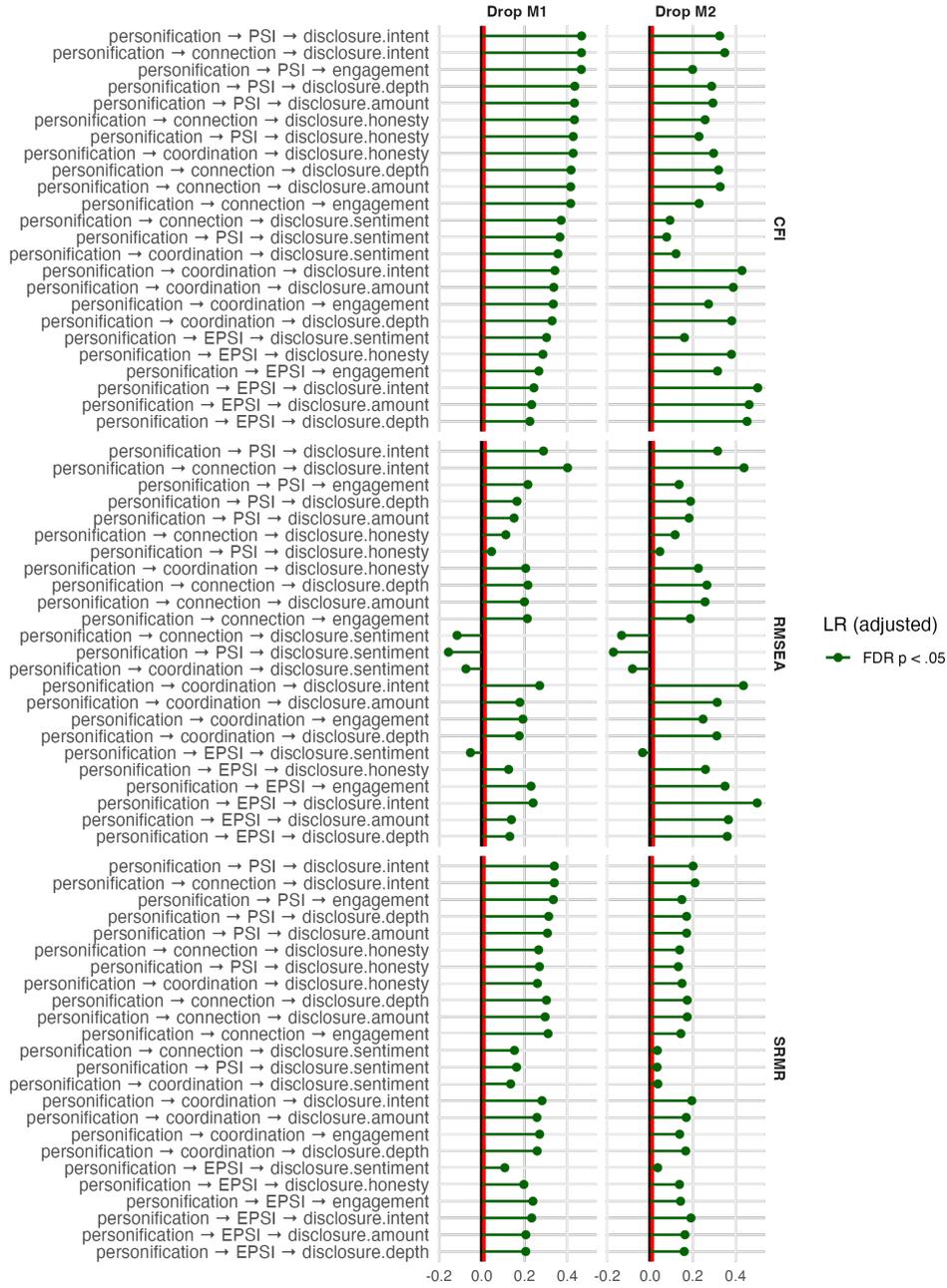

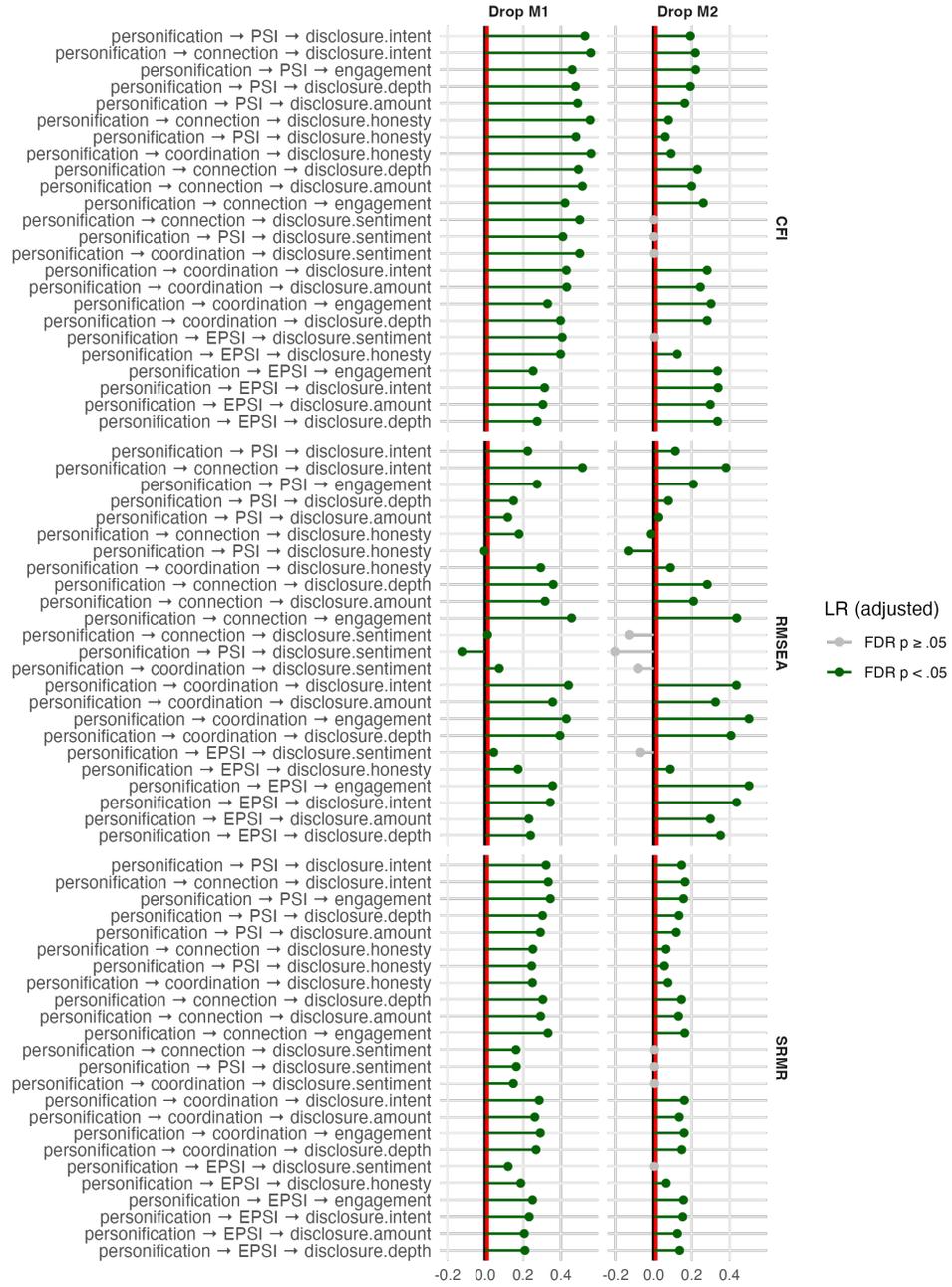

Δ fit: X = personification, Y = salience

Δ fit: X = personification, Y = tolerance

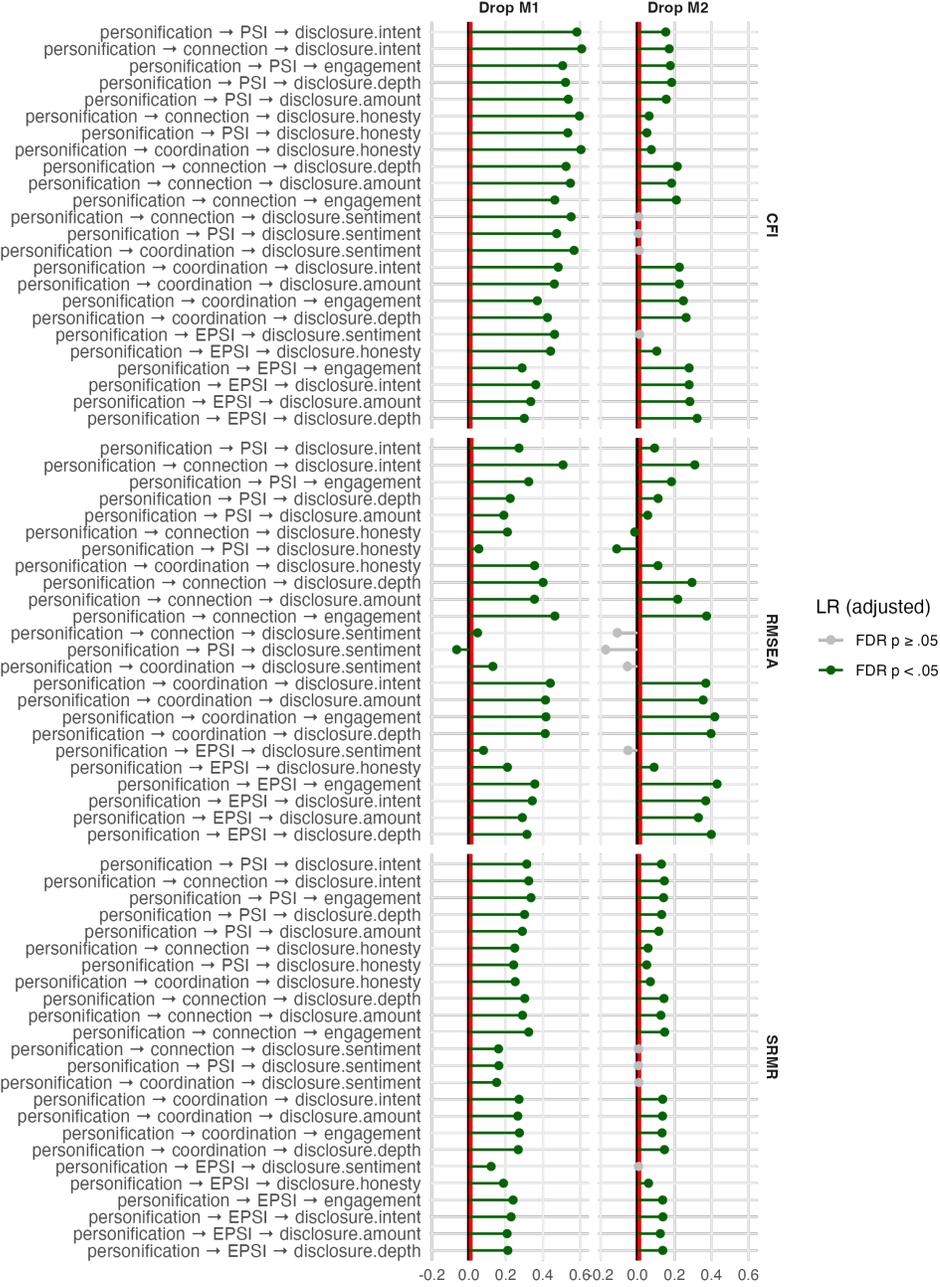

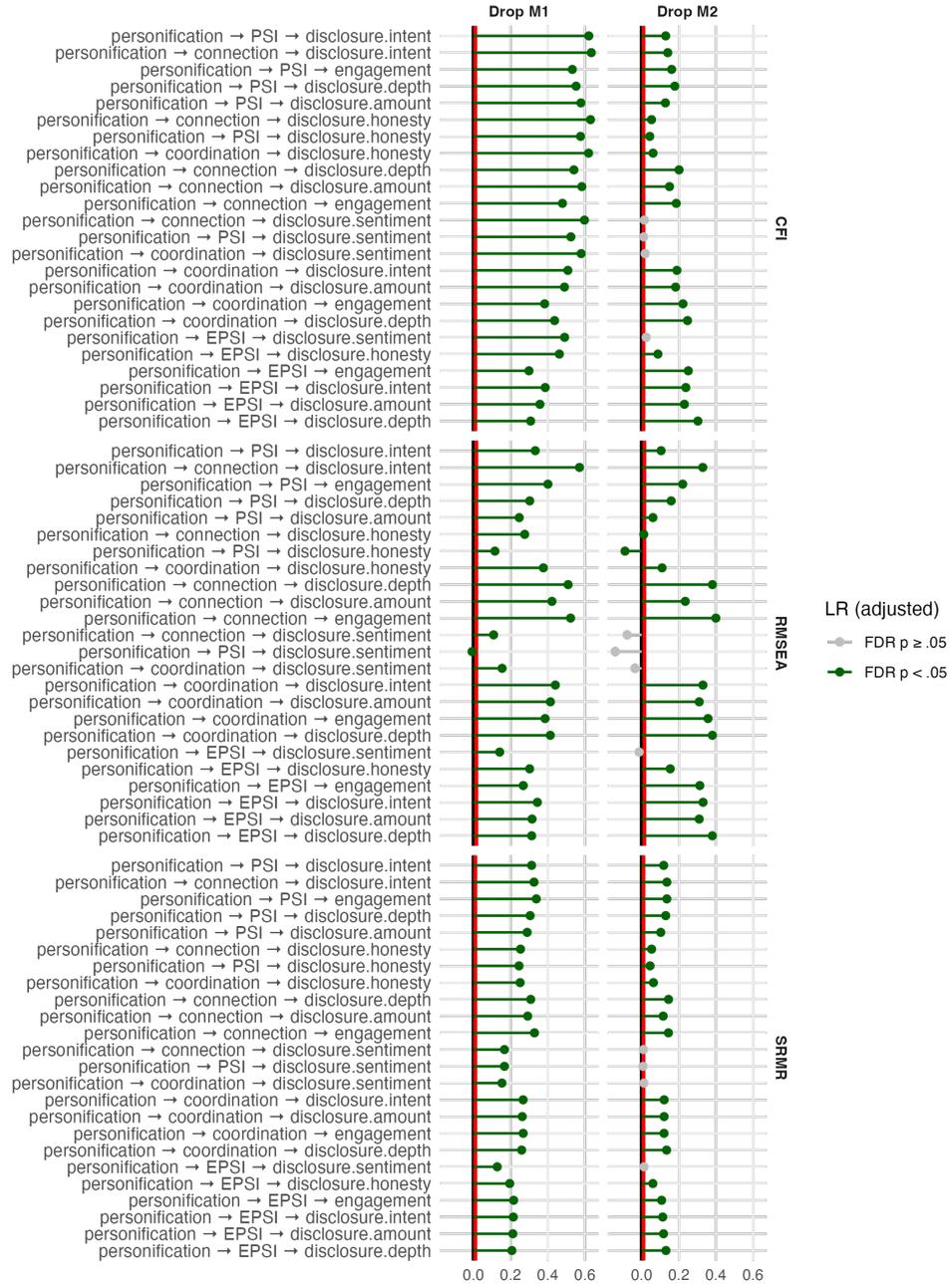

Δ fit: X = personification, Y = withdrawal

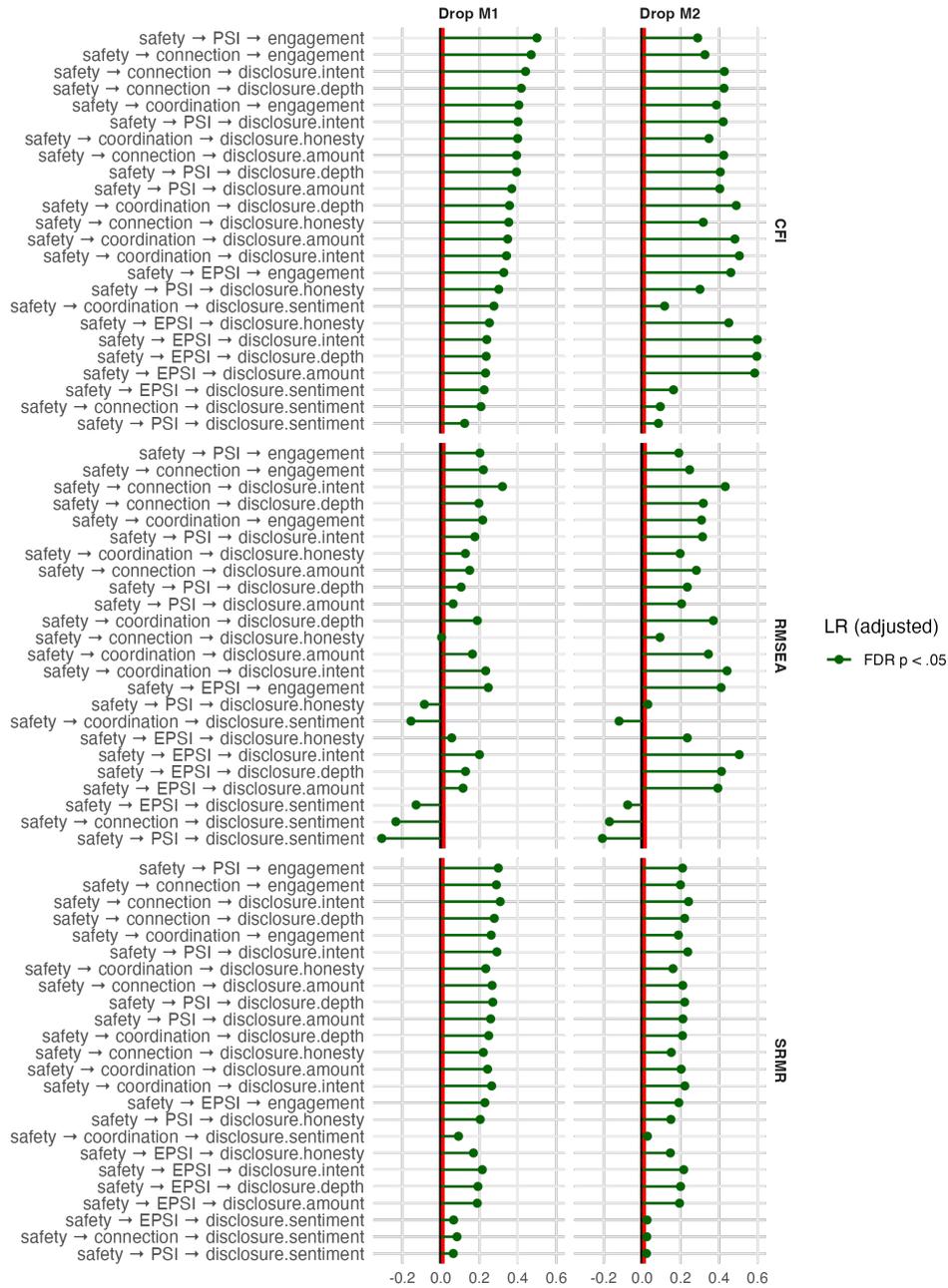

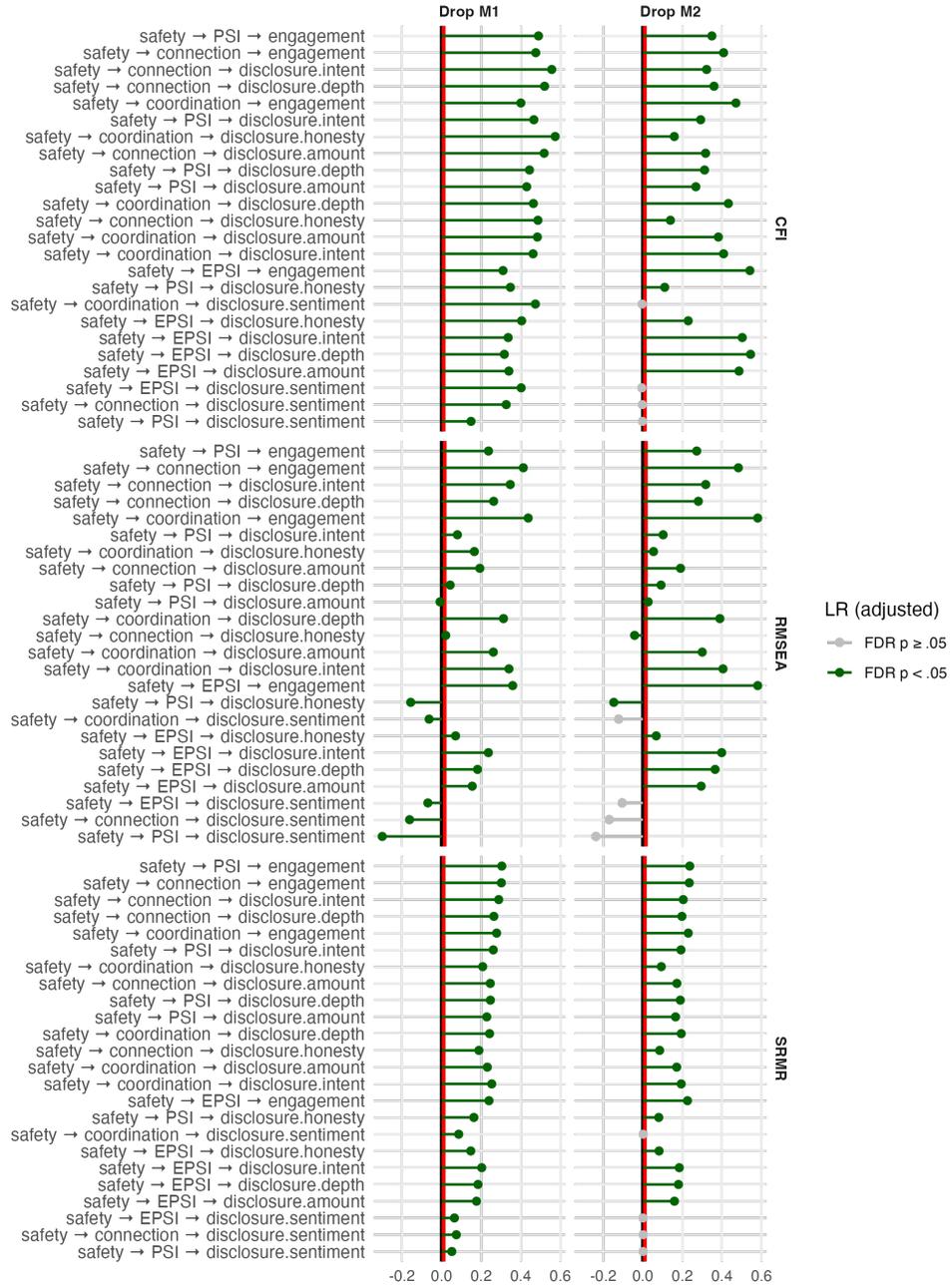

Δ fit: X = safety, Y = tolerance

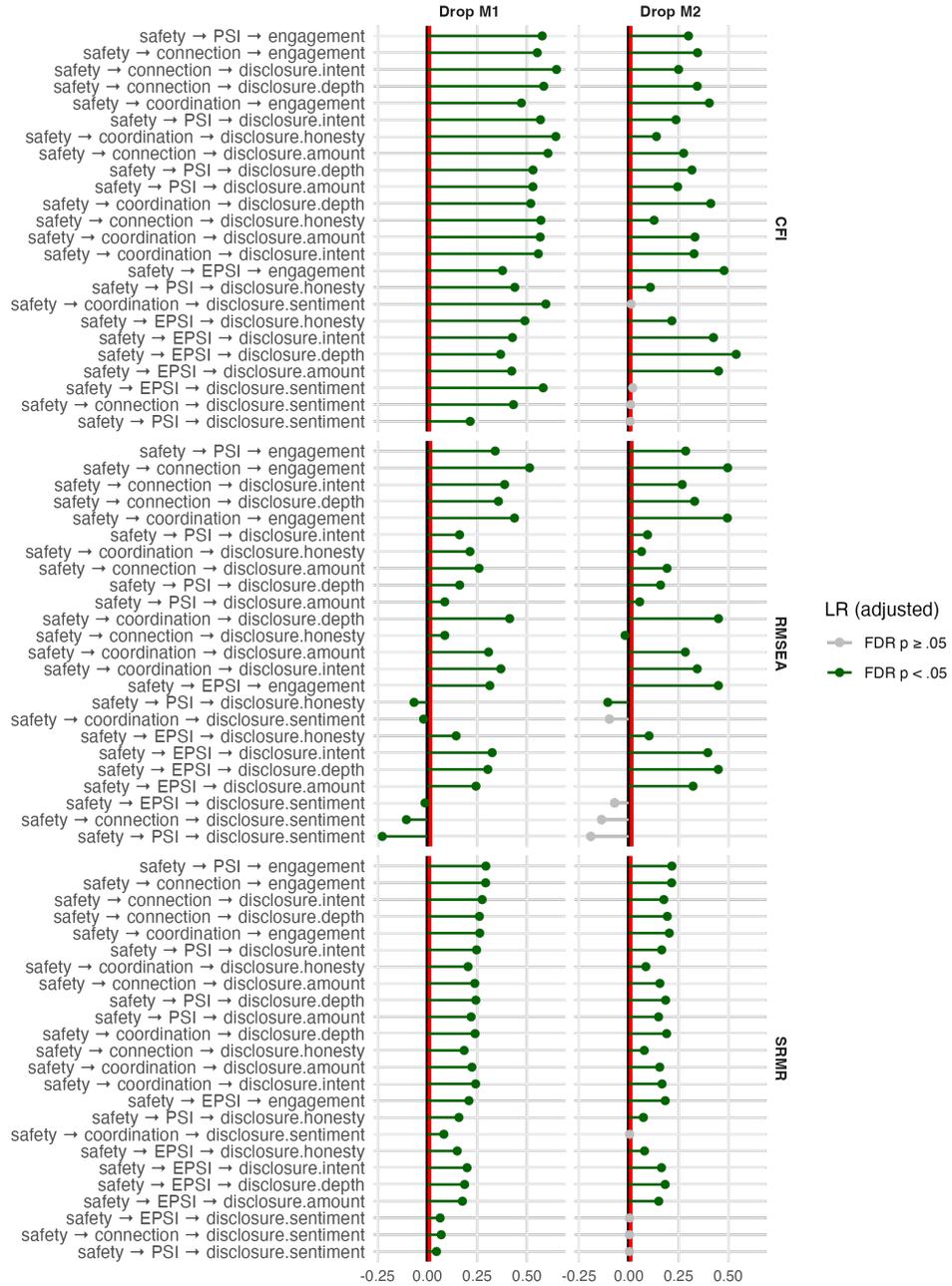